\documentclass{llncs}
\usepackage{amsmath,amsfonts}
\usepackage{xspace}
\usepackage{tabularx}
\usepackage[table]{xcolor}
\usepackage{graphicx}
\usepackage[]{todo}
\usepackage{subfig}
\usepackage{times}
\usepackage{fullpage}
\newcommand{\Props}{AP}
\newcommand{\AP}[1]{\ensuremath{{{{AP}_{#1}}}}}

\newcommand{\true}{\mathrm{true}}
\newcommand{\false}{\mathrm{false}}
\newcommand{\wrt}{w.r.t.\xspace}

\newcommand{\secref}[1]{Sec.~\ref{#1}}

\newcommand{\cL}{\mathcal{L}}

\newcommand{\LTL}{\ensuremath{\mathrm{LTL}}\xspace}

\newcommand{\ltlG}{\mathbf{G}}
\newcommand{\ltlF}{\mathbf{F}}
\newcommand{\ltlU}{\mathbf{U}}
\newcommand{\ltlX}{\mathbf{X}}
\newcommand{\ltlP}{\overline{\mathbf{X}}}

\newenvironment{algo}[2]%
{\par\vspace{1em}\noindent\textbf{Algorithm~#1}~(\emph{#2}).~~}
{}

\DeclareMathOperator{\good}{good}
\DeclareMathOperator{\bad}{bad}
\DeclareMathOperator{\sus}{sus}
\DeclareMathOperator{\ulo}{ulo}
\DeclareMathOperator{\received}{received}
\DeclareMathOperator{\rp}{rp}
\DeclareMathOperator{\lo}{lo}
\DeclareMathOperator{\inlo}{inlo}
\DeclareMathOperator{\kept}{kept}
\DeclareMathOperator{\send}{send}
\DeclareMathOperator{\mou}{mou}
\DeclareMathOperator{\Mon}{Mon}
\DeclareMathOperator{\Prop}{Prop}
\DeclareMathOperator{\merge}{merge}

\newcommand{\toolname}{\textsc{DecentMon}}

\begin{document}
\title{Decentralised LTL monitoring}
\author{Andreas Bauer\inst{1} \and Yli\`{e}s Falcone\inst{2}
\thanks{\scriptsize %
   This author has been supported by an Inria Exploration Grant to visit NICTA, Canberra.}
}

\institute{NICTA
\thanks{\scriptsize NICTA is funded by the Australian
    Government as represented by the Department of Broadband,
    Communications and the Digital Economy and the Australian Research
    Council through the ICT Centre of Excellence
    program.}
 Software Systems Group and Australian National University\\
  \and Laboratoire d'Informatique de Grenoble, UJF Universit\'e Grenoble I, France}

\maketitle
\newcommand{\squishlist}{
 \begin{list}{$\bullet$}
  { \setlength{\itemsep}{0pt}
     \setlength{\parsep}{1pt}
     \setlength{\topsep}{1pt}
     \setlength{\partopsep}{0pt}
     \setlength{\leftmargin}{1.2em}
     \setlength{\labelwidth}{1.5em}
     \setlength{\labelsep}{0.4em} } }
\newcommand{\squishend}{
  \end{list}  }

\pagestyle{plain}
% \vspace{-2em}
\begin{abstract}
  Users wanting to monitor distributed or component-based systems often
  perceive them as monolithic systems which, seen from the outside,
  exhibit a uniform behaviour as opposed to many components displaying
  many local behaviours that together constitute the system's global
  behaviour.  This level of abstraction is often reasonable, hiding
  implementation details from users who may want to specify the system's
  global behaviour in terms of an LTL formula.
  However, the problem that arises then is how such a specification can
  actually be monitored in a distributed system that has no central data
  collection point, where all the components' local behaviours are
  observable.
  In this case, the LTL specification needs to be decomposed into
  sub-formulae which, in turn, need to be distributed amongst the
  components' locally attached monitors, each of which sees only a
  distinct part of the global behaviour.

  The main contribution of this paper is an algorithm for distributing
  and monitoring LTL formulae, such that satisfaction or violation of
  specifications can be detected by local monitors alone.  We present an
  implementation and show that our algorithm introduces only a minimum
  delay in detecting satisfaction/violation of a specification.
  Moreover, our practical results show that the communication overhead
  introduced by the local monitors is considerably lower than the number
  of messages that would need to be sent to a central data collection
  point.
\end{abstract}

% \sloppy
% \vspace{-2.9em}
\section{Introduction}
%
% \vspace{-0.5em}
Much work has been done on monitoring systems \wrt formal specifications
such as linear-time temporal logic (LTL \cite{Pnueli77}) formulae.  For
this purpose, a system is thought of more or less as a ``black box'',
and some (automatically generated) monitor observes its outside visible
behaviour in order to determine whether or not the runtime behaviour satisfies
an LTL formula.  Applications include monitoring programs written in
Java (cf.~\cite{Bodden04,MeredithR10}) or C
(cf.~\cite{Seyster:2010:AIG:1939399.1939433}), monitoring of abstract
Web services (cf.~\cite{HalleV10}), or transactions on typical
e-commerce sites (cf.~\cite{bauer:gore:tiu:ictac09}).

From a system designer's point of view, who defines the overall
behaviour that a system has to adhere to, this ``black box'' view is
perfectly reasonable.
For example, most modern cars have the ability to issue a warning if a
passenger (including the driver) is not wearing a seat belt after the
vehicle has reached a certain speed.  One could imagine using a monitor
to help issue this warning based on the following LTL formalisation,
which captures this abstract requirement:
\[
\begin{array}{lcl}
\varphi = \ltlG \big(\mathit{speed\_low} & \vee & ( (\mathit{pressure\_sensor\_1\_high} \Rightarrow \mathit{seat\_belt\_1\_on})   \\
 & & ~ \wedge\ \ldots \\
 & & ~ \wedge\ (\mathit{pressure\_sensor\_n\_high} \Rightarrow \mathit{seat\_belt\_n\_on}) ) \big)  
\end{array}
\]
The formula $\varphi$ asserts that, at all times, when the car
has reached a certain speed, and the pressure sensor in a seat $i\in[1,n]$ detects
that a person is sitting in it ($\mathit{pressure\_sensor\_i}$ $\mathit{\_high}$), it has to be the case that the
corresponding seat belt is fastened ($\mathit{seat\_belt\_i\_on}$). Moreover, one can build a monitor
for $\varphi$, which receives the respective sensor values and is able to assert whether or not these values constitute a
violation---\emph{but}, only if some central component exists in the
car's network of components, which collects these sensor values and
consecutively sends them to the monitor as input!
In many real-world scenarios, such as the automotive one, this is an
unrealistic assumption mainly for economic reasons, but also because the communication on
a car's bus network has to be kept minimal. Therefore one cannot
continuously send unnecessary sensor information on a bus that is shared
by potentially critical applications where low latency is paramount
(cf.~\cite{DBLP:conf/icse/Broy06}).
In other words, in these scenarios, one has to monitor such a requirement
not based on a single behavioural trace, assumed to be collected by some
global sensor, but based on the many \emph{partial} behavioural traces
of the components which make up the actual system.  We refer to this as
\emph{decentralised LTL monitoring} when the requirement is given in
terms of an LTL formula.

% \vspace{-0.15em}
The main constraint that decentralised LTL monitoring needs to address
is the lack of a global sensor and a central decision
making point asserting whether the system's behaviour has violated or satisfied a
specification. We already pointed out that, from a practical point of
view, a central decision making point (i.e., global sensor) would
require all the individual components to continuously send events over
the network, and thereby negatively affecting the response time for
other potentially critical applications on the network.  Moreover from a
theoretical point of view, a central observer (resp.\ global sensor)
basically resembles the classical LTL monitoring problem, where the
decentralised nature of the system under scrutiny does not play a role.
%

% \vspace{-0.2em}
Arguably, there exist a number of real-world
component-based applications, where the monitoring of an LTL formula can
be realised via global sensors and/or central decision making points,
e.g., when network latency and criticality do not play an important
role. However, here we want to focus on those cases where there exists no global 
trace, no central decision making point, and where the goal is to keep
the communication, required for monitoring the LTL formula, at a
minimum.

% \vspace{-0.2em}
%
% \paragraph{Our setting.}
In the decentralised setting, we assume that the system under scrutiny
consists of a set of $n$ components $\mathcal{C} = \{ C_1, C_2, \ldots,
C_n \}$, communicating on a synchronous bus, each
of which has a local monitor attached to it.
The set of all events is $\Sigma = \Sigma_1 \cup \Sigma_2 \cup \ldots
\cup \Sigma_n$, where $\Sigma_i$ is the set of events visible to the
monitor at component $C_i$.  The global LTL formula, on the other hand,
is specified over a set of propositions, $AP$, such that $\Sigma =
2^{AP}$.  Moreover, we demand for all $i, j \leq n$ with $i \neq j$ that
$\Sigma_i \cap \Sigma_j = \emptyset$ holds, i.e., events are local \wrt
the components where they are monitored.
%; $\Sigma$ is a pair-wise disjoint partition.

At a first glance, the synchronous bus may seem an overly stringent
constraint imposed by our setting.  However, it is by no means
unrealistic, since in many real-world systems, especially
critical ones, communication is synchronous.
For example, the FlexRay bus protocol (cf.\ \cite{Pop:2008:TAF:1348531.1348541}) 
%and its precursor byteflight (cf.\ \cite{byteflight}), both
 used
for safety-critical systems in the automotive domain, allows synchronous
communication.  Similar systems are used in avionics, where synchronous
implementations of control systems have, arguably, played an even
greater role than in the automotive domain due to their deterministic
notion of concurrency and the strong guarantees one can give
concerning their correctness.
%
% \vspace{-1.5em}
\paragraph{Brief overview of the approach.}
Let as before $\varphi$ be an LTL formula formalising a requirement over
the system's global behaviour. Then every local monitor, $M_i$, will at
any time, $t$, monitor its own LTL formula, $\varphi^t_i$, \wrt a
partial behavioural trace, $u_i$.
Let us use $u_i(m)$ to denote the $(m+1)$-th event in a trace $u_i$, and
$\mathbf{u} = (u_1, u_2, \ldots, u_n)$ for the \emph{global trace},
obtained by pair-wise parallel composition of the partial traces, each
of which at time $t$ is of length $t+1$ (i.e., $\mathbf{u} = u_1(0) \cup
u_2(0) \cup \ldots \cup u_n(0) \cdots u_1(t) \cup u_2(t) \cup \ldots
\cup u_n(t)$).
Note that from this point forward we will use $\mathbf{u}$ only when, in
a given context, it is important to consider a global trace.  However,
when the particular type of trace (i.e., partial or global) is
irrelevant, we will simply use $u, u_i$, etc.  We also shall refer to
partial traces as local traces due to their locality to a particular
monitor in the system.

% The relationship between $\varphi^t_i$ and $\varphi$ can now be
% described in terms of our monitoring algorithm's main properties.
%
The decentralised monitoring algorithm evaluates the global trace
$\mathbf{u}$ by considering the locally observed traces $u_i, i\in
[1,n]$ in separation.
In particular, it exhibits the following properties.
\squishlist %\begin{itemize}
\item If a local monitor yields $\varphi^t_i = \bot$ (resp. $\varphi^t_i =
\top$) on some component $C_i$ by observing $u_i$, it implies that
$\mathbf{u}\Sigma^\omega\subseteq\Sigma^\omega\setminus\cL(\varphi)$
(resp. $\mathbf{u}\Sigma^\omega\subseteq\cL(\varphi)$) holds where
$\cL(\varphi)$ is the set of infinite sequences in $\Sigma^\omega$
described by $\varphi$. That is, a locally observed violation
(resp. satisfaction) is, in fact, a global violation
(resp. satisfaction).  Or, in other words, $\mathbf{u}$ is a bad
(resp. good) prefix for $\varphi$.
\item If 
  the monitored trace
  $\mathbf{u}$ is such that
  $\mathbf{u}\Sigma^\omega\subseteq\Sigma^\omega\setminus\cL(\varphi)$
  (resp. $\mathbf{u}\Sigma^\omega\subseteq\cL(\varphi)$), one of the
  local monitors on some
  component $C_i$ yields $\varphi^{t'}_i = \bot$ (resp. $\varphi^{t'}_i
  = \top$), $t'\geq t$, for an observation $u_i'$, an extension of
  $u_i$, the local observation of $\mathbf{u}$ on $C_i$, because of some
  latency induced by decentralised monitoring, as we shall see.
  \squishend %\end{itemize}
However, in order to allow for the local detection of global violations
(and satisfactions), monitors must be able to communicate, since their
traces are only partial \wrt the global behaviour of the system.
Therefore, our second important objective is to also monitor with \emph{minimal communication
overhead} (in comparison with a centralised solution where at any time,
$t$, all $n$ monitors send the observed events to a central decision
making point).
%
% \vspace{-0.9em}
\paragraph{Outline.} 
%The paper is structured as follows.
Section~\ref{sec:prelim} introduces basic notions and
notation.
LTL monitoring by means of formula rewriting (progression), a central
concept to our paper, is discussed in \secref{sec:prog}.
In \secref{sec:dprog}, we lift this concept to the
decentralised setting. %outlined above.
The semantics induced by decentralised LTL monitoring is
outlined in \secref{sec:sem}, whereas \secref{sec:alg} details on how
the local monitors operate in this setting and gives a concrete
algorithm for this purpose.
%
%We also implemented this algorithm.
Experimental results, showing the
feasibility of our approach, are presented in \secref{sec:impl}. 
Section~\ref{sec:conc} concludes and gives pointers to some related
approaches. 
The proofs for all results claimed in this paper are in Appendix~\ref{sec:proofs}.
% For the sake of readability, the proofs for the results of this paper are in Appendix~\ref{sec:proofs}.
%%% Local Variables: 
%%% mode: latex
%%% TeX-master: "main"
%%% End: 

% \vspace{-1em}
\section{Preliminaries}
\label{sec:prelim}
% \vspace{-0.6em}
%
\paragraph{The considered architecture.}
Each component of the system emits events at discrete time instances.
An event $\sigma$ is a set of \emph{actions} denoted by some atomic
propositions from the set $\mathit{AP}$, i.e., $\sigma \in 2^{\mathit{AP}}$. 
We denote $2^{\mathit{AP}}$ by $\Sigma$ and call it 
the \emph{alphabet} (of system events).

As our system operates under the \emph{perfect synchrony hypothesis}
(cf.~\cite{Jantsch:2003:MES:861648}), we assume that its components
communicate with each other in terms of sending and receiving messages
(which, for the purpose of easier presentation, can also be encoded by
actions) at \emph{discrete} instances of time, which are represented using
identifier $t \in \mathbb{N}^{\geq0}$.  Under this hypothesis, it is assumed
that neither computation nor communication take time.  In other words,
at each time $t$, a component may receive up to $n-1$ messages and dispatch up to 1
message, which in the latter case will always be available at the
respective recipient of the messages at time $t+1$.
Note that these assumptions extend to the components' monitors, which
operate and communicate on the same synchronous bus.  The hypothesis of
perfect synchrony essentially abstracts away implementation details of
how long it takes for components or monitors to generate, send, or
receive messages.  As indicated in the introduction, this is a common
hypothesis for certain types of systems, which can be designed and
configured (e.g., by choosing an appropriate duration between time $t$
and $t+1$) to not violate this hypothesis
(cf.~\cite{Jantsch:2003:MES:861648}).

We use a projection function $\Pi_i$ to restrict atomic propositions or events to the local view of
monitor $M_i$, which can only observe those of component $C_i$. For atomic propositions, $\Pi_i: 2^{\mathit{AP}} \rightarrow
2^{\mathit{AP}}$ and we note $\AP{i}=\Pi_i(\AP{})$ for $i\in [1,n]$. For events, $\Pi_i: 2^{\Sigma}\rightarrow
2^\Sigma$ and we note $\Sigma_i = \Pi_i(\Sigma)$, for $i\in [1,n]$. We also assume $\forall i,j \leq n.\ i \neq j \Rightarrow \AP{i}\cap\AP{j}=\emptyset$ and consequently $\forall i,j \leq n.\ i \neq j \Rightarrow \Sigma_i \cap \Sigma_j =
\emptyset$.
Seen over time, each component $C_i$ produces a \emph{trace} of events,
also called its \emph{behaviour}, which for $t$ time steps is encoded as
$u_i = u_i(0)\cdot u_i(1) \cdots u_i(t-1)$ with $\forall t'<t.\ u_i(t') \in \Sigma_i$. Finite traces
over an alphabet $\Sigma$ are elements of the set $\Sigma^\ast$ and
are typically encoded by $u, u', \ldots$, whereas infinite traces
over $\Sigma$ are elements of the set $\Sigma^\omega$ and are
typically encoded by $w, w', \ldots$ The set of all traces is given by
the set $\Sigma^\infty = \Sigma^\ast \cup \Sigma^\omega$. The set $\Sigma^\ast\setminus\{\epsilon\}$ is noted $\Sigma^+$. The finite or infinite sequence $w^t$ is the
\emph{suffix} of the trace $w \in \Sigma^\infty$, starting at time $t$, i.e., $w^t = w(t) \cdot w(t+1) \cdots$.
The system's global behaviour, $\mathbf{u} = (u_1, u_2, \ldots, u_n)$
can now be described as a sequence of pair-wise union of the local events in
component's traces, each of which at time $t$ is of length $t + 1$
i.e., $\mathbf{u} = u(0) \cdots  u(t)$.
%
% \vspace{-1em}
\paragraph{Linear Temporal Logic (LTL).}
We monitor a system \wrt a global specification, expressed as an LTL~\cite{Pnueli77} formula, that does not
state anything about its distribution or the system's architecture. 
Formulae of LTL can be described using the following grammar:
$
\varphi ::= p \mid (\varphi) \mid \neg \varphi \mid \varphi \vee \varphi \mid \ltlX\varphi \mid \varphi \ltlU \varphi,
$
where $p \in \mathit{AP}$. Additionally, we allow the following operators, each
of which is defined in terms of the above ones: $\top = p \vee \neg p$,
$\bot = \neg \top$, $\varphi_1 \wedge \varphi_2 = \neg(\neg \varphi_1
\vee \neg \varphi_2)$, $\ltlF\varphi = \top \ltlU \varphi$, and
$\ltlG\varphi = \neg\ltlF(\neg \varphi)$.  The operators typeset in
bold are the temporal operators.  Formulae without temporal operators
are called \emph{state formulae}.
We describe the set of all LTL formulae over $\mathit{AP}$ by the set $\LTL(\mathit{AP})$,
or just $\LTL$ when the set of atomic propositions is clear from the
context or does not matter.
\begin{table}[t]
% \vspace{-3em}
\caption{LTL semantics over infinite traces}
\[
\begin{array}{lcl}
  w^i \models p                          & \Leftrightarrow & p \in w(i), \text{ for any } p\in \mathit{AP}\\
  w^i \models \neg \varphi               & \Leftrightarrow & w^i \not\models \varphi\\
  w^i \models \varphi_1 \vee \varphi_2    & \Leftrightarrow &  w^i \models \varphi_1 \vee w^i \models \varphi_2\\
  w^i \models \ltlX \varphi              & \Leftrightarrow & w^{i+1} \models \varphi\\
  w^i \models \varphi_1 \ltlU \varphi_2   & \Leftrightarrow &
   \exists{k \in [i, \infty[}.\ w^k \models \varphi_2 \wedge \forall{l \in [i, k[}.\ w^l \models \varphi_1\\
   % \exists{k\geq i}.\ w^k \models \varphi_2 \wedge \forall{i \leq l < k}.\ w^l \models \varphi_1\\
\end{array} 
\]
\label{tab:ltlsemantics}
% \vspace{-3em}
\end{table}
%
% According to \cite{Pnueli77}, the semantics of LTL formulae is defined
% \wrt infinite traces:
The semantics of LTL~\cite{Pnueli77} is defined
\wrt infinite traces:
%
% \vspace{-0.5em}
\begin{definition}
\label{def:ltl_semantics}
Let $w \in \Sigma^\omega$ %be an infinite trace 
and $i \in
\mathbb{N}^{\geq 0}$.  Satisfaction of an \LTL formula by
$w$ at time $i$ is inductively defined as given in
Table~\ref{tab:ltlsemantics}.
%\TODO{AB@YF: We need to use ``your set   notation'' for integer intervals here, too!\\
% YF@AB: do you mean that the second part of the until case should be rewritten like that  $\forall l \in [i, k[.\ w^l \models \varphi_1$ ? However, we have not introduced the open interval notation. It is quite common though. As you want!
% AB@YF: Not just the second part.  Either we use the notation for all intervals or not at all.  Tried to do it.  \\YF@AB: I am not sure the about the notation $\infty$: I find it dangerous wrt nipticking reviewers. I don't think the $\geq i$ is a big issue since it is not really an inconsistent notation.}
\end{definition}
%
% \vspace{-0.5em}
When $w^0 \models \varphi$ holds, we also write $w
\models \varphi$ to denote the fact that \emph{$w$ is a model for
  $\varphi$}.  As such, every formula $\varphi \in \LTL(\mathit{AP})$ describes a
set of infinite traces, called its \emph{language}, and is denoted by
$\cL(\varphi) \subseteq \Sigma^\omega$.  In this paper,
a language describes desired or undesired system
behaviours, formalised by an LTL formula.
%
%%% Local Variables: 
%%% mode: latex
%%% TeX-master: "main"
%%% End: 
% \vspace{-1em}
\section{Monitoring LTL formulae by progression}
\label{sec:prog}
%%%%%%%%%%%%%%%%%%%%%%%%%%%%%%%%%%%%%%%%%%%%%%%%%%%%%%%%%%%%%%%%%%%%%%%%%%%%
% \vspace{-0.5em}
%
Central to our monitoring algorithm is the notion of \emph{good and bad
  prefixes} for an LTL formula or, to be more precise, for the language
it describes:
\begin{definition}%[Good/bad prefixes]
\label{def:pf}
Let $L \subseteq \Sigma^\omega$ be a language.  The set of all
  \emph{good prefixes} (resp. \emph{bad prefixes}) of $L$ is given by $\good(L)$ (resp. $\bad(L)$) and defined as follows:
\vspace{-0.4em}
  \[
\begin{array}{ccc}
  \good(L) = \{ u \in \Sigma^\ast \mid u\cdot\Sigma^\omega \subseteq L \}, &\ \ \ \ \ \ \ &
  \bad(L) = \{ u \in \Sigma^\ast \mid u\cdot\Sigma^\omega \subseteq \Sigma^\omega\setminus L \}.
  % \bad(L) = \{ u \in \Sigma^\ast \mid u\Sigma^\omega \subseteq \overline{L} \}.
  \end{array}
\vspace{-0.4em}
  \]
\end{definition}
To further ease presentation, we will shorten $\good(\cL(\varphi))$ (resp.\
$\bad(\cL(\varphi))$) to $\good(\varphi)$ (resp.\ $\bad(\varphi)$).

Although there exist a myriad of different approaches to monitoring LTL
formulae, based on various finite-trace semantics (cf.\
\cite{bauer:leucker:schallhart:jlc10}), one valid way of looking at the
monitoring problem for some formula $\varphi \in \LTL$ is the following:
The monitoring problem of $\varphi \in \LTL$ is to devise an efficient
monitoring algorithm which, in a stepwise manner, receives events from a system under
scrutiny and states whether or not the trace observed so far
constitutes a good or a bad prefix of $\cL(\varphi)$.  One monitoring
approach along those lines is described in~\cite{bauer:leucker:schallhart:FSTTCS06}.
We do not want to reiterate how in~\cite{bauer:leucker:schallhart:FSTTCS06} a monitor is constructed for
some LTL formula, but rather review an alternative monitoring procedure
based on formula rewriting, which is also known as formula progression,
or just \emph{progression} in the domain of planning with temporally
extended goals (cf.~\cite{Bacchus:1998:PTE:590220.590230}).

Progression splits a formula into a formula expressing what needs to be
satisfied by the current observation and a new formula (referred to as a \emph{future goal} or \emph{obligation}), which has to
be satisfied by the trace in the future.
As progression plays a crucial role in decentralised
LTL monitoring, we recall its definition for the full set of LTL
operators.
%
% \begin{definition}% [LTL progression]
%   Let $\varphi \in \LTL$, and $\sigma \in \Sigma$ be an event. Then, the \emph{progression function} $P:
%   \LTL \times \Sigma \rightarrow \LTL$ is defined as follows:
%   %
%   \[
%   \begin{array}{lcr}
%   \begin{array}{lcl}
%     P(p \in AP, \sigma)                      & = & \top, \hbox{ if } p \in \sigma, \bot \hbox{ otherwise}\\
%     P(\neg \varphi, \sigma)                  & = & \neg P(\varphi, \sigma)\\
%     P(\varphi_1 \vee \varphi_2, \sigma)       & = &  P(\varphi_1, \sigma) \vee P(\varphi_2, \sigma)\\
%     P(\varphi_1 \ltlU \varphi_2, \sigma)      & = & P(\varphi_2, \sigma) \vee P(\varphi_1, \sigma) \wedge \varphi_1 \ltlU \varphi_2
%   \end{array} 
%   & \ \ \ \ \ & 
%   \begin{array}{lcl}
%     P(\ltlG\varphi, \sigma)                  & = & P(\varphi, \sigma) \wedge \ltlG(\varphi)\\
%     P(\ltlF\varphi, \sigma)                  & = & P(\varphi, \sigma) \vee \ltlF(\varphi)    \\
%     P(\ltlX\varphi,\sigma)                   & = & \varphi \\
%                                              &   & 
%   \end{array} 
%   \end{array}
%   \]
% \end{definition}
% %
\begin{definition}% [LTL progression]
\label{def:ltl_progression}
  Let $\varphi,\varphi_1,\varphi_2 \in \LTL$, and $\sigma \in \Sigma$ be an event. Then, the \emph{progression function} $P:
  LTL \times \Sigma \rightarrow LTL$ is inductively defined as follows:
  \[
  \begin{array}{lcr}
  \begin{array}{lcl}
    P(p \in AP, \sigma)                      & = & \top, \hbox{ if } p \in \sigma, \bot \hbox{ otherwise}\\
    P(\varphi_1 \vee \varphi_2, \sigma)       & = &  P(\varphi_1, \sigma) \vee P(\varphi_2, \sigma)\\
    P(\varphi_1 \ltlU \varphi_2, \sigma)      & = & P(\varphi_2, \sigma) \vee P(\varphi_1, \sigma) \wedge \varphi_1 \ltlU \varphi_2\\
    P(\ltlG\varphi, \sigma)                  & = & P(\varphi, \sigma) \wedge \ltlG(\varphi)\\
    P(\ltlF\varphi, \sigma)                  & = & P(\varphi, \sigma) \vee \ltlF(\varphi)    
  \end{array} 
  & \ \ \ \ \ & 
  \begin{array}{lcl}
    P(\top,\sigma) & = &\top\\
    P(\bot,\sigma) &= &\bot\\
    P(\neg \varphi, \sigma)                  & = & \neg P(\varphi, \sigma)\\
    P(\ltlX\varphi,\sigma)                   & = & \varphi
  \end{array} 
  \end{array}
\vspace{-0.4em}
  \]
\end{definition}
Note that monitoring using rewriting with similar rules as above has
been described, for example, in
\cite{rosu-havelund-2005-jase,DBLP:journals/logcom/BarringerRH10},
although not necessarily with the same finite-trace semantics in mind
that we are discussing in this paper. Informally, the progression function ``mimics'' the \LTL semantics on an event $\sigma$, as it is stated by the following lemma.
\begin{lemma}
\label{lem:mimicsevent}
Let $\varphi$ be an \LTL formula, $\sigma$ an event and $w$ an infinite trace, we have $\sigma\cdot w\models \varphi \Leftrightarrow w\models P(\varphi,\sigma)$.
\end{lemma}
% \vspace{-1em}
% \begin{proof}
%  The proof of this lemma relies on a structural induction on $\varphi$. The formal proof can be found in Appendix~\ref{sec:proof:prog}.
% \end{proof}
%
\begin{lemma}
  \label{lem:prog}
  If $P(\varphi, \sigma) = \top$, then $\sigma \in \good(\varphi)$,
  whereas if $P(\varphi, \sigma) = \bot$, then $\sigma \in
  \bad(\varphi)$.%
% The progression function provides a monitoring algorithm for LTL:
% \[
% \forall\varphi\in LTL.\forall\sigma\in\Sigma^*.
% \big(P(\varphi, \sigma) = \top\Rightarrow \sigma \in \good(\varphi)\big)
% \wedge 
%   \big(P(\varphi, \sigma) = \bot \Rightarrow \sigma \in
%   \bad(\varphi)\big).
% \]
\end{lemma}
% 
% \begin{proof}
%  The proof of this lemma relies on a structural induction on the formula $\varphi$ and Lemma~\ref{lem:mimicsevent}. The formal proof can be found in Appendix~\ref{sec:proof:prog}.
% \end{proof}
%
Moreover, from Corollary~\ref{lem:prog} and Definition~\ref{def:pf} it
follows that if $P(\varphi, \sigma) \notin \{ \top, \bot \}$, then
there exist traces $w, w' \in \Sigma^\omega$, such that $\sigma\cdot w
\models \varphi$ and $\sigma\cdot w' \not\models \varphi$ hold.
Let us now get back to \cite{bauer:leucker:schallhart:FSTTCS06}, which
introduces a finite-trace semantics for LTL monitoring called $\LTL_3$.
It is captured by the following definition.
\begin{definition}
  \label{def:ltl3}
  Let $u \in \Sigma^\ast$, the satisfaction relation of $\LTL_3$,
 $\models_3: \Sigma^\ast \times \LTL \rightarrow
  \mathbb{B}_3$, with $\mathbb{B}_3 = \{ \top, \bot, {?} \}$, is defined
  as
% \vspace{-0.5em}
  \[
  \begin{array}{lcl}
    u & \models_3 & \varphi = \left\{ 
      \begin{array}{ll}
        \top & \mbox{ if }  u \in \good(\varphi), \\
        \bot & \mbox{ if }  u \in \bad(\varphi), \\
        {?} & \mbox{ otherwise}.
      \end{array}
    \right.
  \end{array}
  \]
\end{definition}
% \vspace{-0.5em}
%
Based on this definition, it now becomes obvious how progression
\emph{could} serve as a monitoring algorithm for $\LTL_3$.
\begin{theorem}
  \label{thm:prog}
Let $u = u(0)\cdots u(t)\in\Sigma^+$ be a trace, and $v \in \LTL$
  be the verdict, obtained by $t+1$ consecutive applications of the
  progression function of $\varphi$ on $u$, i.e., $v = P( \ldots
  (P(\varphi, u(0)), \ldots, u(t))))$.  The following cases arise:
  If $v = \top$, then $u \models_3 \varphi = \top$ holds.
  If $v = \bot$, then $u \models_3 \varphi = \bot$ holds.
  Otherwise, $u \models_3 \varphi = {?}$ holds.
  % \begin{itemize}
  % \item If $v = \top$, then $u \models_3 \varphi = \top$ holds.
  % \item If $v = \bot$, then $u \models_3 \varphi = \bot$ holds.
  % \item If $v \notin \{ \top, \bot \}$, then $u \models_3 \varphi =
  %   {?}$ holds.
  % \end{itemize}
\end{theorem}
%
% \begin{proof}
% The theorem can be shown by an induction based on Definitions~\ref{def:pf}--\ref{def:ltl3} and Corollary~\ref{lem:prog}. The formal proof can be found in Appendix~\ref{sec:proof:prog}.
% \end{proof}
%
Note that in comparison with the monitoring procedure for $\LTL_3$,
described in \cite{bauer:leucker:schallhart:FSTTCS06}, our algorithm,
implied by this theorem, has the disadvantage that the formula, which is
being progressed, may grow in size relative to the number of events.
However, in practice, the addition of some practical
simplification rules to the progression function usually prevents this
problem from occurring.
%%% Local Variables: 
%%% mode: latex
%%% TeX-master: "main"
%%% End: 

% \vspace{-1em}
\section{Decentralised progression}
\label{sec:dprog}
%
%%%%%%%%%%%%%%%%%%%%%%%%%%%%%%%%%%%%%%%%%%%%%%%%%%%%%%%%%%%%%%%%%%%%%%
% \vspace{-0.5em}
Conceptually, a monitor, $M_i$, attached to component $C_i$, which
observes events over $\Sigma_i \subseteq \Sigma$, is a rewriting engine
that accepts as input an event $\sigma \in \Sigma_i$, and an LTL formula
$\varphi$, and then applies LTL progression rules.
Additionally at each time $t$, in our $n$-component architecture, a monitor can send a message and receive
up to $n - 1$ messages in order to communicate with the other monitors
in the system, using the same synchronous bus that the system's
components communicate on.
The purpose of these messages is to send future or even past obligations
to other monitors, encoded as LTL formulae. In a nutshell, a formula is
sent by some monitor $M_i$, whenever the most urgent outstanding
obligation imposed by $M_i$'s current formula at time $t$,
$\varphi_i^t$, cannot be checked using events from $\Sigma_i$ alone.
Intuitively, the urgency of an obligation is defined by the occurrences
(or lack of) certain temporal operators in it.  For example, in order to
satisfy $p \wedge \ltlX q$, a trace needs to start with
$p$, followed by a $q$.  Hence, the obligation imposed by the subformula
$p$ can be thought of as ``more urgent'' than the one imposed by $\ltlX
q$.  A more formal definition is given later in this section.

When progressing an LTL formula, e.g., in the domain of planning to
rewrite a temporally extended LTL goal during plan search, the rewriting
engine, which implements the progression rules, will progress a state
formula $p \in \AP{}$, with an event $\sigma$ such that $p\notin\sigma$, to
$\bot$, i.e., $P(p, \emptyset) = \bot$ (see\ Definition~\ref{def:ltl_progression}). 
However, doing this in the decentralised setting, could lead to wrong
results.  In other words, we need to make a distinction as to why $p
\notin \sigma$ holds locally, and then to progress accordingly.
Consequently, the progression rule for atomic propositions 
is simply adapted by parameterising it by a local set of atomic propositions $\AP{i}$:
% \vspace{-1em}
\begin{equation}
\begin{array}{lcl}
P(p, \sigma,{\AP{i}}) & = &  \left\{ 
  \begin{array}{ll}
    \top & \mbox{ if }  p \in \sigma, \\
    \bot & \mbox{ if }  p \notin \sigma \wedge p \in \AP{i}, \\ 
    \ltlP p & \mbox{ otherwise},
  \end{array}
\right.
\end{array}\label{eq:p1}
\end{equation}
where for every $w \in \Sigma^\omega$ and $j>0$, we have $w^j \models \ltlP
\varphi$ if and only if $w^{j-1} \models \varphi$.  In other words,
$\ltlP$ is the dual to the $\ltlX$-operator, sometimes referred to as
the ``previously-operator'' in past-time LTL (cf.~\cite{Lichtenstein:1985:GP:648065.747612}).
To ease presentation, the formula $\ltlP^m\varphi$ is a short for
$
m \atop {{\overbrace{\ltlP\ltlP\ldots\ltlP}}} ~{\displaystyle{\varphi}}.
$
% (resp.\ for other operators like $\ltlF$, $\ltlG$).
%
%
Our operator is somewhat different to the standard use of $\ltlP$: it can only precede an atomic proposition or an atomic
proposition which is preceded by further $\ltlP$-operators. Hence, the
restricted use of the $\ltlP$-operator does not give us the full
flexibility (or succinctness gains~\cite{DBLP:journals/eatcs/Markey03})
of past-time LTL.
Using the $\ltlP$-operator, let us now formally define the \emph{urgency} of
an LTL formula $\varphi$ using a pattern matching on $\varphi$ as follows:
% \TODO{YF@AB: put a formal and inductive definition of urgency that corresponds to the one in the tool. CHECK\&KILL}%
%
\begin{definition}% [Urgent formula]
  Let $\varphi$ be an LTL formula, and $\Upsilon: \LTL \rightarrow
  \mathbb{N}^{\geq 0}$ be an inductively defined function assigning a
  level of \emph{urgency} to an LTL formula as follows.
\[
 \begin{array}{rcll}
  \Upsilon(\varphi) & = & \text{match $\varphi$ with}\\
&  & \varphi_1\vee\varphi_2 \mid \varphi_1\wedge\varphi_2 & \rightarrow \max(\Upsilon(\varphi_1),\Upsilon(\varphi_2))\\
& \mid & \ltlP \varphi' & \rightarrow 1+\Upsilon(\varphi') \\
& \mid & \_ & \rightarrow 0
 \end{array}
\]
%   $\Upsilon(\varphi) = 0$, if $\varphi$ is not a state formula, or a
%   state formula preceded by one or many $\ltlP$-operators.
%   $\Upsilon(p) = 1$ for any $p \in AP$, and $\Upsilon(\ltlP\varphi) =
%   \Upsilon(\varphi) + 1$.

  A formula $\varphi$ is said to be \emph{more urgent} than formula $\psi$, if
  and only if $\Upsilon(\varphi) > \Upsilon(\psi)$ holds.  A formula
  $\varphi$ where $\Upsilon(\varphi) = 0$ holds is said to be not
  urgent.
\end{definition}
Moreover, the above modification to the progression rules has obviously
the desired effect: If $p \in \sigma$, then nothing changes, otherwise
if $p \notin \sigma$, we return $\ltlP p$ in case that the monitor
$M_i$ cannot observe $p$ at all, i.e., in case that $p \notin \AP{i}$
holds. This effectively means, that $M_i$ cannot decide whether or not
$p$ occurred, and will therefore turn the state formula $p$ into an
obligation for some other monitor to evaluate rather than produce a
truth-value.
Of course, the downside of rewriting future goals into past goals that
have to be processed further, is that violations or satisfactions of a global goal will
usually be detected \emph{after} they have occurred.  However, since
there is no central observer which records all events at the same time,
the monitors \emph{need} to communicate their respective results to
other monitors, which, on a synchronous bus, occupies one or more time
steps, depending on how often a result needs to be passed on until it
reaches a monitor which is able to actually state a verdict.  We
shall later give an upper bound on these communication times, and show
that our decentralised monitoring framework is, in fact, optimal under
the given assumptions (see Theorem~\ref{theo:maxdelay}).
\begin{example}
  \label{ex:prog}
  Let us assume we have a decentralised system consisting of three
  components, $A, B, C$, such that $\AP{A} = \{ a \}$, $\AP{B} = \{
  b \}$, and $\AP{C} = \{ c \}$, and that a global formula $\varphi =
  \ltlF(a \wedge b \wedge c)$ needs to be monitored in a decentralised
  manner.
  Let us further assume that, initially, $\varphi^0_A = \varphi^0_B =
  \varphi^0_C = \varphi$.
  Let $\sigma = \{ a, b \}$ be the system event at time $0$; that is,
  $M_A$ (resp. $M_B$, $M_C$) observes $\Pi_A(\sigma) = \{ a \}$ (resp. $\Pi_B(\sigma) = \{ b \}$, $\Pi_C(\sigma) =
  \emptyset$) when $\sigma$ occurs.
  The rewriting that takes place in all three monitors to generate the
  next local goal formula, using the modified set of rules, and
  triggered by $\sigma$, is as follows:
  \[
% \vspace{-0.4em}
  \begin{array}{lll}
    \varphi^{1}_A = & P(\varphi, \{ a \}, \{ a \})& = P(a, \{ a \}, \{ a \}) \wedge P(b, \{ a\}, \{ a \}) \wedge P(c, \{ a \}, \{ a \}) \vee \varphi \\
& &= \ltlP b \wedge \ltlP c \vee \varphi \\
    \varphi^{1}_B = & P(\varphi, \{ b \}, \{ b \})& = P(a, \{ b \}, \{ b \}) \wedge P(b, \{ b\}, \{ b \}) \wedge P(c, \{ b \}, \{ b \}) \vee \varphi\\
& & = \ltlP a \wedge \ltlP c \vee \varphi \\
    \varphi^{1}_C = & P(\varphi, \emptyset, \{ c \})& = P(a, \emptyset, \{ c \}) \wedge P(b, \emptyset, \{ c \}) \wedge P(c, \emptyset, \{ c \}) \vee \varphi \\
& &= \ltlP a \wedge \ltlP b \wedge \bot \vee \varphi = \varphi \\
  \end{array}
  \]
% \vspace{-0.4em}
\end{example}

But we have yet to define progression for past goals: For this purpose,
each monitor has local storage to keep a \emph{bounded} number
of past events. The event that occurred at time $t-k$ is referred as $\sigma(-k)$. On a monitor observing $\Sigma_i$, the progression of a past goal $\ltlP^m\varphi$, at time $t\geq m$, is defined as follows:
\begin{equation}
% \vspace{-0.4em}
\begin{array}{lcl}
P(\ltlP^m\varphi, \sigma, \AP{i}) & = &  \left\{ 
  \begin{array}{ll}
    \top & \mbox{ if }  \varphi = p \mbox{ for some } p \in \AP{i} \cap \Pi_i(\sigma(-m)), \\
    \bot & \mbox{ if }  \varphi = p \mbox{ for some } p \in \AP{i} \setminus \Pi_i(\sigma(-m)), \\
    \ltlP^{m+1}\varphi & \mbox{ otherwise},
  \end{array}
\right.
\end{array}\label{eq:p2}
% \vspace{-0.4em}
\end{equation}
where, for $i\in [1,n]$, $\Pi_i$ is the projection function associated to each monitor $M_i$, respectively.
%
% We distinguish between past obligations, which are formulae of the form
% $\ltlP \varphi$, present obligations, which are state formulae, and
% future obligations which are temporal formulae not using the
% $\ltlP$-operator.  
Note that since we do not allow $\ltlP$ for the specification of a
global system monitoring property, our definitions will ensure that the
local monitoring goals, $\varphi_i^t$, will never be of the form
$\ltlP\ltlX\ltlX p$, which is equivalent to a future obligation, despite
the initial $\ltlP$.  In fact, our rules ensure that a formula preceded
by the $\ltlP$-operator is either an atomic proposition, or an atomic proposition
which is preceded by one or many $\ltlP$-operators.  Hence, in rule (\ref{eq:p2}), we do not need to consider any
other cases for $\varphi$.
%%% Local Variables: 
%%% mode: latex
%%% TeX-master: "main"
%%% End: 

% \vspace{-1em}
\section{Semantics}
\label{sec:sem}
%%%%%%%%%%%%%%%%%%%%%%%%%%%%%%%%%%%%%%%%%%%%%%%%%%%%%%%%%%%%%%%%%%%%%%%%%%%%%%%%%%%%%%%%%%%%%%%%
% \vspace{-1em}
In the previous example, we can clearly see that monitors $M_A$ and
$M_B$ cannot determine whether or not $\sigma$, if interpreted as a trace
of length $1$, is a good prefix for the global goal formula
$\varphi$.\footnote{Note that $\cL(\varphi)$, being a \emph{liveness}
  language~\cite{alpern87recognizing}, does not have any bad
  prefixes.}
Monitor $M_C$ on the other hand did not observe an action $c$, and
therefore, is the only monitor after time $0$, which knows that $\sigma$
is not a good prefix, and that, as before, after time $1$,
$\varphi$ is the goal that needs to be satisfied by the system under
scrutiny.
Intuitively, the other two monitors know that if their respective past
goals were satisfied, then $\sigma$ would be a good prefix, but in order
to determine this information, they need to send and receive messages to
and from each other, containing obligations, i.e., LTL formulae.

Before we outline how this is done in our setting, let us discuss the
semantics, we obtain from this decentralised application of progression.
We already said that monitors detect good and bad prefixes for a global
formula.  In other words, if a monitor's progression evaluates to $\top$
(resp. $\bot$), then the trace seen so far is a good (resp. bad) prefix,
and if neither monitor comes to a Boolean truth-value as verdict, we
keep monitoring. This latter case indicates that, so far, the trace is
neither a good nor a bad prefix for the global formula.
\begin{definition}
  \label{def:dltl}
  Let $\mathcal{C} = \{ C_1, \ldots, C_n \}$ be the set of system
  components, $\varphi \in \LTL$ be a global goal, and $\mathcal{M} = \{
  M_1, \ldots, M_n \}$ be the set of component monitors.
  Further, let $\mathbf{u} = u_1(0) \cup \ldots \cup u_n(0) \cdots
  u_1(t) \cup \ldots \cup u_n(t) \in \Sigma^\ast$ be the global
  behavioural trace of the system, obtained by composition of all local
  component traces, at time $t \in \mathbb{N}^{\geq 0}$.
  If for some component $C_i$, with $i \leq n$, containing a local obligation $\varphi_i^t$, $M_i$ reports $P(\varphi_i^t, u_i(t),\AP{i}) = \top$
  (resp.\ $\bot$), then $\mathbf{u} \models_D \varphi = \top$ (resp.\
  $\bot$). Otherwise, we have $\mathbf{u} \models_D \varphi =
  {?}$.
\end{definition}

By $\models_D$ we denote the satisfaction relation on finite traces in
the decentralised setting to differentiate it from $\LTL_3$ as well as
standard $\LTL$ which is defined on infinite traces.
Obviously, $\models_3$ and $\models_D$ both yield values from the same
truth-domain. However, the semantics are not equivalent, since the
modified progression function used in the above definition sometimes
rewrites a state formula into an obligation concerning the past rather
than returning a verdict.
On the other hand, in the case of a one-component system (i.e., all
propositions of a formula can be observed by a single monitor), the
definition of $\models_D$ matches Theorem~\ref{thm:prog}, in particular because our
progression rule \eqref{eq:p1} is then equivalent to the standard case.
Monitoring $\LTL_3$ with progression becomes a special case of
decentralised monitoring, in the following sense:
%
% \vspace{-0.3em}
\begin{corollary}
\label{cor:reducedone}
  If $|\mathcal{M}| = 1$, then
  $\forall u \in \Sigma^\ast.\ \forall\varphi \in \LTL.\ u \models_3 \varphi = u \models_D \varphi$.
  % $|\mathcal{C}| = |\mathcal{M}| = 1 \Rightarrow \forall u \in \Sigma^\ast,\forall\varphi \in \LTL: u \models_3 \varphi = u \models_D \varphi$.
\end{corollary}
% \vspace{-0.2em}
%%% Local Variables: 
%%% mode: latex
%%% TeX-master: "main"
%%% End: 

% \vspace{-2em}
\section{Communication and decision making}
\label{sec:alg}
%%%%%%%%%%%%%%%%%%%%%%%%%%%%%%%%%%%%%%%%%%%%%%%%%%%%%%%%%%%%%%%%%%%%%%
%%%%%%%%%%%%%%%%%%%%%%%%%%%%%%%%%%%%%%%%%%%%%%%%%%%%%%%%%%%%%%%%%%%%%%
% \vspace{-0.8em}
Let us now describe the communication mechanism that enables local
monitors to determine whether a trace is a good or a bad prefix.  Recall
that each monitor only sees a projection of an event to its locally
observable set of actions, encoded as a set of atomic propositions, respectively.

Generally, at time $t$, when receiving an event $\sigma$, a monitor,
$M_i$, will progress its current obligation, $\varphi_i^t$, into
$P(\varphi_i^t,\sigma,\AP{i})$, and send the result to another
monitor, $M_{j\neq i}$, whenever the most urgent obligation, $\psi \in
\sus(P(\varphi_i^t,\sigma,$ $\AP{i}))$, is such that $\Prop(\psi)
\subseteq (\AP{j})$ holds, where $\sus(\varphi)$ is the \emph{set of urgent subformulae} of $\varphi$ and $\Prop: \LTL \rightarrow 2^{\AP{}}$ is
the function which yields the set of occurring propositions of an \LTL
formula. 
% The set of urgent subformuale of an \LTL formula is inpired from the
% classical syntactic closure and optimised in the context of
% decentralised \LTL monitoring.
%
% \vspace{-0.5em}
\begin{definition}
 \label{def:ucl}
The function $\sus:\LTL\rightarrow 2^{\LTL}$ is inductively defined as follows:
% \vspace{-0.5em}
\[
\begin{array}{rrcll}
\sus(\varphi)  =  \text{match $\varphi$ with} & 
& & \varphi_1\vee\varphi_2\mid\varphi_1\wedge\varphi_2 &\rightarrow \sus(\varphi_1)\cup\sus(\varphi_2)\\
&& \mid & \neg\varphi' &\rightarrow \sus(\varphi')\\
&& \mid & \ltlP\varphi' &\rightarrow \{\ltlP\varphi'\} \\
&& \mid & \_ & \rightarrow \emptyset
\end{array}
\]
\end{definition}
%
% \vspace{-1em}
%
The set $\sus(\varphi)$ contains the past sub-formulae of $\varphi$,
i.e., sub-formulae starting with a future temporal operator are
discarded. It uses the fact that, in decentralised progression,
$\ltlP$-operators are only introduced in front of atomic
propositions. Thus, only the cases mentioned explicitly in the pattern
matching need to be considered. Moreover, for formulae of the form
$\ltlP\varphi'$, i.e., starting with an $\ltlP$-operator, it is not
needed to apply {$\sus$} to $\varphi'$ because $\varphi'$ is necessarily
of the form $\ltlP^d p$ with $d\geq0$ and $p\in\AP{}$, and does not
contain more urgent formulae than $\ltlP\varphi'$.

Note that, if there are several equally urgent obligations for distinct monitors, then $M_i$
sends the formula to only one of the corresponding monitors according to a priority order between monitors. Using this order ensures that the delay induced by evaluating the global system specification in a decentralised fashion is bounded, as we shall see in Theorem~\ref{theo:maxdelay}. For simplicity, in the following, for a set of component monitors $\mathcal{M} = \{M_1, \ldots, M_n \}$ the sending order is the natural order on the interval $[1,n]$. This choice of the local monitor to send the obligation is encoded through the function $\Mon:{\cal M}\times 2^\AP\rightarrow {\cal M}$. For a monitor $M_i\in{\cal M}$ and a set of atomic propositions $\AP{}'\in 2^\AP{}$, $\Mon(M_i,\AP{}')$ is the monitor $M_{j_{\min}}$ s.t. $j_{\min}$ is the smallest integer in $[1,n]$ s.t. there is a monitor for an atomic proposition in $\AP{}'$. Formally: $\Mon(M_i,\AP{}')=j_{\min}=\min \{j\in [1,n]\setminus\{i\} \mid \AP{}'\cap\AP{j}\neq\emptyset\}$.

Once $M_i$ has sent %its message at time $t$, containing
$P(\varphi_i^t,\sigma,\AP{i})$, it sets $\varphi_i^{t+1} = \#$, where
$\#\notin\AP{}$ is a special symbol for which we define progression by
\begin{equation}
P(\#, \sigma, \AP{i}) = \#.\label{eq:p3}
\end{equation}
and $\forall \varphi \in \LTL.\ \varphi \wedge \# = \varphi$.  On the
other hand, whenever $M_i$ receives a formula, $\varphi_{j\neq i}$, sent
from a monitor $M_j$, it will add the new formula to its existing
obligation, i.e., its current obligation $\varphi_i^{t}$ will be
replaced by the conjunction $\varphi_i^{t} \wedge \varphi_{j\neq i}$.
Should $M_i$ receive further obligations from other monitors but $j$, it
will add each new obligation as an additional conjunct in the same
manner.
%
% In an \LTL formula, the symbol $\#$ is eliminated by adding the
% following rule to \LTL semantics: $\forall\varphi\in \LTL.\ \varphi
% \wedge \# = \varphi$. Note that no further rule is needed since, in the
% algorithm of local monitors, the symbol $\#$ will never be in the scope of
% a temporal operator and only conjuncts will be added to $\#$.
%
% On the other hand, whenever $M_i$ receives a 

Let us now summarise the above steps in the form of an explicit algorithm
that describes how the local monitors operate and make decisions.
%
% \vspace{-0.5em}
\begin{algo}{L}{Local Monitor}
  Let $\varphi$ be a global system specification, and $\mathcal{M} = \{
  M_1, \ldots, M_n \}$ be the set of component monitors.
  The algorithm Local Monitor, executed on each $M_i$, returns $\top$
  (resp.\ $\bot$), if $\sigma \models_D \varphi_i^t$ (resp.\
  $\sigma\not\models_D \varphi_i^t$) holds, where $\sigma \in
  \Sigma_i$ is the projection of an event to the observable set of
  actions of the respective monitor, and $\varphi_i^t$ the monitor's
  current local obligation.
  % 
% \vspace{-0.5em}
  \begin{description}
  \item[L1.] [Next goal.]
    Let $t \in \mathbb{N}^{\geq 0}$ denote the current time step and
    $\varphi_i^t$ be the monitor's current local obligation.  If $t =
    0$, then set $\varphi_i^t := \varphi$.
  \item[L2.] [Receive event.] Read next $\sigma$.
  \item[L3.] [Receive messages.] Let $\{\varphi_j\}_{j\in [1,n],j\neq i}$ be the set of
    received obligations at time $t$ from other monitors. Set $\varphi_i^t := \varphi_i^t
    \wedge \bigwedge_{j\in [1,n],j\neq i} \varphi_j$.
  \item[L4.] [Progress.] Determine $P(\varphi_i^t, \sigma, \AP{i})$
    and store the result in $\varphi_i^{t+1}$.
  % \item[L4.] [Progress.] Let the rewriting engine determine $P(\varphi_i^t, \sigma, \AP{i})$
  %   and store the result in $\varphi_i^{t+1}$.
  \item[L5.] [Evaluate and return.]
    If $\varphi_i^{t+1} = \top$ return $\top$, if $\varphi_i^{t+1} =
    \bot$ return $\bot$.
  \item[L6.] [Communicate.]
    Set $\psi \in \sus(\varphi_i^{t+1})$ to be the most urgent obligation
    of $\varphi_i^{t+1}$.
    Send $\varphi_i^{t+1}$ to monitor $\Mon(M_i,$ $\Prop(\psi))$. 
    % Then, for every $\psi' \in cl(\varphi_i^{t+1})$ which is equally
    % urgent as $\psi$,
    % send $\varphi_i^{t+1}$ to monitor $Mon(Prop(\psi))$.
  \item[L7.] [Replace goal.]
    If in step L6 a message was sent at all, set $\varphi_i^{t+1} := \#$.
    Then go back to step L1.
    \qed
  \end{description}
\end{algo}
%
% \vspace{-0.5em}
The input to the algorithm, $\sigma$, will usually resemble the latest
observation in a consecutively growing trace, $u_i = u_i(0) \cdots u_i(t)$,
i.e., $\sigma = u_i(t)$.
We then have that $\sigma \models_D \varphi_i^t$ (i.e., the
algorithm returns $\top$) implies that $u \models_D \varphi$ holds (resp.\ for
$\sigma\not\models_D \varphi_i^t$).
\begin{table}[tbp]
%   \vspace{-1em}
  \caption{Decentralised progression of $\varphi = \ltlF (a \wedge b \wedge c)$ in a 3-component system.}
  \scalebox{0.9}{
%     \hspace{-2em}
    \setlength{\extrarowheight}{4pt}
    \begin{tabularx}{\linewidth}{r|l|l|l|l}
      \hline
      $t$: & 0 & 1 & 2 & 3 \\
      \hline
      $\sigma$: & $\{ a, b \}$ & $\{ a, b, c \}$ & $\emptyset$ & $\emptyset$\\
      \hline
      $M_A$: & $\begin{array}{ll}\varphi^1_A  & := P(\varphi,\sigma,\AP{A}) \\ & = \ltlP b \wedge \ltlP c \vee \varphi\end{array}$ & 
               $\begin{array}{ll}\varphi^2_A & := P(\varphi_B^1 \wedge \#, \sigma, \AP{A}) \\ & = \ltlP^2c \vee (\ltlP b \wedge \ltlP c \vee \varphi)\end{array}$ & 
               $\begin{array}{ll}\varphi^3_A & := P(\varphi_C^2 \wedge \#, \sigma, \AP{A}) \\ & = \ltlP^2b \vee (\ltlP b \wedge \ltlP c \vee \varphi)\end{array}$ &
               $\begin{array}{ll}\varphi^4_A & := P(\varphi_C^3 \wedge \#, \sigma, \AP{A}) \\ &  = \ltlP^3b \vee (\ltlP b \wedge \ltlP c \vee \varphi) \end{array}$ \\
      \hline
      $M_B$: & $\begin{array}{ll}\varphi^1_B & := P(\varphi,\sigma, \AP{B}) \\ &= \ltlP a \wedge \ltlP c \vee \varphi\end{array}$ & 
               $\begin{array}{ll}\varphi^2_B & := P(\varphi_A^1 \wedge \#, \sigma, \AP{B}) \\ & = \ltlP^2c \vee (\ltlP a \wedge \ltlP c \vee \varphi)\end{array}$ & 
               $\begin{array}{ll}\varphi^3_B & := P(\#, \sigma, \AP{B}) \\ & = \# \end{array}$ &
               \cellcolor[gray]{0.9}$\begin{array}{ll}\varphi^4_B & := P(\varphi_A^3 \wedge \#, \sigma, \AP{B}) \\ & = \top  \end{array}$ \\
      \hline
      $M_C$: & \cellcolor[gray]{0.9}$\begin{array}{ll}\varphi^1_C & := P(\varphi,\sigma, \AP{C})  \\ & = \varphi\end{array}$ & 
               $\begin{array}{ll}\varphi^2_C & := P(\varphi, \sigma, \AP{C}) \\ & = \ltlP a \wedge \ltlP b \vee \varphi \end{array}$ & 
               $\begin{array}{ll}\varphi^3_C & := P(\varphi_A^2 \wedge \varphi_B^2 \wedge \#, \sigma, \AP{C})  \\ & =\ltlP^2 a \wedge \ltlP^2 b \vee \varphi \end{array}$ &
               $\begin{array}{ll}\varphi_C^4  & := P(\#, \sigma, \AP{C}) \\ & = \# \end{array}$ \\
      \hline
    \end{tabularx}}
%   \vspace{-2em}
  \label{tab:ex}
\end{table}

\begin{example}
  \label{ex:dec}
  To see how this algorithm works, let us continue the decentralised
  monitoring process initiated in Example~\ref{ex:prog}.
  Table~\ref{tab:ex} shows how the situation evolves for all three
  monitors, when the global LTL specification in question is $\ltlF(a
  \wedge b \wedge c)$ and the ordering between components is $A<B<C$.
  An evolution of $M_A$'s local obligation, encoded as $P(\varphi_B^1
  \wedge \#, \sigma,\AP{A})$ (see cell $M_A$ at $t=1$) indicates
  that communication between the monitors has occurred: $M_B$ sent its
  obligation to $M_A$, at the end of step $0$.  Likewise for the other
  obligations and monitors.  The interesting situations are marked in
  grey: In particular at $t=0$, $M_C$ is the only monitor who knows for
  sure that, so far, no good nor bad prefix occurred (see grey cell at
  $t=0$).
  At $t=1$, we have the desired situation $\sigma = \{a, b, c\}$, but
  because none of the monitors can see the other monitors' events, it
  takes another two rounds of communication until both $M_A$ and $M_B$
  detect that, indeed, the global obligation had been satisfied at $t=1$
  (see grey cell at $t=3$).  
\end{example}

% \begin{example}
%   \label{ex:dec}
%   To see how this algorithm works, let us continue the decentralised
%   monitoring process initiated in Example~\ref{ex:prog}.
%   Table~\ref{tab:ex} shows how the situation evolves for all three monitors, when the
%   global LTL specification in question is $\ltlF(a \wedge b \wedge
%   c)$ and the ordering between componenents is $A>B>C$.
% %
%   An evolution of $M_C$'s local obligation, encoded as $P(\varphi_A^1
%   \wedge \varphi, \sigma,\AP{A})$ (see cell $M_C$ at $t=1$) indicates that
%   communication between the monitors has occurred: $M_A$ sent its
%   obligation to $M_C$, at the end of step $0$.  Likewise for the other obligations and monitors.
%   The interesting situations are marked in grey: In particular at $t=0$, $M_C$ is the only monitor who knows for sure that, so far, no good nor bad prefix occurred (see grey cell at $t=0$).
%   %
%   At $t=1$, we have the desired situation $\sigma = \{a, b, c\}$, but
%   because none of the monitors can see the other monitors' events, it
%   takes another two rounds of communication until both $M_A$ and $M_B$
%   detect that, indeed, the global obligation has been satisfied at $t=1$
%   (see grey cells at $t=3$).  
% \end{example}

This example highlights a worst case \emph{delay} between the occurrence
and the detection of a good (resp.\ bad) trace by a good (resp. bad)
prefix, caused by the time it takes for the monitors to communicate
obligations to each other.
This delay directly depends on the number of monitors in the system, and
is also the upper bound for the number of past events each monitor needs
to store locally in order to be able to progress all occurring past
obligations:
\begin{theorem}% [Maximum delay]
\label{theo:maxdelay}
%Let, for any $p \in \Props$, $\ltlP^m p \in \LTL$ be a local obligation
Let, for any $p \in \Props$, $\ltlP^m p$ be a local obligation obtained
by Algorithm~L executed on some monitor $M_i \in \mathcal{M}$.  At any
time $t \in \mathbb{N}^{\geq 0}$, $m \leq \min(|\mathcal{M}|,t+1)$.
% Let, for any $p \in AP$, $\ltlP^m p \in \LTL$ be a local obligation obtained by Algorithm~L executed on some 
% monitor $M_i \in \mathcal{M}$.  In the worst case, $m \leq\min(|\mathcal{M}|,t+1)$ at any time $t \in \mathbb{N}^{\geq 0}$.
\end{theorem}
%
% \vspace{-1em}
\begin{proof}
We provide below a sketch of the proof explaining the intuition on the theorem. The formal proof can be found in Appendix~\ref{proofs:decentmon}.

  Recall that $\ltlP$-operators are only introduced directly in front of
  atomic propositions according to rule
  \eqref{eq:p1} when $M_i$ rewrites a propositional formula $p$
  with $p \notin \AP{i}$.
  Further $\ltlP$-operators can only be added according to rule
  \eqref{eq:p2} when $M_i$ is unable to evaluate an obligation of the form
  $\ltlP^h p$.  
  The interesting situation occurs when a monitor $M_i$ maintains a set of
  urgent obligations of the form $\{\ltlP^h p_1, \ldots, \ltlP^j p_l\}$
  with $h,j \in \mathbb{N}^{\geq 0}$, then, according to step L6 of
  Algorithm~L, $M_i$ will transmit the obligations to one monitor only thereby adding one additional
  $\ltlP$-operator to the remaining obligations: $\{\ltlP^{h+1} p_2,
  \ldots, \ltlP^{j+1} p_l\}$.
  Obviously, a single monitor cannot have more than $|\mathcal{M}| - 1$
  outstanding obligations that need to be sent to the other monitors at
  any time $t$.  So, the worst case delay is initiated during
  monitoring, if at some time \emph{all} outstanding obligations of each monitor $M_i$, $i\in[1,|{\cal M}|]$,
  are of the form $\{ \ltlP p_1, \ldots, \ltlP p_l\}$ with $p_1,
  \ldots, p_l \notin \AP{i}$ (i.e., the obligations are all equally
  urgent), in which case it takes $|\mathcal{M}| - 1$ time steps until
  the last one has been chosen to be sent to its respective monitor $M_j$. Using an ordering between components ensures here that each set of obligations will decrease in size after being transmitted once.
  Finally, a last monitor, $M_j$ will receive an obligation of the
  form $\ltlP^{|\mathcal{M}|} p_k$ with $1 \leq k \leq l$ and $p_k
  \in \AP{j}$.
  \qed
\end{proof}
%
% \vspace{-0.3em}
Consequently, the monitors only need to memorise a \emph{bounded history} of the trace read so far, i.e., the last $|\mathcal{M}|$ events.

Example~\ref{ex:dec} also illustrates the relationship to the $\LTL_3$
semantics discussed earlier in~\secref{sec:prog}. This relationship is formalised by the two following theorems stating the ``soundness and completeness'' of the algorithm.
\begin{theorem}
\label{theo:soundness}
  Let $\varphi \in \LTL$ and $u \in \Sigma^\ast$, then $u \models_D
  \varphi = \top/\bot \Rightarrow u \models_3 \varphi = \top/\bot$, and
  $u \models_3 \varphi = {?}  \Rightarrow u \models_D \varphi = {?}$.
\end{theorem}
%
% \begin{proof}
% The proof is performed by showing that the initial obligation (the global specification) is ``propagated'' along monitors' local obligations. 
% %The formal proof can be found in Appendix~\ref{proofs:decentmon}. \qed
% \end{proof}
%
% \vspace{-0.5em}
In particular, the example shows how the other direction of the theorem
does not necessarily hold.  Consider the trace $u = \{a, b\} \cdot\{a,b,c\}$:
clearly, $u \models_3 \ltlF(a \wedge b \wedge c) = \top$, but we have $u
\models_D \ltlF(a \wedge b \wedge c) = {?}$ in our example.  Again, this
is a direct consequence of the delay introduced in our setting.

However, Algorithm~L detects all verdicts for a specification as if the system was not distributed.
%
% \vspace{-0.5em}
\begin{theorem}
\label{theo:completeness}
Let $\varphi \in \LTL$ and $u \in \Sigma^\ast$, then $u\models_3\varphi = \top/\bot \Rightarrow \exists u'\in\Sigma^\ast.\ |u'|\leq n\wedge u\cdot u'\models_D \varphi = \top/\bot$, where $n$ is the number of components in the system.
\end{theorem}
\section{Experimental results}
\label{sec:impl}
%
% \begin{figure}
% \includegraphics{figs/decentmon-crop.pdf}
% \caption{Overview of the architecture of {\toolname}}
% \label{fig:decentmon}
% \end{figure}
%
% \vspace{-0.7em}
{\toolname} 
%(see Fig.~\ref{fig:decentmon})\TODO{YF@AB: added a figure here} 
is an 
%executable system, 
implementation, simulating the above distributed LTL
monitoring algorithm in 1,800 LLOC, written in the functional programming
language OCaml. It can be freely downloaded and run from~\cite{decenttool2}.
The system takes as input multiple traces (that can be automatically
generated), corresponding to the behaviour of a distributed system, and
an LTL formula.  Then the formula is monitored against the traces in two
different modes: a) by merging the traces to a single, global trace and
then using a ``central monitor'' for the formula (i.e., all local
monitors send their respective events to the central monitor who makes
the decisions regarding the trace), and b) by using the decentralised
approach introduced in this paper (i.e., each trace is read by a
separate monitor).
We have evaluated the two different monitoring approaches (i.e.,
centralised vs.\ decentralised) using two different set-ups
described in the remainder of this section.
% \vspace{-1.8em}
\begin{table}[t]
\setlength{\extrarowheight}{1mm}
\centering
% \vspace{-1em}
\caption{Benchmarks for randomly generated LTL formulae}
\scalebox{1}{%
\begin{tabular}{|c|cc|r|r|r|r|r|r|}
  \hline
  \multicolumn{3}{|c|}{} & \multicolumn{2}{c|}{centralised} & \multicolumn{2}{c|}{decentralised} & \multicolumn{2}{c|}{\emph{diff.\ ratio}}\\
  \hline
  $|\varphi|$ & \multicolumn{2}{c|}{$\Sigma_c$ and $\Sigma_d$} & \textbar{trace}\textbar  & \#msg. & \textbar{trace}\textbar & \#msg.\ & \textbar{trace}\textbar & \#msg. \\
  \hline
  \rowcolor[gray]{0.9}
  $1$ & $\{a,b,c\}$ & $\{a|b|c\}$ & 1.369 & 4.107 & 1.634 & 0.982 & 1.1935 & 0.2391\\
  $2$    &  $\{a,b,c\} $  &   $\{a|b|c\}$  & 2.095   &  6.285  &   2.461   &  1.647  &  1.1747   &  0.262\\
  \rowcolor[gray]{0.9}
  $3$    &  $\{a,b,c\}$   &   $\{a|b|c\}$    &  3.518  & 10.554  &   4.011 & 2.749 & 1.1401 & 0.2604 \\
  $4$    &  $\{a,b,c\}$   &   $\{a|b|c\}$    &  5.889 & 17.667 &  6.4  & 4.61 &  1.0867 & 0.2609 \\
  \rowcolor[gray]{0.9}
  $5$   &  $\{a,b,c\}$   &   $\{a|b|c\}$    &  9.375 & 28.125 & 9.935 & 7.879 & 1.0597 & 0.2801\\
  $6$    &  $\{a,b,c\}$   &   $\{a|b|c\}$    &  11.808   & 35.424  &  12.366   &  9.912  &  1.0472 & 0.2798\\
  \hline
\end{tabular}}
% \vspace{-1.8em}
\label{tab:bench:random}
\end{table}
% \vspace{-0.5em}
\paragraph{Evaluation of randomly generated formulae.}
{\toolname} randomly generated 1,000 LTL formulae of various sizes in the
architecture described in Example~\ref{ex:prog}. How both monitoring
 approaches compared on these formulae can be seen in
Table~\ref{tab:bench:random}.
The first columns show the size of the monitored LTL formulae and the
underlying alphabet(s) of the monitor(s). 
Note that our system measures formula size in terms of the operator
entailment\footnote{Our practical experiments show that this way of
  measuring the size of a formula is more representative of how
  difficult it is to progress it in a decentralised manner.} inside it
(state formulae excluded), e.g., $\ltlG (a\wedge b) \vee \ltlF c$ is of
size $2$. The entry {\textbar{trace}\textbar} denotes the average length
of the traces needed to reach a verdict.
For example, the last line in Table~\ref{tab:bench:random} says that we
monitored 1,000 randomly generated LTL formulae of size 6.  On average,
traces were of length 11.808 when the central monitor came to a verdict,
and of length 12.366 when one of the local monitors came to a verdict.
The difference ratio, given in the second last column then shows the
average delay; that is, on average the traces were 1.0472 times longer
in the decentralised setting than the traces in the centralised setting.
The number of messages, \#msg., in the centralised setting, corresponds
to the number of events sent by the local monitors to the central
monitor (i.e., ${|trace|\times |\Sigma_d|}$), and in the decentralised setting to the number of obligations
transmitted between local monitors.
What is striking here is that the amount of communication needed in the
decentralised setting is ca.\ only 25\% of the communication overhead
induced by central monitoring, where local monitors need to send each
event to a central monitor.
%
% \vspace{-0.5em}

\begin{table}[t]
\setlength{\extrarowheight}{1mm}
\centering
% \vspace{-1em}
\caption{Benchmarks for LTL specification patterns}
\scalebox{1}{%
\begin{tabular}{|c|cc|r|r|r|r|r|r|}
  \hline
  \multicolumn{3}{|c|}{} & \multicolumn{2}{c|}{centralised} & \multicolumn{2}{c|}{decentralised} & \multicolumn{2}{c|}{\emph{diff.\ ratio}}\\
  \hline
  pattern & \multicolumn{2}{c|}{$\Sigma_c$ and $\Sigma_d$} & \textbar{trace}\textbar  & \#msg. & \textbar{trace}\textbar & \#msg.\ & \textbar{trace}\textbar & \#msg. \\
  \hline
  \rowcolor[gray]{0.9}
  absence  &  $\{a,b,c\}$ & $\{a|b|c\}$ & 156.17 & 468.51 & 156.72 & 37.94 & 1.0035 & 0.0809 \\
  existence & $\{a,b,c\}$ & $\{a|b|c\}$ & 189.90 &  569.72 & 190.42 & 44.41 & 1.0027 & 0.0779\\
  \rowcolor[gray]{0.9}
 bounded existence & $\{a,b,c\}$ & $\{a|b|c\}$ & 171.72 & 515.16 & 172.30 & 68.72 & 1.0033 & 0.1334 \\
  universal & $\{a,b,c\}$ & $\{a|b|c\}$ & 97.03 & 291.09 & 97.66 & 11.05 & 1.0065 & 0.0379\\
  \rowcolor[gray]{0.9}
precedence & $\{a,b,c\}$ & $\{a|b|c\}$ & 224.11 & 672.33 & 224.72 & 53.703 & 1.0027 & 0.0798 \\
  response & $\{a,b,c\}$ & $\{a|b|c\}$ & 636.28 & 1,908.86 & 636.54 & 360.33 & 1.0004 & 0.1887 \\
  \rowcolor[gray]{0.9}
  precedence\ chain & $\{a,b,c\}$ & $\{a|b|c\}$ & 200.23 & 600.69 & 200.76 & 62.08 & 1.0026 & 0.1033 \\
  response\ chain & $\{a,b,c\}$ & $\{a|b|c\}$ & 581.20 & 1,743.60 & 581.54 & 377.64 & 1.0005 & 0.2165 \\
  \rowcolor[gray]{0.9}
  constrained\ chain & $\{a,b,c\}$ & $\{a|b|c\}$ & 409.12 & 1,227.35 & 409.62 & 222.84 & 1.0012 & 0.1815 \\
  \hline
\end{tabular}}
% \vspace{-2.5em}
\label{tab:bench:patterns}
\end{table}
% \vspace{-1em}
\paragraph{Evaluation using specification patterns.}
In order to evaluate our approach also at the hand of realistic LTL
specifications, we conducted benchmarks using LTL formulae following the
well-known LTL specification patterns (\cite{302672}, whereas the actual
formulae underlying the patterns are available at this site
\cite{patternswebsite2} and recalled in~\cite{decenttool2}). In this
context, to randomly generate formulae, we proceeded as follows. For a
given specification pattern, we randomly select one of the formulae
associated to it. Such a formulae is ``parametrised'' by some atomic
propositions. To obtain the randomly generated formula, using the
distributed alphabet, we randomly instantiate the atomic propositions.

The results of this test are reported in
Table~\ref{tab:bench:patterns}: for each kind of pattern (absence,
existence, bounded existence, universal, precedence, response, precedence chain, response chain, constrained chain), we
generated again 1,000 formulae, monitored over the same architecture as
used in Example~\ref{ex:prog}.
%
% \vspace{-1.5em}
\paragraph{Summary.}
Both benchmarks certainly substantiate that the decentralised monitoring
of an LTL formula induces a much lower communication overhead compared
to a centralised solution.  In fact, when considering the more realistic
benchmark using the specification patterns, the communication overhead
was significantly lower compared to monitoring randomly generated
formulae. The same is true for the delay: in case of monitoring LTL
formulae corresponding to specification patterns, the delay is almost
negligible; that is, the local monitors detect violation/satisfaction of
a monitored formula at almost the same time as a global monitor with access to all observations at any time.  Note that we have
further benchmarks available on \cite{decenttool2} (omitted for space reasons), also to highlight
the effect of differently sized alphabets and validate the maximal delay (Theorem~\ref{theo:maxdelay}).
Note further that in our tests, we have used continuous
simplification of the goal formulae in order to avoid a formula
explosion problem caused by rewriting.  In {\toolname}, advanced syntactic
simplification rules\footnote{Compared to RuleR~\cite{DBLP:journals/logcom/BarringerRH10}, the state-of-art rule-based runtime verification tool, for LTL specifications, our simplification function produced better results (see~\cite{decenttool2})} were introduced and sufficient for the purpose of our experiments.
% although there may be
% anatomic cases, where this can fail, of course.
%
%%%%%%%%%%%%%%%%%%%%%%%%%%%%%%%%%%%%%%%%%%%%%%%%%%%%%%%%%%%%%%%%%%%%%%%%%%%%%%%%%%%%%%%%%%%%%%%%%%%
%%%%%%%%%%%%%%%%%%%%%%%%%%%%%%%%%%%%%%%%%%%%%%%%%%%%%%%%%%%%%%%%%%%%%%%%%%%%%%%%%%%%%%%%%%%%%%%%%%%
%%%%%%%%%%%%%%%%%%%%%%%%%%%%%%%%%%%%%%%%%%%%%%%%%%%%%%%%%%%%%%%%%%%%%%%%%%%%%%%%%%%%%%%%%%%%%%%%%%%

% \vspace{-1.5em}
\section{Related work and conclusions}
\label{sec:conc}
\vspace{-0.5em}
This work is by no means the first to introduce an approach to
monitoring the behaviour of distributed systems.
For example, \cite{SenVAR06a} introduced M\textsc{t}TL, a temporal logic
for describing properties of asynchronous systems, as well as a
monitoring procedure that, given a partially ordered execution of a
parallel asynchronous system, establishes whether or not there exist
runs in the execution that violate a given M\textsc{t}TL correctness
property.  While at first this may seem to coincide with the work
presented in this paper, there are noteworthy differences:
First, many of the problems addressed in \cite{SenVAR06a} stem from the
fact that the systems to be monitored operate concurrently; that is,
create a partially ordered set of behaviours.  Our application domain
are distributed but synchronous systems.
Second, we take LTL ``off-the-shelf''; that is, we do not add modalities
to express properties concerning the distributed nature of the system
under scrutiny.  On the contrary, our motivation is to enable users to
conceive a possibly distributed system as a single, monolithic system by
enabling them to specify properties over the outside visible behaviour
only---independent of implementation specific-details, such as the
number of threads or components---and to automatically ``distribute the
monitoring'' process for such properties for them.  (Arguably, this also
bears the advantage that users do not need to learn another formalism to
express system properties.)
Finally, we address the fact that in many distributed systems it is not
possible to collect a global trace or insert a global decision making
point, thereby forcing the automatically distributed monitors to
communicate.  But at the same time we try and keep communication at a
minimum; that is, to not transmit the occurrence of every single
observed event, because many such applications would not tolerate this
kind of overhead.  This aspect, on the other hand, does not play a role
in \cite{SenVAR06a} where the implementation was tried on parallel
(Java) programs which are not executed on physically separated CPUs as
in our case, and where one can collect a set of global behaviours to
reason about.

Other recent works like \cite{DBLP:conf/fm/GenonMM06} target physically
distributed systems, but do not focus on the communication overhead that
may be induced by their monitoring.  Similarly, % to \cite{SenVAR06a},
this work also mainly addresses the problem of monitoring systems which
produce partially ordered traces ({\`a} la Diekert and Gastin),
and introduces abstractions to deal with the combinational explosion of
these traces.

To the best of our knowledge, our work is the first to address the
problem of automatically distributing LTL monitors, and to introduce a
decentralised monitoring approach that not only avoids a global point of
observation or any form of central trace collection, but also tries to
keep the number of communicated messages between monitors at a minimum.
What is more, our experimental results show that this approach does not
only ``work on paper'', but that it is feasible to be implemented.
Indeed, even the expected savings in communication overhead could be
observed for the set of chosen LTL formulae and the automatically
generated traces, when compared to a centralised solution in which the
local monitors transmit all observed events to a global monitor.

%%% Local Variables: 
%%% mode: latex
%%% TeX-master: "main"
%%% End: 

\bibliographystyle{unsrt}
% \vspace{-0.5em}
\bibliography{bibliography,longstrings,biblio}

\appendix
\section{Proofs}
\label{sec:proofs}
%%%%%%%%%%%%%%%%%%%%%%%%%%%%%%%%%%%%%%%%%%%%%%%%%%%%%%%%%%%%%%%%%%%%%%%
%%%%%%%%%%%%%%%%%%%%%%%%%%%%%%%%%%%%%%%%%%%%%%%%%%%%%%%%%%%%%%%%%%%%%%
%%%%%%%%%%%%%%%%%%%%%%%%%%%%%%%%%%%%%%%%%%%%%%%%%%%%%%%%%%%%%%%%%%%%%%
%
This section contains the proofs of the results stated in this paper.
%
%%%%%%%%%%%%%%%%%%%%%%%%%%%%%%%%%%%%%%%%%%%%%%%%%%%%%%%%%%%%%%%%%%%%%%
%%%%%%%%%%%%%%%%%%%%%%%%%%%%%%%%%%%%%%%%%%%%%%%%%%%%%%%%%%%%%%%%%%%%%%
\subsection{Proofs for Section~\ref{sec:prog}}
\label{sec:proof:prog}
%%%%%%%%%%%%%%%%%%%%%%%%%%%%%%%%%%%%%%%%%%%%%%%%%%%%%%%%%%%%%%%%%%%%%%
%%%%%%%%%%%%%%%%%%%%%%%%%%%%%%%%%%%%%%%%%%%%%%%%%%%%%%%%%%%%%%%%%%%%%%
%
\paragraph{Proof of Lemma~\ref{lem:mimicsevent}.}
The following inductive proof follows the argument conveyed by
Proposition~3 of \cite{Bacchus:1998:PTE:590220.590230}.  For
completeness sake, here we want to give the complete, formal, detailed proof.

The lemma is a direct consequence of the semantics of \LTL
(Definition~\ref{def:ltl_semantics}) and the definition of progression
(Definition~\ref{def:pf}).  
% It will be useful in the remaining proofs.  % AB@YF: What will be useful? YF@AB: This sentence did not really made sense as it is obvious that a lemma is used to prove a theorem :)
Recall that this lemma states that the progression function ``mimics''
the \LTL semantics on some event $\sigma$.
\begin{proof}
 We shall prove the following statement:
\[
 \forall \sigma\in\Sigma. \forall w\in\Sigma^\omega. \forall \varphi\in\LTL.\ \sigma\cdot w\models \varphi \Leftrightarrow w\models P(\varphi,\sigma).
\]
Let us consider an event $\sigma\in\Sigma$ and an infinite trace $w\in\Sigma^\omega$, the proof is done by a structural induction on $\varphi\in\LTL$.
\newline
\newline
\textbf{Base Case: $\varphi\in\{\top,\bot,p\in\AP{}\}$.}%
\begin{itemize}
 \item Case $\varphi=\top$. This case is trivial since, according to the definition of the progression function, $\forall\sigma\in\Sigma.\ P(\top,\sigma)=\top$. Moreover, according to the \LTL semantics of $\top$, $\forall w\in\Sigma^\omega.\ w\models\top$.
\item Case $\varphi=\bot$. This case is symmetrical to the previous one.
\item Case $\varphi = p\in \AP{}$. Recall that, according to the progression function for atomic propositions, we have $P(p,\sigma)=\top$ if $p\in\sigma$ and $\bot$ otherwise.
\begin{itemize}
\item Let us suppose that $\sigma\cdot w\models p$. According to the \LTL semantics of atomic propositions, it means that $p\in\sigma$, and thus $P(p,\sigma)=\top$. And, due to the \LTL semantics of $\top$, we have $\forall w\in\Sigma^\omega.\ w\models\top$.
\item Let us suppose that $w\models P(p,\sigma)$. Since $P(p,\sigma)\in\{\top,\bot\}$, we have necessarily $P(p,\sigma)=\top$. According to the progression function, $P(p,\sigma)=\top$ amounts to $p\in\sigma$. Using the \LTL semantics of atomic propositions, we deduce that $\sigma\cdot w\models p$.
\end{itemize}
\end{itemize}
\textbf{Induction Case: $\varphi\in\{\neg\varphi',\varphi_1\vee\varphi_2,\varphi_1\wedge\varphi_2,\ltlG\varphi',\ltlF\varphi',\ltlX\varphi',\varphi_1\ltlU\varphi_2\}$.} Our induction hypothesis states that the lemma holds for some formulae $\varphi',\varphi_1,\varphi_2\in\LTL$.
\begin{itemize}
\item Case $\varphi=\neg\varphi'$. On one hand, using the progression function for $\neg$, we have $P(\neg\varphi',\sigma)=\neg P(\varphi',\sigma)$. On the other hand, using the \LTL semantics of operator $\neg$, we have $w\models\varphi\Leftrightarrow w\not\models \neg\varphi$. Thus, we have $\sigma\cdot w\models \neg\varphi'$ iff $\sigma\cdot w\not\models\varphi'$ iff (induction hypothesis on $\varphi'$) $w\not\models P(\varphi',\sigma)$ iff $w\models \neg P(\varphi',\sigma)$ iff $w\models P(\neg\varphi',\sigma)$.
\item Case $\varphi=\varphi_1\vee\varphi_2$. Recall that, according to the progression function for operator $\vee$, we have $P(\varphi_1\vee\varphi_2,\sigma)= P(\varphi_1,\sigma)\vee P(\varphi_2,\sigma)$.
\begin{itemize}
\item Let us suppose that $\sigma\cdot w\models \varphi_1\vee\varphi_2$. We distinguish again two sub-cases: $\varphi_1\vee\varphi_2=\top$ or $\varphi_1\vee\varphi_2\neq\top$. If $\varphi_1\vee\varphi_2=\top$, then this case reduces to the case where $\varphi=\top$, already treated. If $\varphi_1\vee\varphi_2\neq\top$, it means that either $\sigma\cdot w\models \varphi_1$ or $\sigma\cdot w\models\varphi_2$. Let us treat the case where $\sigma\cdot w\models \varphi_1$ (the other case is similar). From $\sigma\cdot w\models \varphi_1$, we can apply the induction hypothesis on $\varphi_1$ to obtain $w\models P(\varphi_1,\sigma)$, then, $w\models P(\varphi_1,\sigma)\vee P(\varphi_1,\sigma) = P(\varphi_1\vee\varphi_2,\sigma)$.
\item Let us suppose that $w\models P(\varphi_1\vee\varphi_2,\sigma)=P(\varphi_1,\sigma)\vee P(\varphi_2,\sigma)$. We distinguish again two sub-cases: $P(\varphi_1\vee \varphi_2,\sigma)=\top$ or $P(\varphi_1\vee \varphi_2,\sigma)\neq\top$.
\begin{itemize}
\item If $P(\varphi_1\vee \varphi_2,\sigma)=\top$, then we again distinguish two sub-cases:
\begin{itemize}
\item If $P(\varphi_1,\sigma)=\top$ or $P(\varphi_2,\sigma)=\top$. Let us treat the case where $P(\varphi_1,\sigma)=\top$ (the other case is similar). Applying the induction hypothesis on $\varphi_1$, we have $\sigma\cdot w\models\varphi_1\Leftrightarrow w\models P(\varphi_1,\sigma)$. Then, consider $w\in\Sigma^\omega$, we have $\sigma\cdot w\models \varphi_1$, and consequently $\sigma\cdot w\models \varphi_1\vee\varphi_2$.
\item  If $P(\varphi_1,\sigma)\neq\top$ and $P(\varphi_2,\sigma)\neq\top$, then we have $P(\varphi_1,\sigma)=\neg P(\varphi_2,\sigma)$. Applying the induction hypothesis on $\varphi_1$ and $\varphi_2$, we obtain $\sigma\cdot w\models\varphi_1\Leftrightarrow\sigma\cdot w\not\models\varphi_2$. Let us consider $w\in\Sigma^\omega$. If $\sigma\cdot w\models\varphi_1$, then we have $\sigma\cdot w\models\varphi_1\vee\varphi_2$. Else ($\sigma\cdot w\not\models\varphi_1$), we have $\sigma\models\varphi_2$, and then $\sigma\cdot w\models \varphi_1\vee\varphi_2$.
\end{itemize}
\item If $P(\varphi_1\vee \varphi_2,\sigma)\neq\top$, then we have either $w\models P(\varphi_1,\sigma)$ or $w\models P(\varphi_2,\sigma)$. Let us treat the case where $w\models P(\varphi_1,\sigma)$ (the other case is similar). From $w\models P(\varphi_1,\sigma)$, we can apply the induction hypothesis on $\varphi_1$ to obtain $\sigma\cdot w\models \varphi_1$, and thus $\sigma\cdot w\models \varphi_1\vee \varphi_2$.
\end{itemize}
\end{itemize}
\item Case $\varphi=\varphi_1\wedge\varphi_2$. This case is similar to the previous one.
\item Case $\varphi=\ltlG\varphi'$. Recall that, according to the progression function for operator $\ltlG$, $P(\ltlG\varphi',\sigma)=P(\varphi',\sigma)\wedge \ltlG\varphi'$.
\begin{itemize}
\item Let us suppose that $\sigma\cdot w\models \ltlG\varphi'$. According to the \LTL semantics of operator $\ltlG$, we have $\forall i\in\mathbb{N}^{\geq 0}.\ (\sigma\cdot w)^i\models \varphi'$. In particular, it implies that $(\sigma\cdot w)^0\models \varphi'$, i.e., $\sigma\cdot w\models \varphi'$ and $\forall i\in\mathbb{N}^{\geq 0}.\ (\sigma\cdot w^1)^i\models\varphi'$, i.e., $(\sigma\cdot w)^1=w\models \ltlG\varphi'$. Using the induction hypothesis on $\varphi'$, from $\sigma\cdot w\models\varphi'$, we obtain $w\models P(\varphi',\sigma)$. As expected, according to the \LTL semantics  of operator $\wedge$, we have $w\models P(\ltlG\varphi',\sigma)\wedge\ltlG\varphi'=P(\ltlG\varphi',\sigma)$.
\item Let us suppose that $w\models P(\ltlG\varphi',\sigma)=P(\varphi',\sigma)\wedge \ltlG\varphi'$. It follows that $w\models P(\varphi',\sigma)$, and thus, using the induction hypothesis on $\varphi'$, $\sigma\cdot w\models \varphi'$. Using the \LTL semantics of operator $\ltlG$, from $\sigma\cdot w\models \varphi'$ and $w\models \ltlG\varphi'$, we deduce $\forall i\in\mathbb{N}^{\geq 0}.\ w^i\models \varphi'$, and then $\forall i\in\mathbb{N}.\ (\sigma\cdot w)^i\models \varphi'$, i.e., $\sigma\cdot w\models \ltlG\varphi'$.
\end{itemize}
\item Case $\varphi=\ltlF\varphi'$. This case is similar to the previous one.
\item Case $\varphi=\ltlX\varphi'$. On one hand, using the progression function for $\ltlX$, we have $P(\ltlX\varphi',\sigma)=\varphi'$. On the other hand, using the \LTL semantics of operator $\ltlX$, we have $\sigma\cdot w\models \ltlX\varphi'$ iff $w\models \varphi'$. Thus, we have $\sigma\cdot w\models \ltlX\varphi'$ iff $w\models \varphi'$ iff (induction hypothesis on $\varphi'$) $w\models P(\ltlX\varphi',\sigma)$.
\item Case $\varphi=\varphi_1\ltlU\varphi_2$.  Recall that, according to the progression function for operator $\ltlU$, $P(\varphi_1\ltlU\varphi_2,\sigma)=P(\varphi_2,\sigma)\vee (P(\varphi_1,\sigma)\wedge\varphi_1\ltlU\varphi_2)$.
\begin{itemize}
\item Let us suppose that $\sigma\cdot w\models \varphi_1\ltlU\varphi_2$. According to the \LTL semantics of operator $\ltlU$, we have $\exists i\in\mathbb{N}^{\geq 0}.\ (\sigma\cdot w)^i\models \varphi_2 \wedge \forall 0\leq l <i.\ (\sigma\cdot w)^l\models\varphi_1$. Let us distinguish two cases: $i=0$ and $i>0$.
\begin{itemize}
\item If $i=0$, then we have $\sigma\cdot w\models\varphi_2$. Applying the induction hypothesis on $\varphi_2$, we have $w\models P(\varphi_2,\sigma)$, and consequently $w\models P(\varphi_1\ltlU\varphi_2,\sigma)$.
\item Else ($i>0$), we have $\forall 0\leq l<i.\ (\sigma\cdot w)^l\models \varphi_1$. Consequently, we have $(\sigma\cdot w)^0 \models \varphi_1$, and thus $\sigma\cdot w\models \varphi_1$. Moreover, from $\forall 0\leq l<i.\ (\sigma\cdot w)^l\models \varphi_1$, we deduce $\forall 0\leq l<i-1.\ w^l\models \varphi_1$. From $(\sigma\cdot w)^i\models\varphi_2$, we deduce $w^{i-1}\models\varphi_2$. From $w^{i-1}\models\varphi_2$ and $\forall 0\leq l<i.\ (\sigma\cdot w)^l\models \varphi_1$, we deduce $w\models\varphi_1\ltlU\varphi_2$. Applying, the induction hypothesis on $\varphi_1$, from $\sigma\cdot w\models\varphi_1$, we obtain $w\models P(\varphi_1,\sigma)$. Finally, from $w\models\varphi_1\ltlU\varphi_2$ and $w\models P(\varphi_1,\sigma)$, we obtain $w\models P(\varphi_1\ltlU\varphi_2,\sigma)$.
\end{itemize}
\item Let us suppose that $w\models P(\varphi_1\ltlU\varphi_2,\sigma)$.

 We distinguish two cases: $P(\varphi_1\ltlU\varphi_2,\sigma)=\top$ and $P(\varphi_1\ltlU\varphi_2,\sigma)\neq\top$.
\begin{itemize}
\item If $P(\varphi_1\ltlU\varphi_2,\sigma)=P(\varphi_2,\sigma)\vee (P(\varphi_1,\sigma)\wedge\varphi_1\ltlU\varphi_2)=\top$. We distinguish again two sub-cases.
\begin{itemize}
\item If $P(\varphi_2,\sigma)=\top$ or $P(\varphi_1,\sigma)\wedge\varphi_1\ltlU\varphi_2=\top$. If $P(\varphi_2,\sigma)=\top$, then applying the induction hypothesis on $\varphi_2$, we have $\sigma\cdot w\models \varphi_2\Leftrightarrow w\models \top$. Then, from $\sigma\cdot w\models \varphi_2$, we obtain, according to the \LTL semantics of operator $\ltlU$, $\sigma\cdot w\models\varphi_1\ltlU\varphi_2$. If $P(\varphi_1,\sigma)\wedge\varphi_1\ltlU\varphi_2=\top$, we directly deduce that $\varphi_1\ltlU\varphi_2=\top$, and then this case reduces to the case where $\varphi=\top$, already treated.
\item If $P(\varphi_2,\sigma)\neq\top$ and $P(\varphi_1,\sigma)\wedge\varphi_1\ltlU\varphi_2\neq\top$, then we have $P(\varphi_2,\sigma)=\neg (P(\varphi_1,\sigma)\wedge\varphi_1\ltlU\varphi_2)=\neg P(\varphi_1,\sigma)\vee\neg (\varphi_1\ltlU\varphi_2)$. Applying the induction hypothesis on $\varphi_1$ and $\varphi_2$, we have $\sigma\cdot w\models\varphi_1\Leftrightarrow w\models P(\varphi_1,\sigma)$, and $\sigma\cdot w\models\varphi_2\Leftrightarrow w\models P(\varphi_2,\sigma)$, and thus $\sigma\cdot w\models\varphi_2\Leftrightarrow (\sigma\cdot w\not\models \varphi_1\vee w\not\models\varphi_1\ltlU\varphi_2)$. Let us now follow the \LTL semantics of operator $\ltlU$ and consider the two cases: $\sigma\cdot w\models \varphi_2$ or $\sigma\cdot w\not\models \varphi_2$. If $\sigma\cdot w\models \varphi_2$, thus $\sigma\cdot w\models \varphi_1\ltlU\varphi_2$ (according to the \LTL semantics of $\ltlU$). Else ($\sigma\cdot w\not\models \varphi_2$), then $\sigma\cdot w\models \varphi_1$ and $w\models\varphi_1\ltlU\varphi_2$, and thus $\sigma\cdot w\models \varphi_1\ltlU\varphi_2$.
\end{itemize}
\item  If $P(\varphi_1\ltlU\varphi_2,\sigma)\neq\top$, it means that either $w\models P(\varphi_2,\sigma)$ or $w\models P(\varphi_1,\sigma)$ $\wedge \varphi_1\ltlU\varphi_2$.
\begin{itemize}
\item If $w\models P(\varphi_2,\sigma)$, then applying the induction hypothesis on $\varphi_2$, we have $\sigma\cdot w\models\varphi_2$. Then, following the \LTL semantics of operator $\ltlU$, we obtain $\sigma\cdot w\models \varphi_1\ltlU\varphi_2$.
\item If $w\models P(\varphi_1,\sigma)\wedge\varphi_1\ltlU\varphi_2$, then we have $w\models P(\varphi_1,\sigma)$ and $w\models \varphi_1\ltlU\varphi_2$. Applying the induction hypothesis on $\varphi_1$, we have $\sigma\cdot w\models\varphi_1$. From $w\models \varphi_1\ltlU\varphi_2$, we have $\exists i\in\mathbb{N}^{\geq 0}.\ w^i\models\varphi_2\wedge \forall 0\leq l<i.\ w^l\models\varphi_1$. It implies that $(\sigma\cdot w)^{i+1}\models\varphi_2$ and $\forall 0<l< i+1.\ (\sigma\cdot w)^l\models\varphi_1$. Using, $\sigma\cdot w\models\varphi_1$, i.e., $(\sigma\cdot w)^0\models\varphi_1$ and the \LTL semantics of operator $\ltlU$, we finally obtain $\sigma\cdot w\models\varphi_1\ltlU\varphi_2$.
\end{itemize}
\end{itemize}
\end{itemize}
\end{itemize}
\qed
\end{proof}

%%% Local IspellDict: "british"
%%% Local Variables: 
%%% mode: latex
%%% TeX-master: "../main"
%%% End: 

%
%
%%%%%%%%%%%%%%%%%%%%%%%%%%%%%%%%%%%%%%%%%%%%%%%%%%%%%%%%%%%%%%%%%%%%%%
\paragraph{Proof of Lemma~\ref{lem:prog}.}
%%%%%%%%%%%%%%%%%%%%%%%%%%%%%%%%%%%%%%%%%%%%%%%%%%%%%%%%%%%%%%%%%%%%%%
%
We shall prove the following statement.
 \[
\begin{array}{lll}
\forall\varphi\in LTL. \forall\sigma\in\Sigma.& & P(\varphi, \sigma) = \top\Rightarrow \sigma \in \good(\varphi)\\
& \wedge &  P(\varphi, \sigma) = \bot \Rightarrow \sigma \in \bad(\varphi).
\end{array}
\]
The proof uses the definition of the \LTL semantics (Definition~\ref{def:ltl_semantics}), the definition of good and bad prefixes (Definition~\ref{def:pf}), the progression function (Definition~\ref{def:ltl_progression}), and Lemma~\ref{lem:mimicsevent}.
\begin{proof}
According to Lemma~\ref{lem:mimicsevent}, we have $\forall\sigma\in\Sigma.\forall w\in\Sigma^\omega.\ \sigma\cdot w\models \varphi \Leftrightarrow w\models P(\varphi,\sigma)$. Consequently, we have $\forall\sigma\in\Sigma.\forall w\in\Sigma^\omega.\ \sigma\cdot w\models \varphi \Leftrightarrow \forall\sigma\in\Sigma.\forall w\in\Sigma^\omega.\ w\models P(\varphi,\sigma)$ and $\forall\sigma\in\Sigma.\forall w\in\Sigma^\omega.\ \sigma\cdot w\not\models \varphi \Leftrightarrow \forall\sigma\in\Sigma.\forall w\in\Sigma^\omega.\ w\not\models P(\varphi,\sigma)$. Consequently, when $P(\varphi,\sigma)=\top$, we have $\forall\sigma\in\Sigma.\forall w\in\Sigma^\omega.\ \sigma\cdot w\models \varphi$, i.e., $\sigma\in\good(\varphi)$. Similarly, when $P(\varphi,\sigma)=\bot$, we have $\forall\sigma\in\Sigma.\forall w\in\Sigma^\omega.\ \sigma\cdot w\not\models \varphi$, i.e., $\sigma\in\bad(\varphi)$.

The proof can also be obtained in a more detailed manner as shown below. Let us consider $\sigma\in\Sigma$ and $\varphi\in \LTL$. The proof is performed by a structural induction on $\varphi$.
\newline
\newline
\textbf{Base Case: $\varphi\in\{\top,\bot,p\in\AP{}\}$.}
\begin{itemize}
\item Case $\varphi=\top$. In this case, the proof is trivial since $P(\top,\sigma)=\top$ and, according to the \LTL semantics of $\top$ and the definition of good prefixes, $\good(\top)=\Sigma^\ast$.
\item Case $\varphi=\bot$. Similarly, in this case, the proof is trivial since $P(\bot,\sigma)=\bot$ and $\bad(\bot)=\Sigma^\ast$.
\item Case $\varphi=p\in AP$.
\newline
Let us suppose that $P(\varphi,\sigma)=\top$. According to the progression function, it means that $p\in\sigma$. Moreover, since $\varphi=p$, according to the \LTL semantics of atomic propositions, for any $w\in\Sigma^\omega$, we have $\sigma\cdot w \models \varphi$. According to the definition of good prefixes, it means that $\sigma\in\good(\varphi)$.
\newline
The proof for $P(\varphi,\sigma)=\bot\Rightarrow \sigma\in\bad(\varphi)$ is similar.
\end{itemize}
\textbf{Induction Case: $\varphi\in\{\neg\varphi',\varphi_1\vee\varphi_2,\varphi_1\wedge\varphi_2,\ltlG\varphi',\ltlF\varphi',\ltlX\varphi',\varphi_1\ltlU\varphi_2\}$.} Our induction hypothesis states that the lemma holds for some formulae $\varphi',\varphi_1,\varphi_2\in\LTL$.
\begin{itemize}
\item Case $\varphi=\neg\varphi'$. In this case, the result is obtained by using the induction hypothesis on $\varphi'$ and the equality's $\bot=\neg\top$ and $\neg(\neg\varphi)=\varphi$.
\item Case $\varphi=\varphi_1\vee\varphi_2$. Recall that, according to the progression function for operator $\vee$, $P(\varphi_1\vee\varphi_2,\sigma)=P(\varphi_1,\sigma)\vee P(\varphi_2,\sigma)$.
\newline
Let us suppose that $P(\varphi,\sigma)=\top$. We distinguish two cases:
\begin{itemize}
\item If $P(\varphi_1,\sigma)=\top$ or $P(\varphi_2,\sigma)=\top$. Let us treat the case where $P(\varphi_1,\sigma)=\top$. Using the induction hypothesis on $\varphi_1$, we have $\sigma\in\good(\varphi_1)$. According to the definition of good prefixes, we have $\forall w\in\Sigma^\omega.\ \sigma\cdot w\models \varphi_1$. We easily deduce, using the \LTL semantics of operator $\vee$,  that $\forall w\in\Sigma^\omega.\ \sigma\cdot w \models \varphi_1\vee\varphi_2$, that is, $\sigma\in\good(\varphi_1\vee\varphi_2)$.
\item If $P(\varphi_1,\sigma)\neq\top$ and $P(\varphi_2,\sigma)\neq\top$. Since $P(\varphi,\sigma)=\top$, we have  $P(\varphi_1,\sigma)=\neg P(\varphi_2,\sigma)$. Using Lemma~\ref{lem:mimicsevent}, we have $\forall w\in\Sigma^\omega.\ \sigma\cdot w\models \varphi_1\Leftrightarrow w\models P(\varphi_1,\sigma)$ and $\forall w\in\Sigma^\omega.\ \sigma\cdot w\models \varphi_2\Leftrightarrow w\models P(\varphi_2,\sigma)$. We deduce that $\forall w\in\Sigma^\omega.\ \sigma\cdot w\models \varphi_1 \Leftrightarrow \sigma\cdot w\not\models \varphi_2$. Let us consider $w\in\Sigma^\omega$. If $\sigma\cdot w\models\varphi_1$, we have $\sigma\cdot w\models\varphi_1\vee\varphi_2$. Else ($\sigma\cdot w\not\models\varphi_1$), we have $\sigma\cdot w\models\varphi_2$, and then $\sigma\cdot w\models\varphi_2\vee\varphi_1$. That is, $\forall w\in\Sigma^\omega.\ \sigma\cdot w\models \varphi_1\vee\varphi_2$, i.e., $\sigma\in\good(\varphi_1\vee\varphi_2)$.
\end{itemize}
Let us suppose that $P(\varphi,\sigma)=\bot$. In this case, we have $P(\varphi_1,\sigma)=\bot$ and $P(\varphi_2,\sigma)=\bot$. Similarly, we can apply the induction hypothesis on $\varphi_1$ and $\varphi_2$ to find that $\sigma$ is bad prefix of both $\varphi_1$ and $\varphi_2$, and is thus a bad prefix of $\varphi_1\vee\varphi_2$ (using the \LTL semantics of operator $\vee$).
\item Case $\varphi=\varphi_1\wedge\varphi_2$. This case is symmetrical to the previous one.
\item Case $\varphi=\ltlG\varphi'$. Recall that, according to the progression function for operator $\ltlG$, $P(\ltlG\varphi',\sigma)=P(\varphi',\sigma) \wedge \ltlG\varphi'$.
\newline
Let us suppose that $P(\varphi,\sigma)=\top$. It means that $P(\varphi',\sigma)=\top$ and $\ltlG\varphi'=\top$. This case reduces to the case where $\varphi=\top$.
\newline
Let us suppose that $P(\varphi,\sigma)=\bot$. We distinguish two cases.
\begin{itemize}
\item If $P(\varphi',\sigma)=\bot$ or $\ltlG\varphi'=\bot$. We distinguish again two sub-cases.
\begin{itemize}
\item Sub-case $P(\varphi',\sigma)=\bot$. Using the induction hypothesis on $\varphi'$, we deduce that $\sigma\in\bad(\varphi')$, i.e., $\forall w\in\Sigma^\omega.\ \sigma\cdot w\not\models\varphi'$. Following the \LTL semantics of operator $\ltlG$, we deduce that $\forall w\in\Sigma^\omega.\ \sigma\cdot w\not\models\ltlG\varphi'$, i.e., $\sigma\in\bad(\ltlG \varphi')$.
\item Sub-case $\ltlG\varphi'=\bot$. This case reduces to the case where $\varphi=\bot$.
\end{itemize}
\item If $P(\varphi',\sigma)\neq\bot$ and $\ltlG\varphi'\neq\bot$. From $P(\varphi',\sigma) \wedge \ltlG\varphi'=\bot$, we deduce that $P(\varphi',\sigma) = \neg \ltlG\varphi'$. Using Lemma~\ref{lem:mimicsevent} on $\varphi'$, we have $\forall w\in\Sigma^\omega.\ \sigma\cdot w\models \varphi' \Leftrightarrow w\models P(\varphi',\sigma)$. Thus $\forall w\in\Sigma^\omega.\ \sigma\cdot w\models \varphi' \Leftrightarrow w\not\models \ltlG\varphi'$. Let us consider $w\in\Sigma^\omega$. If $\sigma\cdot w\models \varphi'$, then we have $w\not\models \ltlG\varphi'$. According to the \LTL semantics of operator $\ltlG$, it means that $\exists i\in\mathbb{N}^{\geq0}.\ w^i\not\models\varphi'$. Thus, still following the \LTL semantics of operator $\ltlG$, $(\sigma\cdot w)^{i+1}\not\models\varphi'$, and, consequently $\sigma\cdot w\not\models\ltlG\varphi'$. Else ($\sigma\cdot w\not\models \varphi'$), we have directly $\sigma\cdot w\not\models \ltlG\varphi'$.
\end{itemize}
\item Case $\varphi=\ltlF\varphi'$. Recall that, according to the progression function for operator $\ltlF$, $P(\ltlF\varphi',\sigma)=P(\varphi',\sigma)\vee \ltlF\varphi'$.
\newline
Let us suppose that $P(\varphi,\sigma)=\top$. We distinguish two cases.
\begin{itemize}
\item If $P(\varphi',\sigma)=\top$ or $\ltlF\varphi'=\top$.
\begin{itemize}
\item Sub-case $P(\varphi',\sigma)=\top$. Following the previous reasoning, using the induction hypothesis on $\varphi'$, the \LTL semantics of operator $\ltlF$, and the definition of good prefixes, we obtain the expected result.
\item Sub-case $\ltlF\varphi'=\top$. This case reduces to the case where $\varphi=\top$.
\end{itemize}
\item If $P(\varphi',\sigma)\neq\top$ and $\ltlF\varphi'\neq\top$. From $P(\varphi',\sigma) \vee \ltlG\varphi'=\bot$, we deduce that $P(\varphi',\sigma) = \neg \ltlF\varphi'$. Using Lemma~\ref{lem:mimicsevent} on $\varphi'$, we have $\forall w\in\Sigma^\omega.\ \sigma\cdot w\models \varphi' \Leftrightarrow w\models P(\varphi',\sigma)$. We thus have $\forall w\in\Sigma^\omega.\ \sigma\cdot w\models \varphi' \Leftrightarrow w\not\models \ltlF\varphi'$. Let us consider $w\in\Sigma^\omega$. If $\sigma\cdot w\models \varphi'$, using the \LTL semantics of  operator $\ltlF$, we have directly $\sigma\cdot w\models \ltlF\varphi'$. Else ($\sigma\cdot w\not\models \varphi'$), we have $w\models \ltlF\varphi'$. According to the \LTL semantics of operator $\ltlF$, it means that $\exists i\in\mathbb{N}^{\geq0}.\ w^i\models\varphi'$, and thus $(\sigma\cdot w)^{i+1}\models\varphi'$. Consequently $\sigma\cdot w\models\ltlF\varphi'$. That is, $\sigma\in\good(\ltlF\varphi')$.
\end{itemize}
Let us suppose that $P(\varphi,\sigma)=\bot$. It means that $P(\varphi',\sigma)=\bot$ and $\ltlF\varphi'=\bot$. A similar reasoning as the one used for the case $\varphi=\ltlG\varphi'$ and $P(\varphi,\sigma)=\top$ can be applied to obtain the expected result.
\item Case $\varphi=\ltlX\varphi'$. Recall that, according to the progression function for operator $\ltlX$, $P(\ltlX\varphi',\sigma)=\varphi'$.
\newline
Let us suppose that $P(\varphi,\sigma)=\top$. It means that $\varphi'=\top$. According to the \LTL semantics of $\top$, we have $\forall w\in\Sigma^\omega.\ w\models\varphi'$. Then, $\forall w\in\Sigma^\omega.\ \sigma\cdot w\models\ltlX\varphi'=\varphi$. That is, $\sigma\in\good(\ltlX\varphi')$.
\newline
Let us suppose that $P(\varphi,\sigma)=\bot$. It means that $\varphi'=\bot$. According to the \LTL semantics of $\bot$, we have $\forall w\in\Sigma^\omega.\ w\not\models\varphi'$. Then, $\forall w\in\Sigma^\omega.\ \sigma\cdot w\not\models\ltlX\varphi'=\varphi$. That is, $\sigma\in\bad(\ltlX\varphi')$.
\item Case $\varphi=\varphi_1 \ltlU \varphi_2$. Recall that, according to the progression function for operator $\ltlU$, $P(\varphi_1\ltlU\varphi_2,\sigma)=P(\varphi_2,\sigma)\vee  (P(\varphi_1,\sigma)\wedge \varphi_1 \ltlU\varphi_2)$.
\newline
Let us suppose that $P(\varphi,\sigma)=\top$. We distinguish two cases.
\begin{itemize}
\item If $P(\varphi_2,\sigma)=\top$ or $P(\varphi_1,\sigma)\wedge \varphi_1 \ltlU\varphi_2=\top$.
\begin{itemize}
\item Sub-case $P(\varphi_2,\sigma)=\top$. Using the induction hypothesis on $\varphi_2$, we have $\sigma\in\good(\varphi_2)$. Let us consider $w\in\Sigma^\omega$, we have $\sigma\cdot w\in \cL(\varphi_2)$, i.e., $(\sigma\cdot w)^0\models \varphi_1\ltlU\varphi_2$. According to the \LTL semantics of $\ltlU$, we have $\sigma\cdot w\models \varphi_1\vee\varphi_2$, i.e., $\sigma\cdot w \in \cL(\varphi_1\ltlU\varphi_2)$. We deduce that $\sigma\in\good(\varphi_1 \ltlU \varphi_2)$.
\item Sub-case $P(\varphi_1,\sigma) \wedge \varphi_1\ltlU\varphi_2=\top$. Necessarily, $\varphi_1\ltlU\varphi_2=\top$ and this case reduces to the first one already treated.
\end{itemize}
\item If $P(\varphi_2,\sigma)\neq\top$ and $P(\varphi_1,\sigma)\wedge \varphi_1 \ltlU\varphi_2\neq\top$. From $P(\varphi_1\ltlU\varphi_2,\sigma)=\top$, we deduce that $P(\varphi_2,\sigma) = \neg (P(\varphi_1,\sigma)\wedge \varphi_1 \ltlU\varphi_2)$. Applying Lemma~\ref{lem:mimicsevent} to $\varphi_2$, we obtain $\forall w\in\Sigma^\omega.\ \sigma\cdot w\models \varphi_2 \Leftrightarrow w\models P(\varphi_2,\sigma)$. We thus have $\forall w\in\Sigma^\omega.\ \sigma\cdot w\models \varphi_2 \Leftrightarrow w \not\models P(\varphi_1,\sigma)\wedge \varphi_1 \ltlU\varphi_2$. Let us consider $w\in\Sigma^\omega$. Let us distinguish two cases. If $\sigma\cdot w\models\varphi_2$, according to the \LTL semantics of $\ltlU$, we have $\sigma\cdot w\models \varphi_1\ltlU\varphi_2$. Else ($\sigma\cdot w\not\models\varphi_2$), it implies that  $\sigma\cdot w\models P(\varphi_1,\sigma)\wedge \varphi_1 \ltlU\varphi_2$, and, in particular $\sigma\cdot w\models \varphi_1 \ltlU\varphi_2$. That is, in both cases, $\sigma\in\good(\varphi_1\ltlU\varphi_2)$.
\end{itemize}
\end{itemize}
\end{proof}
\paragraph{Additional notation.}
For the remaining proofs, we define ${\cal P}$,  the extended progression function on traces that consists in applying successively the progression function defined so far to each event in order.
\begin{definition}
\label{def:epf}
Given a formula $\varphi\in\LTL$ and a trace $u=u(0)\cdots u(t-1)\in\Sigma^+$, the application of extended progression function ${\cal P}$ to $\varphi$ and $u$ is defined as:
\[
 {\cal P}(\varphi, u(0) \cdots u(t-1))={\cal P}(\varphi,u)=P( \ldots  (P(\varphi, u(0)), \ldots, u(t-1))))
\]
\end{definition}
For the sake of readability, in the remainder, we overload the notation of the progression function on events to traces, i.e., ${\cal P}(\varphi,u)$ is denoted $P(\varphi,u)$.
\paragraph{Some intermediate lemmas.}
Based on the previous introduced notation and the definition of the progression function (Definition~\ref{def:pf}), we extend the progression function to traces. The following lemma states some equality's that directly follow from an inductive application of the definition of the progression function on events.
\begin{lemma}
\label{lem:prog_trace}
Given some formulae $\varphi,\varphi_1,\varphi_2\in\LTL$, and a trace
$u\in\Sigma^+$, the progression function can be extended to the trace
$u$ by successively applying the previously defined progression function
to each event of $u$ in order. Moreover, we have:
$\forall\varphi,\varphi_1,\varphi_2\in\LTL.\forall u\in\Sigma^+$.
\[
 \begin{array}{rcl}
P(\top,u) &=& \top,\\
P(\bot,u) &=& \bot,\\
P(p\in\AP,u) &=& \top\ \text{if}\ p\in u(0), \bot\ \text{otherwise}, \\
P(\neg\varphi,u)&=&\neg P(\varphi,u),\\
P(\varphi_1\vee\varphi_2,u)&=&P(\varphi_1,u)\vee P(\varphi_2,u),\\
P(\varphi_1\wedge\varphi_2,u)&=&P(\varphi_1,u)\wedge P(\varphi_2,u),\\
P(\ltlG\varphi,u)&=& \bigwedge_{i=0}^{|u|-1} P(\varphi,u^i)\wedge \ltlG\varphi,\\
P(\ltlF\varphi,u)&=& \bigvee_{i=0}^{|u|-1} P(\varphi,u^i)\vee \ltlF\varphi,\\
P(\ltlX\varphi,u)&=&
\left\{
\begin{array}{ll}
 \varphi & \text{if $|u|=1$}\\
P(\varphi,u^1) & \text{otherwise}
\end{array}\right.\\
P(\varphi_1\ltlU\varphi_2,u) &=& 
\left\{
\begin{array}{ll}
 P(\varphi_2,u) \vee P(\varphi_1,u)\wedge \varphi_1\ltlU\varphi_2 & \text{if $|u|=1$}\\
\bigvee_{i=0}^{|u|-1} \big(P(\varphi_2,u^i) \wedge \bigwedge_{j=0}^{i-1} P(\varphi_1,u^j)\big) \vee \bigwedge_{i=0}^{|u|-1} P(\varphi_1,u^i)\wedge \varphi_1\ltlU\varphi_2 & \text{otherwise}
\end{array}\right.\\

 \end{array}
\]
\end{lemma}
\begin{proof}
The proof is done by two inductions: an induction on the length of the trace $u$ (which is also the number of times the progression function is applied) and a structural induction on $\varphi\in\LTL$.
\newline
\textbf{Base Case: $u=\sigma\in\Sigma, |u|=1$.}
\newline
In this case, the result holds thanks to the definition of the progression function.
\newline
\textbf{Induction case:}
\newline
Let us suppose that the lemma holds for any trace $u\in\Sigma^+$ of some length $t\in\mathbb{N}$ and let us consider the trace $u\cdot\sigma\in\Sigma^+$, we perform a structural induction on $\varphi\in\LTL$.
\newline
\textit{Structural Base case: $\varphi\in\{\top,\bot,p\in\AP{}\}$.}
\begin{itemize}
 \item Case $\varphi=\top$. In this case the result is trivial since we have:
\[
\begin{array}{rcll}
P(\top,u\cdot\sigma) &= &P(P(\top,u),\sigma) & \text{(extended progression)}\\
&=&P(\top,\sigma) & \text{(induction hypothesis on $u$)}\\
&=&\top & \text{(progression on events)}
\end{array}
\]
\item Case $\varphi=\bot$. This case is symmetrical to the previous one.
\item Case $\varphi = p\in \AP{}$. Let us distinguish two cases: $p\in u(0)$ or $p\notin u(0)$.
\begin{itemize}
 \item If $p\in u(0)$, we have:
\[
\begin{array}{rcll}
P(p,u\cdot\sigma) &= &P(P(p,u),\sigma) & \text{(extended progression)}\\
&=&P(\top,\sigma) & \text{(induction hypothesis on $u$)}\\
&=&\top & \text{(progression on events)}
\end{array}
\]
 \item If $p\notin u(0)$, we have:
\[
\begin{array}{rcll}
P(p,u\cdot\sigma) &= &P(P(p,u),\sigma) & \text{(extended progression)}\\
&=&P(\bot,\sigma) & \text{(induction hypothesis on $u$)}\\
&=&\bot & \text{(progression on events)}
\end{array}
\]
\end{itemize}
\end{itemize}
\textit{Induction Case: $\varphi\in\{\neg\varphi',\varphi_1\vee\varphi_2,\varphi_1\wedge\varphi_2,\ltlG\varphi',\ltlF\varphi',\ltlX\varphi',\varphi_1\ltlU\varphi_2\}$.} Our induction hypothesis states that the lemma holds for some formulae $\varphi',\varphi_1,\varphi_2\in\LTL$.
\begin{itemize}
\item Case $\varphi=\neg\varphi'$. We have:
\[
\begin{array}{rcll}
P(\neg\varphi',u\cdot\sigma) &= &P(P(\neg\varphi',u),\sigma) & \text{(extended progression)}\\
&=& P(\neg  P(\varphi',u),\sigma) & \text{(induction hypothesis on $u$ and $\varphi'$)}\\
&=&\neg P( P(\varphi',u),\sigma) & \text{(progression on events)}\\
&=&\neg P(\varphi',u\cdot\sigma) & \text{(extended progression)}
\end{array}
\]
\item Case $\varphi=\ltlX\varphi'$. We have:
\[
\begin{array}{rcll}
P(\ltlX\varphi',u\cdot\sigma) & = & P(P(\ltlX\varphi',u),\sigma) & \text{(extended progression)}\\
&=& P(P(\varphi',u^1),\sigma) & \text{(induction hypothesis on $u$ and $\varphi'$)}\\
&=&P(\varphi',u^1\sigma) & \text{(extended progression)}\\
&=&P(\varphi',(u\cdot\sigma)^1) &\\
\end{array}
\]
\item Case $\varphi=\varphi_1\vee\varphi_2$. We have:
\end{itemize}
\[
\begin{array}{rcll}
P(\varphi_1\vee\varphi_2,u\cdot\sigma) &= &P(P(\varphi_1\vee\varphi_2,u),\sigma) & \text{(extended progression)}\\
&=& P(P(\varphi_1,u)\vee P(\varphi_2,u),\sigma) & \text{(induction hypothesis on $u$ and $\varphi_1,\varphi_2$)}\\
&=& P(P(\varphi_1,u),\sigma) \vee P(P(\varphi_2,u),\sigma) & \text{(progression on events)}\\
&=& P(\varphi_1,u\cdot\sigma) \vee P(\varphi_2,u\cdot\sigma) & \text{(extended progression)}\\
\end{array}
\]
\begin{itemize}
\item Case $\varphi=\varphi_1\wedge\varphi_2$. This case is similar to the previous one.
\item Case $\varphi=\ltlG\varphi'$. We have:
\end{itemize}
\[
\begin{array}{ll}
P(\ltlG\varphi',u\cdot\sigma) & \\
\quad = P(P(\ltlG\varphi',u),\sigma) & \text{(extended progression)}\\
\quad = P(\bigwedge_{i=0}^{|u|-1} P(\varphi',u^i)\wedge \ltlG\varphi',\sigma) & \text{(induction hypothesis on $u$ and $\varphi'$)}\\
\quad = P(\bigwedge_{i=0}^{|u|-1} P(\varphi',u^i),\sigma) \wedge P( \ltlG\varphi',\sigma)  & \text{(progression on events for $\wedge$)}\\
\quad = \bigwedge_{i=0}^{|u|-1} P(P(\varphi',u^i),\sigma) \wedge P( \ltlG\varphi',\sigma)  & \text{(extended progression for $\wedge$)}\\
\quad = \bigwedge_{i=0}^{|u|-1} P(\varphi',u^i\cdot\sigma) \wedge P( \ltlG\varphi',\sigma)  & \text{(extended progression)}\\
\quad = \bigwedge_{i=0}^{|u|-1} P(\varphi',u^i\cdot\sigma) \wedge P(\varphi',\sigma)\wedge  \ltlG\varphi'  & \text{(progression on events for $\ltlG$)}\\
\quad = \bigwedge_{i=0}^{|u\cdot\sigma|-2} P(\varphi',(u\cdot\sigma)^i) \wedge P(\varphi',(u\cdot\sigma)^{|u\cdot\sigma|-1})\wedge  \ltlG\varphi'  & \text{($u^i\cdot\sigma=(u\cdot\sigma)^i$ and $\sigma=(u\cdot\sigma)^{|u\cdot\sigma|-1}$)}\\
\quad = \bigwedge_{i=0}^{|u\cdot\sigma|-1} P(\varphi',(u\cdot\sigma)^i)\wedge \ltlG\varphi' & 
\end{array}
\]
\begin{itemize}
\item Case $\varphi=\ltlF\varphi'$. We have:
\[
\begin{array}{ll}
P(\ltlF\varphi',u\cdot\sigma) & \\
\quad = P(P(\ltlF\varphi',u),\sigma) & \text{(extended progression)}\\
\quad = P(\bigvee_{i=0}^{|u|-1} P(\varphi',u^i)\vee \ltlF\varphi',\sigma) & \text{(induction hypothesis on $u$ and $\varphi'$)}\\
\quad = P(\bigvee_{i=0}^{|u|-1} P(\varphi',u^i),\sigma) \vee P( \ltlF\varphi',\sigma)  & \text{(progression on events)}\\
\quad = \bigvee_{i=0}^{|u|-1} P(\varphi',u^i\cdot\sigma) \vee P( \ltlF\varphi',\sigma)  & \text{(extended progression for $\vee$)}\\
\quad = \bigvee_{i=0}^{|u|-1} P(\varphi',u^i\cdot\sigma) \vee P(\varphi',\sigma)\vee  \ltlF\varphi'  & \text{(progression on events for $\ltlF$)}\\
\quad = \bigvee_{i=0}^{|u\cdot\sigma|-2} P(\varphi',(u\cdot\sigma)^i) \vee P(\varphi',(u\cdot\sigma)^{|u\cdot\sigma|-1}) \vee  \ltlF\varphi'  & \text{($u^i\cdot\sigma=(u\cdot\sigma)^i$ and $\sigma=(u\cdot\sigma)^{|u\cdot\sigma|-1}$)}\\
\quad = \bigvee_{i=0}^{|u\cdot\sigma|-1} P(\varphi',(u\cdot\sigma)^i)\vee \ltlF\varphi' & 
\end{array}
\]
\item Case $\varphi=\varphi_1\ltlU\varphi_2$. We have:
\end{itemize}
\[
\begin{array}{l}
P(\varphi_1\ltlU\varphi_2,u\cdot\sigma)  \\
\text{(extended progression)}\\
\quad = P(P(\varphi_1\ltlU\varphi_2,u),\sigma)\\
\text{(induction hypothesis on $u$, and structural induction hypothesis on $\varphi_1$ and $\varphi_2$)}\\
\quad = P\Big(\bigvee_{i=0}^{|u|-1} \big(P(\varphi_2,u^i) \wedge \bigwedge_{j=0}^{i-1} P(\varphi_1,u^j)\big) \vee \bigwedge_{i=0}^{|u|-1} P(\varphi_1,u^i)\wedge \varphi_1\ltlU\varphi_2,\sigma\Big)\\
\text{(progression on events for $\vee$)}\\
\quad = P\Big(\bigvee_{i=0}^{|u|-1} \big(P(\varphi_2,u^i) \wedge \bigwedge_{j=0}^{i-1} P(\varphi_1,u^j)\big),\sigma) \vee P(\bigwedge_{i=0}^{|u|-1} P(\varphi_1,u^i)\wedge \varphi_1\ltlU\varphi_2,\sigma\Big)\\
\text{(progression on events for $\wedge$ and $\vee$)}\\
\quad = \bigvee_{i=0}^{|u|-1} \big(P(P(\varphi_2,u^i),\sigma) \wedge \bigwedge_{j=0}^{i-1} P(P(\varphi_1,u^j),\sigma)\big)  \vee \bigwedge_{i=0}^{|u|-1} P(P(\varphi_1,u^i),\sigma)\wedge P(\varphi_1\ltlU\varphi_2,\sigma)\\
\text{(extended progression)}\\
\quad = \bigvee_{i=0}^{|u|-1} \big(P(\varphi_2,u^i\cdot\sigma) \wedge \bigwedge_{j=0}^{i-1} P(\varphi_1,u^j\cdot\sigma)\big) \vee \bigwedge_{i=0}^{|u|-1} P(\varphi_1,u^i\cdot\sigma)\wedge P(\varphi_1\ltlU\varphi_2,\sigma)\\
\end{array}
\]
Moreover:
\[
\begin{array}{l}
\bigwedge_{i=0}^{|u|-1} P(\varphi_1,u^i\cdot\sigma)\wedge P(\varphi_1\ltlU\varphi_2,\sigma) \\
\text{(progression on events for $\ltlU$)}\\
\quad = \bigwedge_{i=0}^{|u|-1} P(\varphi_1,u^i\cdot\sigma)\wedge (P(\varphi_2,\sigma)\vee P(\varphi_1,\sigma)\wedge \varphi_1\ltlU\varphi_2) \\
\text{(distribution of $\wedge$ over $\vee$)}\\
\quad = \big(\bigwedge_{i=0}^{|u|-1} P(\varphi_1,u^i\cdot\sigma) \wedge P(\varphi_2,\sigma) \big)\vee \big(\bigwedge_{i=0}^{|u|-1} P(\varphi_1,u^i\cdot\sigma)\wedge P(\varphi_1,\sigma)\wedge \varphi_1\ltlU\varphi_2\big) \\
\text{($\sigma=(u\cdot\sigma)^{|u\cdot\sigma|-1}$ and elimination of $P(\varphi_1,\sigma)$)}\\
\quad = \big(\bigwedge_{i=0}^{|u|-1} P(\varphi_1,u^i\cdot\sigma) \wedge P(\varphi_2,\sigma) \big)\vee \big(\bigwedge_{i=0}^{|u\cdot\sigma|-1} P(\varphi_1,u^i\cdot\sigma)\wedge \varphi_1\ltlU\varphi_2\big) \\
\end{array}
\]
Furthermore:
\[
\begin{array}{l}
\bigvee_{i=0}^{|u|-1} \big(P(\varphi_2,u^i\cdot\sigma) \wedge \bigwedge_{j=0}^{i-1} P(\varphi_1,u^j\cdot\sigma)\big)\vee \big(\bigwedge_{i=0}^{|u|-1} P(\varphi_1,u^i\cdot\sigma) \wedge P(\varphi_2,\sigma) \big)\\
\text{(variable renaming)}\\
\quad =\bigvee_{i=0}^{|u|-1} \big(P(\varphi_2,u^i\cdot\sigma) \wedge \bigwedge_{j=0}^{i-1} P(\varphi_1,u^j\cdot\sigma)\big)\vee \big( P(\varphi_2,\sigma)\wedge \bigwedge_{j=0}^{|u|-1} P(\varphi_1,u^j\cdot\sigma) \big)\\
\text{($\sigma=(u\cdot\sigma)^{|u\cdot\sigma|-1}$)}\\
\quad = \bigvee_{i=0}^{|u\cdot\sigma|-2} \big(P(\varphi_2,(u\cdot\sigma)^i) \wedge \bigwedge_{j=0}^{i-1} P(\varphi_1,u^j\cdot\sigma)\big)\vee \big( P(\varphi_2,(u\cdot\sigma)^{|u\cdot\sigma|-1})\wedge \bigwedge_{j=0}^{|u\cdot\sigma|-2} P(\varphi_1,u^j\cdot\sigma) \big)\\
\quad = \bigvee_{i=0}^{|u\cdot\sigma|-1} \big(P(\varphi_2,(u\cdot\sigma)^i) \wedge \bigwedge_{j=0}^{i-1} P(\varphi_1,(u\cdot\sigma)^j)\big)
\end{array}
\]
Finally:
\[
\begin{array}{l}
P(\varphi_1\ltlU\varphi_2,u\cdot\sigma)  \\
\quad = \bigvee_{i=0}^{|u|-1} \big(P(\varphi_2,u^i\cdot\sigma) \wedge \bigwedge_{j=0}^{i-1} P(\varphi_1,u^j\cdot\sigma)\big) \vee \big(\bigwedge_{i=0}^{|u|-1} P(\varphi_1,u^i\cdot\sigma) \wedge P(\varphi_2,\sigma) \big)\\
\qquad\qquad \vee \big(\bigwedge_{i=0}^{|u\cdot\sigma|-1} P(\varphi_1,u^i\cdot\sigma)\wedge \varphi_1\ltlU\varphi_2\big)\\
\quad =\bigvee_{i=0}^{|u\cdot\sigma|-1} \big(P(\varphi_2,u^i\cdot\sigma) \wedge \bigwedge_{j=0}^{i-1} P(\varphi_1,u^j\cdot\sigma)\big) \vee \big(\bigwedge_{i=0}^{|u\cdot\sigma|-1} P(\varphi_1,u^i\cdot\sigma)\wedge \varphi_1\ltlU\varphi_2\big)\\
\quad =\bigvee_{i=0}^{|u\cdot\sigma|-1} \big(P(\varphi_2,(u\cdot\sigma)^i) \wedge \bigwedge_{j=0}^{i-1} P(\varphi_1,(u\cdot\sigma)^j)\big) \\
\qquad\qquad \vee \big(\bigwedge_{i=0}^{|u\cdot\sigma|-1} P(\varphi_1,(u\cdot\sigma)^i)\wedge \varphi_1\ltlU\varphi_2\big)
\end{array}
\]
\qed
\end{proof}

%%% Local IspellDict: "british"
%%% Local Variables: 
%%% mode: latex
%%% TeX-master: "../main"
%%% End: 

%
We introduce another intermediate lemma, which is a consequence of the
definition of the \LTL semantics (Definition~\ref{def:ltl_semantics}) and
the definition of the progression function
(Definition~\ref{def:epf}). This lemma will be useful in the remaining
proofs. This lemma states that the progression function ``mimics'' the
semantics of \LTL on a trace $u\in\Sigma^+$.
\begin{lemma}
\label{lem:mimicstrace}
Let $\varphi$ be an \LTL formula, $u\in\Sigma^+$
a non-empty trace and $w\in\Sigma^\omega$ an infinite trace, we have $u\cdot
w\models \varphi \Leftrightarrow w\models P(\varphi,u)$.
\end{lemma}
\begin{proof}
 We shall prove the following statement:
\[
 \forall u\in\Sigma^+. \forall w\in\Sigma^\omega. \forall \varphi\in\LTL.\ u \cdot w\models \varphi \Leftrightarrow w\models P(\varphi,u).
\]
Let us consider $u\in\Sigma^+$, the proof is done by a structural induction on $\varphi\in\LTL$.
\newline
\textbf{Base case: $\varphi\in\{\top,\bot,p\in\AP{}\}$.}
\begin{itemize}
 \item Case $\varphi=\top$. This case is trivial since, using Lemma~\ref{lem:prog_trace} on $\top$ and $u$, we have $P(\top,u)=\top$. Moreover, according to the \LTL semantics of $\top$, $\forall w\in\Sigma^\omega.\ u\cdot w\models\top$.
\item Case $\varphi=\bot$. This case is symmetrical to the previous one.
\item Case $\varphi = p\in \AP{}$.
\begin{itemize}
\item Let us suppose that $u\cdot w\models p$. By applying Lemma~\ref{lem:prog_trace} on $\top$ and $u$, we have $P(u,p)=\top$. Moreover, due to the \LTL semantics of $\top$, we have $\forall w\in\Sigma^\omega.\ w\models\top=P(u,p)$.
\item Let us suppose that $w\models P(p,u)$. Since $P(p,u)\in\{\top,\bot\}$, we have necessarily $P(p,u)=\top$. According to the progression function, $P(p,u)=\top$ necessitates that $p\in u(0)$. Using the \LTL semantics of atomic propositions, we deduce that $(u\cdot w)^0\models p$, i.e., $u\cdot w\models p$.
\end{itemize}
\end{itemize}
\textbf{Induction Case: $\varphi\in\{\neg\varphi',\varphi_1\vee\varphi_2,\varphi_1\wedge\varphi_2,\ltlG\varphi',\ltlF\varphi',\ltlX\varphi',\varphi_1\ltlU\varphi_2\}$.} Our induction hypothesis states that the lemma holds for some formulae $\varphi',\varphi_1,\varphi_2\in\LTL$.
\begin{itemize}
\item Case $\varphi=\varphi_1\vee\varphi_2$. Recall that, by applying Lemma~\ref{lem:prog_trace} on $\varphi_1\vee\varphi_2$ and $u$, we have $P(\varphi_1\vee\varphi_2,u)=P(\varphi_1,u)\vee P(\varphi_2,u)$.
\begin{itemize}
\item Let us suppose that $u\cdot w\models \varphi_1\vee\varphi_2$. Let us distinguish two cases: $\varphi_1\vee\varphi_2=\top$ and $\varphi_1\vee\varphi_2\neq\top$. If $\varphi_1\vee\varphi_2=\top$, then this case reduces to the case where $\varphi=\top$ already treated. If $\varphi_1\vee\varphi_2\neq\top$, it means that either $u\cdot w\models \varphi_1$ or $u\cdot w\models\varphi_2$. Let us treat the case where $u\cdot w\models \varphi_1$ (the other case is similar). From $u\cdot w\models \varphi_1$, we can apply the structural induction hypothesis on $\varphi_1$ to obtain $w\models P(\varphi_1,u)$, and  then, $w\models P(\varphi_1,u)\vee P(\varphi_2,u) = P(\varphi_1\vee\varphi_2,u)$.
\item Let us suppose that $w\models P(\varphi_1\vee\varphi_2,u)$. Let us again distinguish two cases. If $P(\varphi_1,u)\vee P(\varphi_2,u)=\top$, then it reduces to the case where $\varphi=\top$ already treated. If $P(\varphi_1,u)\vee P(\varphi_2,u)\neq\top$, then we have either $w\models P(\varphi_1,u)$ or $w\models P(\varphi_2,u)$. Let us treat the case where $w\models P(\varphi_1,u)$ (the other case is similar). From $w\models P(\varphi_1,u)$, we can apply the structural induction hypothesis on $\varphi_1$ to obtain $u\cdot w\models \varphi_1$, and thus, using the \LTL semantics of $\vee$, $u\cdot w\models \varphi_1\vee \varphi_2$.
\end{itemize}
\item Case $\varphi=\varphi_1\wedge\varphi_2$. This case is similar to the previous one.
\item Case $\varphi=\ltlG\varphi'$. Recall that, by applying Lemma~\ref{lem:prog_trace} on $\ltlG\varphi'$ and $u$, we have $ P(\ltlG\varphi',$ $u)=\bigwedge_{i=0}^{|u|-1} P(\varphi',u^i)\wedge \ltlG\varphi'$.
\begin{itemize}
\item Let us suppose that $u\cdot w\models \ltlG\varphi'$. From the \LTL semantics of operator $\ltlG$, we have $\forall i\in\mathbb{N}^{\geq 0}.\ (u\cdot w)^i\models \varphi'$. In particular, it implies that $\forall 0\leq i\leq |u|-1.\ u^i \cdot w\models\varphi'$ and $\forall i\geq 0.\ ((u\cdot w)^{|u|-1})^i \models \varphi'$. Using, $\forall 0\leq i\leq |u|-1.\ u^i\cdot w\models\varphi'$ and applying the structural induction hypothesis on $\varphi'$ and the $u_i$'s, we obtain $\forall 0\leq i\leq |u|-1.\ w\models P(\varphi',u^i)$, and thus $w\models \bigwedge_{i=0}^{|u|-1} P(\varphi',u^i)$. Using $\forall i\geq 0.\ w^i=((u\cdot w)^{|u|-1})^i \models \varphi'$, we obtain $w\models \ltlG\varphi'$. As expected, according to the \LTL semantics of $\wedge$, we have $w\models \bigwedge_{i=0}^{|u|-1} P(\varphi',u^i)\wedge \ltlG\varphi' =P(\ltlG\varphi',u)$.
\item Let us suppose that $w\models P(\ltlG\varphi',u)$. We have $\forall 0\leq i\leq |u|-1.\ w\models P(\varphi',u^i)$ and $w\models\ltlG\varphi'$. Using the structural induction hypothesis on $\varphi'$ and the $u^i$'s, it follows that $\forall 0\leq i\leq |u|-1.\ u^i\cdot w=(u\cdot w)^i\models \varphi'$. Using the semantics of operator $\ltlG$, from $w\models \ltlG \varphi'$ and $\forall 0\leq i\leq |u|-1.\ u^i\cdot w=(u\cdot w)^i\models \varphi'$, we deduce $u\cdot w\models \ltlG\varphi'$.
\end{itemize}
\item Case $\varphi=\ltlF\varphi'$. This case is similar to the previous one.
\item Case $\varphi=\ltlX\varphi'$. Recall that, by applying Lemma~\ref{lem:prog_trace} on $u$ and $\ltlX\varphi'$, we have $P(\ltlX\varphi',u)=P(\varphi',u^1\cdot\sigma)$. Using the \LTL semantics of $\ltlX$, we have $u\cdot w\models \ltlX\varphi'$ iff $u^1\cdot w\models \varphi'$. Thus we have $u\cdot w\models \ltlX\varphi'$ iff $u^1\cdot\sigma\cdot w\models \varphi'$ iff (induction hypothesis on $\varphi'$) $w\models P(\varphi',u^1\cdot\sigma)=P(\ltlX\varphi',u)$.
\item Case $\varphi=\neg\varphi'$. Recall that, by applying Lemma~\ref{lem:prog_trace} on $u$ and $\neg\varphi'$, we have $P(\neg\varphi',u)=\neg P(\varphi',u)$. Using the \LTL semantics of operator $\neg$, we have $\forall\varphi\in\LTL. \forall w\in\Sigma^\omega.\ w\models\varphi\Leftrightarrow w\not\models \neg\varphi$. Thus, we have $u\cdot w\models \neg\varphi'$ iff $u\cdot w\not\models\varphi'$ iff (induction hypothesis on $\varphi'$) $w\not\models P(\varphi',u)$ iff $w\models \neg P(\varphi',u)$ iff $w\models P(\neg\varphi',u)$.
\item Case $\varphi=\varphi_1\ltlU\varphi_2$. Recall that, by applying Lemma~\ref{lem:prog_trace} on $u$ and $\varphi_1\ltlU\varphi_2'$, we have 
\[
P(\varphi_1\ltlU\varphi_2',u)=\bigvee_{i=0}^{|u|-1} \big(P(\varphi_2,u^i) \wedge \bigwedge_{j=0}^{i-1} P(\varphi_1,u^j)\big) \vee \bigwedge_{i=0}^{|u|-1} P(\varphi_1,u^i)\wedge \varphi_1\ltlU\varphi_2.
\]
\begin{itemize}
\item Let us suppose that $u\cdot w\models \varphi_1\ltlU\varphi_2$. According to the \LTL semantics of operator $\ltlU$, $\exists k\in\mathbb{N}^{\geq 0}.\ (u\cdot w)^k\models\varphi_2 \wedge \forall 0\leq l < k.\ (u\cdot w)^l\models\varphi_1$. Let us distinguish two cases: $k>|u|$ and $k\leq |u|$.
\begin{itemize}
\item If $k>|u|$, then we have in particular $\forall 0\leq l \leq |u|-1.\ u^l\cdot w\models\varphi_1$. Applying the structural induction hypothesis on $\varphi_1$ and the $u^l$'s, we find $\forall 0\leq l \leq |u|.\ w\models P(\varphi_1,u^l)$, i.e., $w\models \bigwedge_{l=0}^{|u|-1} P(\varphi_1,u^l)$. From $(\sigma\cdot w)^k\models \varphi_2$ and $k>|u|-1$, we deduce that $\exists k'\geq 0.\ w^{k'}\models \varphi_2$ and $k'=k-|u|+1$. Furthermore, we have $\forall 0\leq i\leq k'.\ ((u\cdot w)^{|u|-1})^{k'}= w\models P(\varphi_1,u)$, i.e., $w\models \bigwedge_{i=0}^{k'} P(\varphi_1,u^i)$. Finally, $w\models P(\varphi_1\ltlU\varphi_2,u)$.
\item If $k\leq|u|-1$, then from $(u\cdot w)^k\models\varphi_2$, we have $u^k\cdot w\models\varphi_2$. Using the induction hypothesis on $\varphi_2$ and $u^k$, we have $w\models P(\varphi_2,u^k)$. Moreover, using $\forall l\leq |k|.\ (u\cdot w)^l=u^l \cdot w\models\varphi_1$ and the induction hypothesis on $\varphi_1$ and the $u^l$'s, we obtain $\forall l\leq |k|.\ (u\cdot w)^l=w\models P(\varphi_1,u^l)$. Finally, we have $w\models \bigwedge_{l=0}^k \models P(\varphi_1,u^l) \wedge P(\varphi_2,u^k)$, and thus $w\models P(\varphi_1\ltlU\varphi_2,u)$.
\end{itemize}
\item Let us suppose that $w\models P(\varphi_1\ltlU\varphi_2,u)$. We distinguish two sub-cases:

 $P(\varphi_1\ltlU\varphi_2,u)=\top$ and $P(\varphi_1\ltlU\varphi_2,u)\neq\top$.
\begin{itemize}
\item Sub-case $P(\varphi_1\ltlU\varphi_2,u)=\top$. We distinguish again three sub-cases:
\begin{itemize}
\item Sub-case $\bigvee_{i=0}^{|u|-1} \big(P(\varphi_2,u^i) \wedge \bigwedge_{j=0}^{i-1} P(\varphi_1,u^j)\big) =\top$. Necessarily, we have $\exists 0\leq i\leq |u|-1.\ P(\varphi_2,u^i) \wedge \bigwedge_{j=0}^{i-1} P(\varphi_1,u^j)=\top$. Otherwise, that would mean that $\exists i_1,i_2\in[0,|u|-1].\ P(\varphi_2,u^{i_1}) \wedge \bigwedge_{j=0}^{i_1-1} P(\varphi_1,u^j) = \neg P(\varphi_2,u^{i_2}) \wedge \bigwedge_{j=0}^{i_2-1} P(\varphi_1,u^j)$ and we would obtain a contradiction. From $P(\varphi_2,u^i) \wedge \bigwedge_{j=0}^{i-1} P(\varphi_1,u^j)=\top$, we have $P(\varphi_2,u^i) =\top$ and $\bigwedge_{j=0}^{i-1} P(\varphi_1,u^j)=\top$. Using the induction hypothesis on $\varphi_1$ and $\varphi_2$, we obtain $u^i \cdot w\models \varphi_2$ and $\forall 0\leq j< i.\ u^j \cdot w\models \varphi_1$. According to the \LTL semantics of operator $\ltlU$, it means $u\cdot w\models\varphi_1\ltlU\varphi_2$.
\item Sub-case $\bigwedge_{i=0}^{|u|-1} P(\varphi_1,u^i)\wedge \varphi_1\ltlU\varphi_2=\top$. In this case, we have necessarily $\varphi_1\ltlU\varphi_2=\top$, and this case reduces to the case where $\varphi=\top$.
\item Sub-case $\bigvee_{i=0}^{|u|-1} \big(P(\varphi_2,u^i) \wedge \bigwedge_{j=0}^{i-1} P(\varphi_1,u^j)\big) \neq\top$ and $\bigwedge_{i=0}^{|u|-1} P(\varphi_1,$ $u^i)\wedge \varphi_1\ltlU\varphi_2\neq\top$. We have then \[\bigvee_{i=0}^{|u|-1} \big(P(\varphi_2,u^i) \wedge \bigwedge_{j=0}^{i-1} P(\varphi_1,u^j)\big) = \neg \Big(\bigwedge_{i=0}^{|u|-1} P(\varphi_1,u^i)\wedge \varphi_1\ltlU\varphi_2 \Big).\] Let us suppose that $\forall i\in\mathbb{N}^{\geq 0}.\ (u\cdot\sigma)\not\models\varphi_2$. Following the induction hypothesis on $\varphi_2$, it means in particular that $\forall 0\leq i\leq |u|-1.\ w\not\models P(\varphi_2,u^i)$. Then, since $w\models P(\varphi_2\ltlU\varphi_2)$, it would imply that $w\models \bigwedge_{i=0}^{|u|-1} P(\varphi_1,u^i)\wedge \varphi_1\ltlU\varphi_2$. But, from $w\models\varphi_1\ltlU\varphi_2$, we would obtain a contradiction according to the \LTL semantics. Hence, let us consider $i$ the minimal $k\in\mathbb{N}^{\geq0}$ s.t. $(u\cdot w)^k\models \varphi_2$. If $i>|u|-1$, then similarly we have $w\models \bigwedge_{i=0}^{|u|-1} P(\varphi_1,u^i)\wedge \varphi_1\ltlU\varphi_2$. It follows that $\forall 0\leq l\leq |u|-1.\ u^l\cdot w\models \varphi_1$ and $\forall |u|-1\leq l <i.\ (u\cdot w)^l\models \varphi_1$, and thus $u\cdot w\models\varphi_1\ltlU\varphi_2$. Else ($i\leq|u|-1$), we can follow a similar reasoning to obtain the expected result.
\end{itemize}
\item Sub-case $P(\varphi_1\ltlU\varphi_2,u)=\top$. Similarly, in this case, we can show that $\exists k\in\mathbb{N}^{\geq 0}.\ (u\cdot w)^k\models\varphi_2$. Then we consider $k_{min}$ the minimal $k$ s.t. $(u\cdot w)^k\models\varphi_2$. Then, we can show that $\forall k'<k_{min}.\ (u\cdot w)^{k'} \models\varphi_1$. And then $u\cdot w\models\varphi_1\ltlU\varphi_2$.
\end{itemize}
\end{itemize}
\end{itemize}
\qed
\end{proof}

%%% Local IspellDict: "british"
%%% Local Variables: 
%%% mode: latex
%%% TeX-master: "../main"
%%% End: 

%
%%%%%%%%%%%%%%%%%%%%%%%%%%%%%%%%%%%%%%%%%%%%%%%%%%%%%%%%%%%%%%%%%%%%%%
\paragraph{Proof for Theorem~\ref{thm:prog}.}
%%%%%%%%%%%%%%%%%%%%%%%%%%%%%%%%%%%%%%%%%%%%%%%%%%%%%%%%%%%%%%%%%%%%%%
%
We shall prove the following statement:
\newline
\[
\begin{array}{lll}
 \forall u\in\Sigma^+.\forall \varphi\in\LTL. & & v = P(\varphi, u)\\
&\Rightarrow& ( v=\top \Rightarrow u\models_3 \varphi=\top ) \wedge (v=\bot \Rightarrow u\models_3 \varphi=\bot).
% && \wedge\  v\notin\{\top,\bot\} \Rightarrow u\models_3 \varphi=?)
\end{array}
 \]
The proof uses the definition of the \LTL semantics (Definition~\ref{def:ltl_semantics}), the definition of good and bad prefixes (Definition~\ref{def:epf}), the progression function (Definition~\ref{def:ltl_progression}), and Lemma~\ref{lem:mimicsevent}.
\begin{proof}
According to Lemma~\ref{lem:mimicstrace}, we have $\forall u\in\Sigma^+.\forall w\in\Sigma^\omega.\ u\cdot w\models \varphi \Leftrightarrow w\models P(\varphi,u)$. Consequently, we have $\forall u\in\Sigma^+.\forall w\in\Sigma^\omega.\ u\cdot w\models \varphi \Leftrightarrow \forall u\in\Sigma^+.\forall w\in\Sigma^\omega.\ w\models P(\varphi,u)$ and $\forall u\in\Sigma^+.\forall w\in\Sigma^\omega.\ u\cdot w\not\models \varphi \Leftrightarrow \forall u\in\Sigma^+.\forall w\in\Sigma^\omega.\ w\not\models P(\varphi,u)$. Consequently, when $P(\varphi,u)=\top$, we have $\forall u\in\Sigma^+.\forall w\in\Sigma^\omega.\ u\cdot w\models \varphi$, i.e., $u\in\good(\varphi)$. Also, when $P(\varphi,u)=\bot$, we have $\forall u\in\Sigma^+.\forall w\in\Sigma^\omega.\ u\cdot w\not\models \varphi$, i.e., $u\in\bad(\varphi)$.
\qed

\end{proof}
%
%%%%%%%%%%%%%%%%%%%%%%%%%%%%%%%%%%%%%%%%%%%%%%%%%%%%%%%%%%%%%%%%%%%%%%
%%%%%%%%%%%%%%%%%%%%%%%%%%%%%%%%%%%%%%%%%%%%%%%%%%%%%%%%%%%%%%%%%%%%%%
\subsection{Proofs for Section~\ref{sec:sem}}
%%%%%%%%%%%%%%%%%%%%%%%%%%%%%%%%%%%%%%%%%%%%%%%%%%%%%%%%%%%%%%%%%%%%%%
%%%%%%%%%%%%%%%%%%%%%%%%%%%%%%%%%%%%%%%%%%%%%%%%%%%%%%%%%%%%%%%%%%%%%%
%
\paragraph{Proof of Corrolary~\ref{cor:reducedone}.}
We shall prove the following statement:
\[
|\mathcal{M}| = 1 \Rightarrow \forall u \in \Sigma^\ast. \forall\varphi \in \LTL.\ u \models_3 \varphi = u \models_D \varphi
\]
\begin{proof}
The proof is trivial, since in case of one component in the system, the extended progression rule~(\ref{eq:p1}) is reduced to its initial definition in the centralised case, i.e., $\forall p\in\AP{}.\forall\sigma\in\Sigma.\ P(p, \sigma,{\AP{1}})= P(p, \sigma)$. Moreover, no past goal is generated, i.e., the extended progression rule~(\ref{eq:p2}) is never applied.
\qed
\end{proof}
%
%%%%%%%%%%%%%%%%%%%%%%%%%%%%%%%%%%%%%%%%%%%%%%%%%%%%%%%%%%%%%%%%%%%%%%
%%%%%%%%%%%%%%%%%%%%%%%%%%%%%%%%%%%%%%%%%%%%%%%%%%%%%%%%%%%%%%%%%%%%%%
\subsection{Proofs for Section~\ref{sec:alg}}
\label{proofs:decentmon}
%%%%%%%%%%%%%%%%%%%%%%%%%%%%%%%%%%%%%%%%%%%%%%%%%%%%%%%%%%%%%%%%%%%%%%
%%%%%%%%%%%%%%%%%%%%%%%%%%%%%%%%%%%%%%%%%%%%%%%%%%%%%%%%%%%%%%%%%%%%%%
%
Let us first formalize a bit more Algorithm~L by introducing some additional notation.
\begin{itemize}
 \item $\send(i,t,j)\in\{\true,\false\}$ is a predicate indicating whether or not the monitor $i$ sends a formula to monitor $j$ at time $t$ with $i\neq j$. %\TODO{AB@YF: Use $\top$/$\bot$ for $true$/$false$?\\YF@AB: No I don't think so because I wanted to distinguish Boolean predicates from predicates/function returning an LTL formula. CHECK\&KILL.}
 \item $\send(i,t)\in\{\true,\false\}$ is a predicate indicating whether or not the monitor $i$ sends a formula to some monitor at time $t$.
\item $\kept(i,t)\in\LTL$ is the local obligation kept by monitor $i$ at time $t$ for the next round (time $t+1$).
\item $\received(i,t,j)\in\LTL$ is the obligation received by monitor $i$ at time $t$ by monitor $j$ with $i\neq j$.
\item $\received(i,t)\in\LTL$ is the obligation received by monitor $i$ at time $t$ from all monitors.
\item $\inlo(i,t,\varphi)\in\LTL$ is the local obligation of monitor $i$ at time $t$ when monitoring the global specification formula $\varphi$, before applying the progression functioni.e, after applying step L3 of Algorithm~L.
\item $\lo(i,t,\varphi)\in\LTL$ is the local obligation of monitor $i$ at time $t$ when monitoring the global specification formula $\varphi$ after applying the progression function, i.e, after applying step L4 of Algorithm~L.
\item $\mou(\varphi)\in\sus(\varphi)$ is the most urgent formula belonging to the set of urgent subformulae of $\varphi$.
\item $\ulo(i,t,\varphi) = \sus\big(\lo(i,t,\varphi)\big)$ is the set of urgent local obligation of monitor $i$ at time $t$ when monitoring the global specification formula $\varphi$.
\end{itemize}
Based on the previous notation and Algorithm L, we have the following relations:
\begin{itemize}
 \item $\send(i,t,j)$ is $\true$ if monitor $M_j$ is the first monitor containing the most urgent obligation contained in the local obligation of $M_i$, according to the order in $[1,m]$. Formally:
\[ 
\begin{array}{lcl}
\send(i,t,j) & = &  \left\{ 
\begin{array}{ll}
\true & \mbox{if }  M_j=\Mon\big(M_i,\Prop(\ulo(i,t,\varphi))\big)\wedge \ulo(i,t,\varphi)\neq\emptyset\\
\false   & \mbox{otherwise}
\end{array}
\right.
\end{array}
\]
 \item $\send(i,t)$ is $\true$ if monitor $M_i$ sends his local obligation to some monitor. Formally: $\send(i,t)=\exists j\in [1,n]\setminus\{i\}.\ \send(i,t,j)$.
\item $\kept(i,t)\in\LTL$ is either \# if $M_i$ sends its local obligation to some monitor at time $t-1$ or its local obligation at time $t-1$ otherwise. Formally:
\[ 
\begin{array}{lcl}
\kept(i,t) & = &  \left\{ 
\begin{array}{ll}
\# & \mbox{if } \exists j\in [1,n]\setminus\{i\}.\send(i,t-1,j) \\
\lo(i,t-1,\varphi)   & \mbox{else}
\end{array}
\right.
\end{array}
\]
\item $\received(i,t,j)$ is the local obligation of $M_j$ received by $M_i$ at time $t$ if $t\geq 1$ and $M_j$ sends actually something to $M_i$. Formally:
\[ 
\begin{array}{lcl}
\received(i,t,j)& = &  \left\{ 
\begin{array}{ll}
\lo(j,t-1,\varphi) & \mbox{if }  \exists j\in [1,n]\setminus\{i\}.\ \send(j,t-1,i) \wedge t\geq 1\\
\#   & \mbox{else}
\end{array}
\right.
\end{array}
\]
\item $\received(i,t)$ is the conjunction of all obligations received by monitor $i$ from all other monitors at time $t$. Formally:
\[
\received(i,t)=\bigwedge_{j=1,j\neq i}^{|\cal M|} \received(i,t,j)
\]
\item $\inlo(i,t,\varphi)$ is
\begin{itemize}
\item at time $t\geq 1$ what was kept by $M_i$ at time $t-1$ and the received obligation at time $t$;
\item at time $t=0$ the initial obligation, i.e., the global specification $\varphi$.
\end{itemize}
Formally:
\[
 \begin{array}{lcl}
\inlo(i,t,\varphi) & = &  \left\{ 
\begin{array}{ll}
\varphi & \mbox{if }  t=0\\
\kept(i,t-1)\wedge \received(i,t) & \mbox{else}
\end{array}
\right.
\end{array}
\]
\item $\lo(i,t,\varphi)$ is
\begin{itemize}
\item at time $t\geq 1$ the result of progressing what was kept by $M_i$ at time $t-1$ and the received obligation at time $t$ with the current local event $u_i(t)$;
\item at time $t=0$ the result of progressing the initial obligation, i.e., the global specification with the current local event $u_i(0)$.
\end{itemize}
Formally:
\[
 \begin{array}{lcl}
\lo(i,t,\varphi) & = &  \left\{ 
\begin{array}{ll}
P(\varphi,u_i(0),\AP{i}) & \mbox{if }  t=0\\
P(\kept(i,t-1)\wedge \received(i,t),u_i(t),\AP{i})   & \mbox{else}
\end{array}
\right.
\end{array}
\]
\end{itemize}
Now, we can clearly state the theorem:
\[
\forall t\in \mathbb{N}^{\geq 0}.\forall\varphi\in\LTL.\forall i\in [1,n].\forall \ltlP^d p\in \ulo(i,t,\varphi).\ d\leq \min(n, t+1)
\]
% %
\paragraph{Preliminaries to the proof.}
Let us first start with some remarks. At step L3 in Algorithm L, the local obligation of a monitor $M_i$ is defined to be $\varphi_i^t  \wedge \bigwedge_{j\in [1,m],j\neq i} \varphi_j$ where $\varphi_j$ is an obligation received from monitor $M_j$ and $\varphi_i^t$ is the local obligation kept from time $t-1$ (if $t=0$,  $\varphi_i^t=\varphi$). Let us note that the local obligation kept by the monitor from time $t-1$ to time $t$, with $t\geq 1$, are not urgent. The result should thus be established on the \emph{urgent} local obligations transmitted and rewritten by local monitors. More formally, this is stated by the following lemma.
\begin{lemma}
\label{lem:ulo}
According to Algorithm~L, we have:
\[
 \ulo(i,t,\varphi) =\bigcup_{j=1,j\neq i}^{|\cal M|} \sus\big(P(\received(i,t),u_i(t),\AP{i})\big)
\]
\end{lemma}
\begin{proof}
First let us notice that the formulae kept by any monitor $M_i$ at any time $t$ are not urgent. Indeed, we have: $\forall i\in[1,n].\forall t\in\mathbb{N}^{\geq 0}.$
\[ 
\begin{array}{lcl}
\sus(\kept(i,t)) & = &  \left\{ 
\begin{array}{ll}
\sus(\#) & \mbox{if }  \exists j\in [1,n]\setminus\{i\}.\ \send(i,t,j) \\
\sus(\lo(i,t-1,\varphi))   & \mbox{if } \sus(\lo(i,t-1,\varphi))=\emptyset 
\end{array}
\right.
\end{array}
\]
That is $\forall i\in [1,n]. \forall t\geq 0.\ \sus(\kept(i,t))=\emptyset$. Thus, $\forall i\in[1,n].\forall t\in\mathbb{N}^{\geq 0}.\forall\varphi\in\LTL.$
\[
\begin{array}{ll}
\ulo(i,t,\varphi) \\
\quad= \sus\big(P(\received(i,t),u_i(t),\AP{i})\big)\\
\quad = \sus\big(P(\bigwedge_{j=1,j\neq i}^{|\cal M|} \received(i,t,j),u_i(t),\AP{i})\big) & \text{(definition of $\received(i,t,j)$)}\\
\quad = \sus\big((\bigwedge_{j=1,j\neq i}^{|\cal M|} P(\received(i,t),u_i(t),\AP{i})) & \text{(progression on events)}\\
\quad= \bigcup_{j=1,j\neq i}^{|\cal M|} \sus\big(P(\received(i,t),u_i(t),\AP{i})\big) & \text{(definition of $\sus$)}\\
\end{array}
\]
\qed
\end{proof}

Another last lemma will be needed before entering specifically into the proof. This lemma states that if a past obligation $\ltlP^dp$ is part of a progressed formula, then the past obligation $\ltlP^{d-1}p$ is part of its un-progressed form. More formally, this is stated by the following lemma.
\begin{lemma}
\label{lem:pastoblig}
Let us consider ${\cal M}=\{M_1,\ldots,M_n\}$ where each monitor $M_i$ has a set of local atomic propositions $\AP{i}=\Pi_{i}(\AP{})$ and observes the set of events $\Sigma_i$, we have:
\[
\forall i\in[1,n]. \forall\sigma\in\Sigma_i.\forall \varphi\in\LTL. \forall \ltlP^d\in\sus\big(P(\varphi,\sigma,\AP{i})\big).\ d>1\Rightarrow \ltlP^{d-1}p\in\sus(\varphi)
\]
\end{lemma}
\begin{proof}
Let us consider $\sigma\in\Sigma,\Sigma_i\subseteq\Sigma$. The proof is done by a structural induction on $\varphi\in\LTL$.

\textit{Base Case: $\varphi\in\{\top,\bot,p'\in\AP{}\}$}
\begin{itemize}
 \item Case $\varphi=\top$. In this case, the proof is trivial since $P(\top,\sigma,\AP{i})=\top$ and $\sus(\top)=\emptyset$.
\item Case $\varphi=\bot$. This case is similar to the previous one.
\item Case $\varphi=p'\in\AP{}$. If $p'\in\AP{i}$, then $P(p',\sigma,\AP{i})\in\{\top,\bot\}$ and $\sus(P(p',\sigma,$ $\AP{i}))=\emptyset$. Else ($p'\notin \AP{i}$), $P(p',\sigma,\AP{i})=\ltlP p'$ and $\sus\big(P(p',\sigma,\AP{i})\big)=\emptyset$.
\end{itemize}
\textit{Induction Case: $\varphi\in\{\neg\varphi',\varphi_1\vee\varphi_2,\varphi_1\wedge\varphi_2,\ltlP^{d'}p',\ltlG\varphi',\ltlF\varphi',\ltlX\varphi',\varphi_1\ltlU\varphi_2\}$.} Our induction hypothesis states that the result holds for some formulae $\varphi',\varphi_1,\varphi_2\in\LTL$.
\begin{itemize}
 \item Case $\varphi=\neg\varphi'$. On one hand, we have
\[
\begin{array}{ll}
\sus\big(P(\neg\varphi',\sigma,\AP{i})\big)&=\sus\big(\neg P(\varphi',\sigma,\AP{i})\big)\\
&=\sus\big(P(\varphi',\sigma,\AP{i})\big).
\end{array}
\]
 On the other hand, we have $\sus(\neg\varphi')=\sus(\varphi')$. Thus, by applying directly the induction hypothesis on $\varphi'$, we obtain the expected result.
\item Case $\varphi=\varphi_1\vee\varphi_2$. On one hand, we have
\[
\begin{array}{ll}
\sus\big(P(\varphi_1\vee\varphi_2,\sigma,\AP{i})\big) &= \sus\big(P(\varphi_1,\sigma,\AP{i})\vee P(\varphi_2,\sigma,\AP{i})\big)\\
&=\sus\big(P(\varphi_1,\sigma,\AP{i})\big)\cup\sus\big(P(\varphi_2,\sigma,\Sigma_i)\big).
\end{array}
\]
Thus, $\ltlP^d\in\sus\big(P(\varphi_1\wedge\varphi_2,\sigma,\AP{i})\big)$ implies that $\ltlP^dp\in\sus\big(P(\varphi_1,\sigma,\AP{i})\big)$ or $\ltlP^dp\in\sus\big(P(\varphi_2,\sigma,\AP{i})\big)$. On the other hand, $\sus(\varphi_1\wedge\varphi_2)=\sus(\varphi_1)\cup\sus(\varphi_2)$. Hence, the result can be obtained by applying the induction hypothesis on either $\varphi_1$ or $\varphi_2$ depending on whether $\ltlP^dp\in\sus\big(P(\varphi_1,\sigma,\AP{i})\big)$ or $\ltlP^dp\in\sus\big(P(\varphi_2,\sigma,\AP{i})\big)$.
\item Case $\varphi=\ltlP^{d'}p'$ for some $d'\in\mathbb{N}$ and $p'\in\AP{}$. One one hand, if $p'\in\AP{i}$, then it implies that $P(\ltlP^{d'}p',\sigma,\AP{i})\in\{\top,\bot\}$. Else ($p'\notin\AP{i}$), we have $P(\ltlP^{d'}p',\sigma,$ $\AP{i})=\ltlP^{d'+1}p'$. On the other hand, we have $\sus(\ltlP^{d'}p')=\{\ltlP^{d'}p'\}$.
\item Case $\varphi=\ltlG\varphi'$. By definition of the progression rule for $\ltlG$ and the definition of $\sus$, we have 
\[
\begin{array}{l} 
\sus\big(P(\ltlG\varphi',\sigma,\AP{i})\big)\\
\qquad =\sus\big(P(\varphi',\sigma,\AP{i})\wedge\ltlG\varphi'\big)\\
\qquad =\sus\big(P(\varphi',\sigma,\AP{i})\big). 
\end{array}
\]
Since $\varphi'$ is behind a future temporal operator, the only case where $\sus\big(P(\varphi',\sigma,$ $\AP{i})\big)\neq\emptyset$ is when $\varphi'$ is a state-formula. In that case, we have $\ltlP^dp\in\sus\big(P(\varphi',\sigma,$ $\AP{i})\big)$ implies that $d=1$.
\item Cases $\varphi\in\{\ltlF\varphi',\ltlX\varphi',\varphi_1\ltlU\varphi_2\}$. These cases are similar to the previous one.
\end{itemize}
\qed
\end{proof}

%%% Local IspellDict: "british"
%%% Local Variables: 
%%% mode: latex
%%% TeX-master: "../main"
%%% End: 

%
%
\paragraph{Back to the proof of Theorem~\ref{theo:maxdelay}.}
We have to prove that for any $\ltlP^m p \in \LTL$, a local obligation
of some monitor $M_i \in \mathcal{M}$, $m\leq\min(|\mathcal{M}|,t+1)$ at
any time $t \in \mathbb{N}^{\geq 0}$. We will suppose that there are at
least two components in the system (otherwise, the proof is trivial),
i.e., $|{\cal M}|\geq 2$. The proof is done by distinguishing three
cases according to the value of $t\in\mathbb{N}^{\geq 0}$.
% \TODO{AB@YF:
%   Fixed this paragraph up to the best of my abilities, in particular,
%   replaced $d$ with $m$ to match proof.  CHECK\&KILL!}

% %%%%%%%%%%%%%%%%%%%%%%%%%%%%%%%%%%%%%%%%%%%%%%%%%%%%%%%%%%%%%%%%%%%%%%%%%%
% OLD VERSION:
% %%%%%%%%%%%%%%%%%%%%%%%%%%%%%%%%%%%%%%%%%%%%%%%%%%%%%%%%%%%%%%%%%%%%%%%%%%
% We have to prove that for any $\ltlP^d p \in \LTL$ a local obligation of
% some monitor $M_i \in \mathcal{M}$.\TODO{AB@YF: Sentence incomplete:
%   what do we have to prove?}  In the worst case\TODO{AB@YF: This is not
%   the worst-case, this is ALWAYS the case.},
% $d\leq\min(|\mathcal{M}|,t+1)$ at any time $t \in \mathbb{N}^{\geq
%   0}$. We will suppose that there are at least two components in the
% system (otherwise, the proof is trivial), i.e., $|{\cal M}|\geq 2$. The
% proof is done by distinguishing three cases according to the value of
% $t\in\mathbb{N}^{\geq 0}$.
% %%%%%%%%%%%%%%%%%%%%%%%%%%%%%%%%%%%%%%%%%%%%%%%%%%%%%%%%%%%%%%%%%%%%%%%%%%

\paragraph{\textbf{First case: $t=0$.}} In this case, we shall prove
that $m\leq 1$. The proof is done by a structural induction on
$\varphi\in\LTL$. Recall that for this case, where $t=0$, we have
$\forall i\in[1,|{\cal M}|].\ \lo(i,0,\varphi)=
P(\varphi,u_i(0),\AP{i})$.
\paragraph{Base case: $\varphi\in\{\top,\bot,p\in\AP{}\}$.}
\begin{itemize}
 \item Case $\varphi=\top$. In this case we have $\forall i\in [1,|{\cal M}|].\ \lo(i,0,\top)=P(\top,u_i(0),\AP{i})=\top$. Moreover, $\sus(\top)=\emptyset$.
\item Case $\varphi=\bot$. This case is symmetrical to the previous one.
\item Case $\varphi=p\in\AP{}$. We distinguish two cases: $p\in\AP{i}$ and $p\notin\AP{i}$. If $p\in\AP{i}$, then $\lo(i,0,p)\in\{\top,\bot\}$ and $\sus\big(\lo(i,0,p)\big)=\emptyset$. Else ($p\notin\AP{i}$), we have $\lo(i,0,p)=\ltlP p$, and $\sus\big(\lo(i,0,p)\big)=\{\ltlP p\}=\{\ltlP^1 p\}$.
\end{itemize}
\paragraph{Structural Induction Case: $\varphi\in\{\neg\varphi',\varphi_1\vee\varphi_2,\varphi_1\wedge\varphi_2,\ltlG\varphi',\ltlF\varphi',\ltlX\varphi',\varphi_1\ltlU\varphi_2\}$.} Our induction hypothesis states that the result holds for some formulae $\varphi',\varphi_1,\varphi_2\in\LTL$. 
\begin{itemize}
\item Case $\varphi=\varphi_1\vee\varphi_2$. We have:
\[
\begin{array}{ll}
 \lo(i,0,\varphi_1\vee\varphi_2) \\
\quad = P(\varphi_1\vee\varphi_2,u_i(0),\AP{i})& \text{($\lo$ definition for $t=0$)} \\
\quad = P(\varphi_1,u_i(0),\AP{i}) \vee P(\varphi_2,u_i(0),\AP{i}) & \text{(progression on events)}\\
\quad = \lo(i,0,\varphi_1) \vee \lo(i,0,\varphi_2) & \text{($\lo$ definition for $t=0$)}\\
\sus\big(\lo(i,0,\varphi_1\vee\varphi_2)\big) \\
\quad =\sus\big( \lo(i,0,\varphi_1) \vee \lo(i,0,\varphi_2)\big) &\\
\quad = \sus\big( \lo(i,0,\varphi_1)\big)\cup \sus\big(\lo(i,0,\varphi_2)\big) & \text{($\sus$ definition)}\\
\end{array}
\]
We can apply the induction hypothesis on $\varphi_1$ and $\varphi_2$ to obtain successively:
\[
\begin{array}{l}
 \forall t\geq \mathbb{N}^{\geq 0}.\forall\varphi\in\LTL.\forall \ltlP^m p\in\sus\big(\lo(i,t,\varphi_1)\big).\ m\leq 1\\
\forall t\geq \mathbb{N}^{\geq 0}.\forall\varphi\in\LTL.\forall \ltlP^m p\in\sus\big(\lo(i,t,\varphi_2)\big).\ m\leq 1 \\
\forall t\geq \mathbb{N}^{\geq 0}.\forall\varphi\in\LTL.\forall \ltlP^m p\in \sus\big(\lo(i,t,\varphi_1)\big)\cup\sus\big(\lo(i,t,\varphi_2)\big).\ m\leq 1 \\
\end{array}
\]
\item Case $\varphi=\neg\varphi'$. We have:
\[
\begin{array}{rll}
 \lo(i,0,\neg\varphi') & = P(\neg\varphi',u_i(0),\AP{i}) & \text{($\lo$ definition)}\\
& = \neg P(\varphi',u_i(0),\AP{i}) & \text{(progression on events)}\\
\sus\big(\lo(i,0,\neg\varphi')\big) & =\sus\big(\neg P(\varphi',u_i(0),\AP{i})\big) \\
& =\sus\big(P(\varphi',u_i(0),\AP{i})\big)& \text{($\sus$ definition)} \\
& = \sus\big( \lo(i,0,\varphi')\big)\\
\end{array}
\]
\item Case $\varphi=\ltlX\varphi'$. We have:
\[
\begin{array}{rll}
 \lo(i,0,\ltlX\varphi') & = P(\ltlX\varphi',u_i(0),\AP{i}) & \text{($\lo$ definition)} \\
& = \varphi' & \text{(progression on events)} \\
\sus\big(\lo(i,0,\ltlX\varphi')\big) & =\sus(\varphi') \\
\end{array}
\]
Since $\varphi'$ is behind a future temporal operator, we have $\sus(\varphi')=\emptyset$.
\item Case $\varphi=\ltlG\varphi'$. We have:
\[
\begin{array}{rll}
 \lo(i,0,\ltlG\varphi') & = P(\ltlG\varphi',u_i(0),\AP{i}) & \text{($\lo$ definition)}\\
& = P(\varphi',u_i(0),\AP{i}) \wedge  \ltlG\varphi'  & \text{(progression on events)}\\
& = \lo(i,0,\varphi') \wedge \ltlG\varphi' & \text{($\lo$ definition for $\varphi'$)} \\
\sus\big(\lo(i,0,\ltlG\varphi')\big) & =\sus\big(\lo(i,0,\varphi') \wedge \ltlG\varphi'\big)   \\
& = \sus\big( \lo(i,0,\varphi')\big)\cup \sus(\ltlG \varphi') & \text{($\sus$ definition)}\\
& = \sus\big( \lo(i,0,\varphi')\big)& \text{($\sus(\ltlG\varphi')=\emptyset$)} \\
\end{array}
\]
\item Case $\varphi=\ltlF\varphi'$. This case is similar to the previous one.
\item Case $\varphi=\varphi_1\ltlU\varphi_2$. We have:
\[
\begin{array}{l}
 \lo(i,0,\varphi_1\ltlU\varphi_2)\\
 \text{($\lo$ definition)}\\
\quad = P(\varphi_1\ltlU\varphi_2,u_i(0),\AP{i})\\
 \text{(progression on events)}\\
\quad = P(\varphi_2,u_i(0),\AP{i}) \vee \big(P(\varphi_1,u_i(0),\AP{i})\wedge \varphi_1\ltlU\varphi_2\big)\\
 \text{($\lo$ definition for $\varphi_1$ and $\varphi_2$)}\\
\quad= \lo(i,0,\varphi_2) \vee \lo(i,0,\varphi_1) \wedge \varphi_1\ltlU\varphi_2
\end{array}
\]
\[
\begin{array}{l}
\sus\big(\lo(i,0,\varphi_1\ltlU\varphi_2)\big) \\
\quad =\sus\big(  \lo(i,0,\varphi_1) \vee \lo(i,0,\varphi_2) \wedge \varphi_1\ltlU\varphi_2\big) \\
 \text{($\sus$ definition)}\\
 \quad = \sus\big( \lo(i,0,\varphi_2)\big)\cup \sus\big(\lo(i,0,\varphi_1)\big)\cup  \sus(\varphi_1\ltlU\varphi_2)\\
\text{($\sus(\varphi_1\ltlU\varphi_2=\emptyset$)}\\
\quad = \sus\big( \lo(i,0,\varphi_2)\big)\cup \sus\big(\lo(i,0,\varphi_1)\big)\\
\end{array}
\]
\end{itemize}
For $t\geq 1$, the proof is done by \emph{reductio ad absurdum}. Let us consider some $t\in\mathbb{N}$ and suppose that the theorem does not hold at time $t$. It means that:
\[
\exists\varphi\in\LTL.\exists i\in [1,|{\cal M}|]. \exists \ltlP^d p\in \ulo(i,t,\varphi).\ d> \min(|{\cal M}|, t+1).
\]
According to Lemma~\ref{lem:ulo}, since $\ulo(i,t,\varphi)=\bigcup_{j=1,j\neq i}^{|\cal M|} \sus\big(P(\received(i,t),u_i(t))\big)$, it means that $\exists j_1\in[1,|{\cal M}|]\setminus\{i\}. \ltlP^d p\in \sus\big(P(\received(i,t,j_1),u_i(t),\AP{i})\big)$. Using Lemma~\ref{lem:pastoblig}, we have $\ltlP^{d-1} p\in\sus(\received(i,t,j_1))$. It implies that $\send(j_1,t-1,i)=\true$ and $M_i=\Mon\big(M_{j_1},\Prop(\ulo(j_1,t-1,\varphi))\big)$. We deduce that $i=\min\big\{j\in[1,|{\cal M}|]\setminus\{j_1\}\mid \exists p\in\Prop(\ulo(j,t-1,\varphi)).\ p\in\AP{i}\big\}$. Moreover, from $\ltlP^{d}p\in\ulo(i,t,\varphi)$, we find $p\notin\AP{{i'}}$, with $i<i'$.

We can apply the same reasoning on $\ltlP^{d-1}p$ to find that $i<j_1<i'$ and $p\notin\Pi_{j_1}(\AP{})$. Following the same reasoning and using Lemma~\ref{lem:pastoblig}, we can find a set of indexes $\{j_1,\ldots,j_d\}$ s.t.
\[
\begin{array}{ll}
&\{j_1,\ldots,j_d\}\supseteq [1,|{\cal M}|]\\
\wedge & \forall j\in \{j_1,\ldots,j_d\}.\ p\notin \AP{j}\wedge j\in [1,|{\cal M}|]
\end{array}
\]
Moreover, due to the ordering between components, we know that $\forall k_1,k_2 \in [1,d].\ k_1 < k_2 \Rightarrow j_{k_1} < j_{k_2}$.
\paragraph{\textbf{Case $0<t< |{\cal M}|$.}}In this case we have $d >t+1$, and thus, we have $\ltlP^{d'} p\in\sus\big(\lo(j_t,0,\varphi)\big)$ with $d'>1$ which is a contradiction with the result shown for $t=0$.
\paragraph{\textbf{Case $t\geq |{\cal M}|$.}}  In this case, $\forall k_1,k_2 \in [1,d].\ k_1 < k_2 \Rightarrow j_{k_1} < j_{k_2}$ implies that $\forall j_{k_1},j_{k_2}\in \{j_1,\ldots,j_d\}.\ k_1\neq k_2\Rightarrow j_{k_1}\neq j_{k_2}$. Hence, we have $p\notin \bigcup_{j=j_1}^{j_d} \AP{j} \supseteq \AP{}$. This is impossible.
\qed

\paragraph{Proof of Theorem~\ref{theo:soundness}.} We shall prove that the decentralised monitoring algorithm is sound, i.e., whenever the decentralised monitoring algorithm yields a verdict for a given trace, then the corresponding centralized algorithm yields the same verdicts.
\paragraph{Some intermediate lemmas.}Before proving the main result of this paper, we introduce some intermediate lemmas. The following lemma extends Lemma~\ref{lem:mimicsevent} to the decentralised case, i.e., it states that the progression function mimics \LTL semantics in the decentralised case.

\begin{lemma}
\label{lem:progmimicsltlsem}
Let $\varphi$ be an \LTL formula, $\sigma\in\Sigma$ an event, $\sigma_i$ a local event observed by monitor $M_i$, and $w$ an infinite trace, we have $\sigma\cdot w\models \varphi\Leftrightarrow(\sigma\cdot w)^1\models P(\varphi,\sigma_i,\Sigma_i)$.
\end{lemma}
\begin{proof}
We shall prove that:
 \[
\begin{array}{l}
\forall i\in [1,n].\ \forall\varphi\in\LTL.\forall\sigma\in\Sigma.\forall\sigma_i\in\Sigma_i.\forall w\in\Sigma^\omega.\\
\qquad\qquad\qquad\qquad\qquad\qquad \sigma\cdot w\models \varphi\Leftrightarrow(\sigma\cdot w)^1\models P(\varphi,\sigma_i,\AP{i}).
\end{array} 
\]
The proof is done by induction on the formula $\varphi\in\LTL$. Notice that when $\varphi$ is not an atomic proposition, the lemma reduces to Lemma~\ref{lem:mimicsevent}. Thus, we just need to treat the case $\varphi=p\in\AP{}$.

If $\varphi=p\in\AP{}$. We have $\sigma\cdot w\models p\Leftrightarrow p\in\sigma$. Let us consider $i\in[1,n]$, according to the definition of the progression function (\ref{eq:p1}):
\[
\begin{array}{lcl}
P(p, \sigma_i,\AP{i}) & = &  \left\{ 
  \begin{array}{ll}
    \top & \mbox{ if }  p \in \sigma_i, \\
    \bot & \mbox{ if }  p \notin \sigma_i \wedge p \in \AP{i}, \\ 
    \ltlP p & \mbox{ otherwise},
  \end{array}
\right.
\end{array}
\]
Let us distinguish three cases.
\begin{itemize}
\item Suppose $p\in\sigma_i$. On one hand, we have $p\in\sigma$ and then $\sigma\cdot w\models p$. On the other hand, we have $P(p, \sigma_i,\AP{i})=\top$ and thus $w\models P(p, \sigma_i,\AP{i})$.
\item Suppose $p\notin\sigma_i$ and $p\in\AP{i}$. One one hand, we have $p\in\sigma$, and, because $p\in\AP{i}$ we have $p\notin\sigma$; and thus $\sigma\cdot w\not\models p$. On the other hand, we have $P(p, \sigma_i,\AP{i})=\bot$.
\item Suppose $p\notin\sigma_i$ and $p\notin\AP{i}$, we have $(\sigma\cdot w)^1\models\ltlP p\Leftrightarrow \big((\sigma\cdot w)^{-1}\big)^1\models\ltlP p\Leftrightarrow\sigma\cdot w\models p$.
\end{itemize}
\qed
\end{proof}
The following lemma states that ``the satisfaction of an LTL formula'' is propagated by the decentralised monitoring algorithm.
\begin{lemma}
\label{lem:satisfactionpropagated}
 \[
\begin{array}{l}
\forall t\in\mathbb{N}^{\geq0}.\forall i\in[1,n].\forall\varphi\in\LTL.\forall w\in\Sigma^\omega.\ \\
~~~~~~~~~~~~~~~~~~\inlo(i,t,\varphi)\neq\#\Rightarrow w\models\varphi \Leftrightarrow w^t\models \inlo(i,t,\varphi)
\end{array}
 \]
\end{lemma}
\begin{proof}
 The proof is done by induction on $t\in\mathbb{N}^{\geq 0}$.
\begin{itemize}
 \item For $t=0$, the proof is trivial since $\forall i\in [1,n].\forall\varphi\in\LTL.\ \inlo(i,0,\varphi)=\varphi$ and $w^0=w$.
\item Let us consider some $t\in\mathbb{N}^{\geq 0}$ and suppose that the lemma holds. Let us consider $i\in[1,n]$, we have:
\[
 \inlo(i,t+1,\varphi) = \kept(i,t)\wedge \received(1,t+1).
\]
Let us now distinguish four cases according to the communication performed by local monitors at the end of time $t$, i.e., according to $\send(i,t)$ and $\send(j,t,i)$, for $j\in[1,n]\setminus\{i\}$.
\begin{itemize}
\item If $\send(i,t)=\false$ and $\exists j\in[1,n]\setminus\{i\}.\ \send(j,t,i)=\true$. Then, we have:
\[
 \inlo(i,t+1,\varphi) = P\big( \inlo(i,t,\varphi) \wedge\bigwedge_{j\in J} \inlo(j,t,\varphi),u_i(t+1),\Sigma_i\big).
\]
where $\forall j\in J.\ \send(j,t,i)=\true$. Applying the definition of the progression function, we have:
\[
\begin{array}{l}
 \inlo(i,t+1,\varphi) \\
\quad= P\big( \inlo(i,t,\varphi),u_i(t+1),\Sigma_i\big) \wedge\bigwedge_{j\in J} P\big(\inlo(j,t,\varphi),u_i(t+1),\Sigma_i\big).
\end{array}
\]
Now, we have:
\[
\begin{array}{l}
 w^{t+1}\models \inlo(i,t+1,\varphi) \\
\Leftrightarrow\\
\Big(w^{t+1}\models P\big( \inlo(i,t,\varphi),u_i(t+1),\Sigma_i\big)\Big)\\
\quad\quad \wedge \Big({\forall j\in J}.\ w^{t+1}\models P\big(\inlo(j,t,\varphi),u_i(t+1),\Sigma_i\big)\Big)
\end{array}
\]
With:
\[
\begin{array}{ll}
w^{t+1}\models P\big(\inlo(i,t,\varphi),u_i(t+1),\Sigma_i\big)\\
  \quad\Leftrightarrow (w^t)^1\models P\big(\inlo(i,t,\varphi),u_i(t+1),\Sigma_i\big) & (w^{t+1} = (w^t)^1)\\
\quad \Leftrightarrow  (w(t)\cdot w^{t+1})^1\models P\big(\inlo(i,t,\varphi),u_i(t+1),\Sigma_i\big) & ((w^t)^1=(w(t)\cdot w^{t+1})^1)\\
\quad \Leftrightarrow  w^t\models \inlo(i,t,\varphi) & (\text{Induction Hypothesis})\\
\end{array}
\]
And similarly:
\[
{\forall j\in J}.\ w^{t+1}\models P\big(\inlo(j,t,\varphi),u_i(t+1),\Sigma_i\big) \Leftrightarrow w^t\models \inlo(j,t,\varphi) \\
\]
It follows that:
\[
 w^{t+1}\models \inlo(i,t+1,\varphi) \Leftrightarrow  w^t\models \inlo(i,t,\varphi) \wedge\bigwedge_{j\in J}w^t\models \inlo(i,t,\varphi).
\]
And finally:
\[
  w^{t+1}\models \inlo(i,t+1,\varphi) \Leftrightarrow  w^t\models \inlo(i,t,\varphi).
\]
\item If $\send(i,t)=\true$ and $\exists j\in[1,n]\setminus\{i\}.\ \send(j,t,i)=\true$. Then, we have:
\[
\begin{array}{rcl}
 \inlo(i,t+1,\varphi) &=& P\big(\#\wedge \bigwedge_{j\in J} \inlo(j,t,\varphi),u_i(t+1),\Sigma_i)\\
&=& P\big(\bigwedge_{j\in J} \inlo(j,t,\varphi),u_i(t+1),\Sigma_i)\\
\end{array}
\]
where $\forall j\in J.\ \send(j,t,i)=\true$. The previous reasoning can be followed in the same manner to obtain the expected result.
\item If $\send(i,t)=\false$ and $\forall j\in[1,n]\setminus\{i\}.\ \send(j,t,i)=\false$. Then, we have:
\[
 \inlo(i,t+1,\varphi) = P\big(\inlo(i,t,\varphi) ,u_i(t+1),\Sigma_i).
\]
The previous reasoning can be followed in the exact same manner to obtain the expected result.
\item If $\send(i,t)=\true$ and $\forall j\in[1,n]\setminus\{i\}.\ \send(j,t,i)=\true$. Then, we have:
\[
 \inlo(i,t+1,\varphi) = P(\#,u_i(t+1),\Sigma_i) = \#
\]
In this case, the result holds vacuously.
\end{itemize}
\end{itemize}
\qed
\end{proof}
\paragraph{Back to the proof of Theorem~\ref{theo:soundness}.} The soundness of Algorithm~L is now a straightforward consequence of the two previous lemmas (Lemmas~\ref{lem:progmimicsltlsem} and \ref{lem:satisfactionpropagated}). Indeed, let us consider $u\in\Sigma^*$ s.t. $|u|=t$. We have $u\models_D\varphi=\top$ implies that $\exists i\in[1,n].\ \lo(i,t,\varphi)=\top$ and then $\inlo(i,t+1,\varphi) = \top$. It implies that $\forall w\in\Sigma^\omega.\ w\models \inlo(i,t+1,\varphi)$. Since $|u|=t$, it follows that $\forall w\in\Sigma^\omega.\ (u\cdot w)^t\models \inlo(i,t+1,\varphi)$. Applying Lemma~\ref{lem:satisfactionpropagated}, we have $\forall w\in\Sigma^\omega.\ u\cdot w \models \varphi$, i.e., $u\models_3\varphi=\top$.

The proof for $u\models_D\varphi=\top\Rightarrow u\models_3\varphi=\top$ is similar.
\qed

\paragraph{Proof of Theorem~\ref{theo:completeness}.}
Let us first define some notations. Consider $\varphi\in\LTL, u\in\Sigma^+, i\in[1,|{\cal M}|]$:
\begin{itemize}
 \item $\rp(\varphi,u)$ is the formula $\varphi$ where past sub-formulas are removed and replaced by their evaluations using the trace $u$. Formally:
\[
 \begin{array}{ll}
  \rp(\varphi,u,i) &= \mathtt{match}\; \varphi\; \mathtt{with}\\
&
\begin{array}{ll}
 \mid \ltlP^d p &\rightarrow \left\{
\begin{array}{ll}
\top & \text{ if } p \in u(|u|-d)\\
\bot &\text{ otherwise }\\
\end{array}
\right.\\
 \mid\varphi_1 \wedge \varphi_2& \rightarrow \rp(\varphi_1,u) \wedge \rp(\varphi_2,u)\\
 \mid\varphi_1 \vee \varphi_2& \rightarrow \rp(\varphi_1,u) \vee \rp(\varphi_2,u)\\
\mid \neg\varphi' & \rightarrow \neg \rp(\varphi',u)\\
\mid \_ & \rightarrow \varphi
 \end{array}
\end{array}
\]
\item $\rp(\varphi,u,i)$ is the formula $\varphi$ where past sub-formulas are removed (if possible) and replaced by their evaluations using only the sub-trace $u_i$ of $u$.
\[
 \begin{array}{ll}
  \rp(\varphi,u,i) &= \mathtt{match}\; \varphi\; \mathtt{with}\\
&
\begin{array}{ll}
 \mid \ltlP^d p & \rightarrow \left\{
\begin{array}{ll}
\top & \text{ if } p \in u(|u|-d)\\
\bot &\text{ if } p \notin u(|u|-d) \text{ and } p\in \AP{i} \\
\ltlP^d p & \text{ otherwise }
\end{array}
\right. \\
 \mid\varphi_1 \wedge \varphi_2& \rightarrow \rp(\varphi_1,u,i) \wedge \rp(\varphi_2,u,i)\\
 \mid\varphi_1 \vee \varphi_2& \rightarrow \rp(\varphi_1,u,i) \vee \rp(\varphi_2,u,i)\\
\mid \neg\varphi' & \rightarrow \neg \rp(\varphi',u,i)\\
\mid \_ & \rightarrow \varphi
 \end{array}
\end{array}
\]
\end{itemize}
The following lemma exhibits some straightforward properties of the function $\rp$.
\begin{lemma}
\label{lem:proprp}
 Let $\varphi$ be an \LTL formula, $u\in\Sigma^+$ be a trace of length $t+1$, $i\in[1,|{\cal M}|]$ a monitor of one of the component, $u_i(t)\in\Sigma_i$ the last event of $u$ on component $i$, we have:
\begin{enumerate}
 \item $\rp\big(P(\varphi,\sigma_i,\AP{i}), u\big) = \rp\big(P(\rp(\varphi,u(0)\cdots u(t-1)),\sigma_i,\AP{i}),u\big)$;
\item $\rp\big(P(\varphi,\sigma_i,\AP{i}),u\big) = P(\varphi,u(t),\AP{})$;
\item $P(\varphi,u_i(t),\AP{i}) = P\big(\rp(\varphi,u(0)\cdots u(t-1),i), u_i(t),\AP{i}\big)$;
\item $\bigcup_{\varphi'\in\sus(\varphi)} \Prop(\varphi') \subseteq \AP{i} \Rightarrow \rp(\varphi,u,i) = \rp(\varphi,u)$.
\item For $\{i_1,\ldots,i_n\}= [1,|{\cal M}|]$: $\rp(\rp(\ldots \rp(\varphi,u,i_1),\ldots),u,i_n) = \rp(\varphi,u)$.
\end{enumerate}
\end{lemma}
\begin{proof}
 The proofs of these properties can be done by induction on $\varphi\in\LTL$ and follow directly from the definitions of $\rp$ and the progression function.
\qed
\end{proof}

\begin{lemma}
\label{lem:removingpastinlocaloblig}
Any current local obligation where past sub-formulas have been evaluated using the trace read so far is equal to the initial obligation progressed with this same trace read so far. Formally:
 \[
\begin{array}{l}
  \forall u\in\Sigma^+. \forall i\in[1,|{\cal M}|].\forall t\in\mathbb{N}^*.\\ 
\qquad\qquad\qquad |u|=t+1\wedge \lo(i,t,\varphi)\neq\# \Rightarrow \rp(\lo(i,t,\varphi),u) = P(\varphi,u).
\end{array} 
\]
\end{lemma}
\begin{proof}
 We shall prove this lemma by induction on $t\in\mathbb{N}^*$. Let us consider some component $M_i$ where $i\in [1,|{\cal M}|]$.
\begin{itemize}
 \item For $t=0$. In this case, $|u|=1$ and we have $\rp(\lo(i,0,\varphi),u) = \rp\big(P(\varphi,\sigma_i,\AP{i})\big)$ where $\sigma_i = \Pi(u(0))$. We can obtain the expected result by doing an induction on $\varphi\in\LTL$ where the only case interesting case is $\varphi=p\in \AP{}$. According to the definition of the progression function, we have:
\newline
$\begin{array}{lcl}
P(p, \sigma_i,{\AP{i}}) & = &  \left\{ 
  \begin{array}{ll}
    \top & \mbox{ if }  p \in \sigma_i, \\
    \bot & \mbox{ if }  p \notin \sigma_i \wedge p \in \AP{i}, \\ 
    \ltlP p & \mbox{ otherwise},
  \end{array}
\right.
\end{array}
$
\newline
Moreover, $p\in\sigma_i$ implies $p\in u(0)$ and $p\notin\sigma_i$ with $p\in \AP{i}$ implies $\forall j\in[1,|{\cal M}|].\ p\notin \Pi_j(u(0))$, i.e., $p\notin u(0)$.

On one hand, according to the definition of $\rp$, we have:\newline
$
\begin{array}{lcl}
\rp(\ltlP p, u(0)) & = &  \left\{ 
  \begin{array}{ll}
    \top & \mbox{ if }  p \in u(0), \\
    \bot & \mbox{ if }  p \notin u(0). \\ 
    \end{array}
\right.
\end{array}
$
\newline

Thus, we have:
\newline
$\begin{array}{lcl}
\rp\big(P(p, \sigma_i,{\AP{i}})\big) & = &  \left\{ 
  \begin{array}{ll}
    \top & \mbox{ if }  p \in u(0), \\
    \bot & \mbox{ if }  p \notin u(0). \\ 
  \end{array}
\right.
\end{array}
$
\newline
On the other hand, according to the definition of the progression function, we have:\newline
$
\begin{array}{lcl}
P(\varphi, u(0)) & = &  \left\{ 
  \begin{array}{ll}
    \top & \mbox{ if }  p \in u(0), \\
    \bot & \mbox{ if }  p \notin u(0). \\ 
    \end{array}
\right.
\end{array}
$
\item Let us consider some $t\in\mathbb{N}^*$ and suppose that the property holds. We have:
\[
 \lo(i,t+1,\varphi)= P\big(\kept(i,t)\wedge \received(i,t),u_i(t+1),\AP{i}\big).
\]
Similarly to the proof of Lemma~\ref{lem:satisfactionpropagated}, let us distinguish four cases according to the communication that occurred at the end of time $t$.
\begin{itemize}
\item If $\send(i,t)=\false$ and $\forall j\in[1,|{\cal M}|]\setminus\{i\}.\ \send(j,t,i)=\false$. Then, we have:
\[
\lo(i,t+1,\varphi) = P(\lo(i,t\varphi),u_i(t+1),\AP{i})
\]
Let us now compute $\rp(\lo(i,t+1,\varphi),u(0)\cdots u(t+1))$:
\[
 \begin{array}{ll}
  \rp(\lo(i,t+1,\varphi),u(0)\cdots u(t+1)) & = \rp(P(\lo(i,t,\varphi), u_i(t+1),\AP{i}),u(0)\cdots u(t+1))\\
&\text{(Lemma~\ref{lem:proprp}, item 1)}\\
&= \rp(P(\rp(\lo(i,t,\varphi),u(0)\cdots u(t)), u_i(t+1),\AP{i}),u(0)\cdots u(t+1)) \\
& \text{(induction hypothesis)}\\
&= \rp(P(P(\varphi,u(0)\cdots u(t)),u_i(t+1),\AP{i}),u(0)\cdots u(t+1))\\
& \text{(Lemma~\ref{lem:proprp}, item 2)}\\
& =P(P(\varphi,u(0)\cdots u(t)),u(t+1),\AP{})\\
&\text{($P(\varphi,u(0)\cdots u(t))$ is a future formula)}\\
& = P(\varphi,u(0)\cdots u(t+1))
 \end{array}
\]
\item If $\send(i,t)=\true$ and $\exists j\in[1,|{\cal M}|]\setminus\{i\}.\ \send(j,t,i)=\true$. Then, we have:
\[
\lo(i(i,t+1,\varphi) = P(\bigwedge_{j\in J}\lo(j,t\varphi),u_i(t+1),\AP{i})
\]
s.t. $\forall j\in J.\ \send(j,t,i)=\true$. Then:
\[
\begin{array}{l}
 \rp(\lo(i,t+1,\varphi),u(0)\cdots u(t+1)) \\
\quad= \rp (P(\bigwedge_{j\in J}\lo(j,t\varphi),u_i(t+1),\AP{i}),u(0)\cdots u(t+1)) \\
\text{(definition of the progression function)} \\
\quad = \rp (\bigwedge_{j\in J} P(\lo(j,t\varphi),u_i(t+1),\AP{i}),u(0)\cdots u(t+1)) \\
 \text{(definition of $\rp$)}\\
 \quad= \bigwedge_{j\in J} \rp ( P(\lo(j,t\varphi),u_i(t+1),\AP{i}),u(0)\cdots u(t+1))\\
\text{(Lemma~\ref{lem:proprp}, item 1)} \\
 \quad= \bigwedge_{j\in J} \rp ( P(\rp(\lo(j,t\varphi),u(0)\cdots u(t)),u_i(t+1),\AP{i}),u(0)\cdots u(t+1)) \\
\text{(induction hypothesis)}\\
 \quad= \bigwedge_{j\in J} \rp ( P(P(\varphi,u(0)\cdots u(t)),u_i(t+1),\AP{i}),u(0)\cdots u(t+1)) \\
\text{(Lemma~\ref{lem:proprp}, item 2)}\\
\quad = \bigwedge_{j\in J} \rp ( P(\varphi,u(0)\cdots u(t)\cdot u(t+1) ) ) \\
\text{($P(\varphi,u(0)\cdots u(t+1))$ is a future formula)} \\
\quad =  \bigwedge_{j\in J} P(\varphi,u(0)\cdots u(t+1)) 
\quad = P(\varphi,u(0)\cdots u(t+1))
\end{array}
\]
\item If $\send(i,t)=\false$ and $\exists j\in[1,|{\cal M}|]\setminus\{i\}.\ \send(j,t,i)=\true$. Then, we have:
\[
\begin{array}{l}
\lo(i,t+1,\varphi) = P\big(\lo(i,t,\varphi)\wedge \bigwedge_{i\in J} \lo(j,t,\varphi), u_i(t+1),\AP{i}\big)\\
\quad = P\big(\lo(i,t,\varphi), u_i(t+1),\AP{i}\big) \wedge P\big(\bigwedge_{i\in J} \lo(j,t,\varphi), u_i(t+1),\AP{i}\big)
\end{array}
\]
where $\forall j\in J.\ \send(j,t,i)=\true$. The proof this case is just a combination of the proofs of the two previous cases.
\item If $\send(i,t)=\true$ and $\forall j\in[1,|{\cal M}|]\setminus\{i\}.\ \send(j,t,i)=\false$. Then, we have: $\lo(i,t+1,\varphi) =\#$. The result holds vacuously.
\end{itemize}
\end{itemize}
\qed
\end{proof}
\paragraph{Back to the proof of Theorem~\ref{theo:completeness}.}
The remainder of the proof consists intuitively in showing that in a given architecture, we can take successively two components and merge them to obtain an equivalent architecture in the sense that they produce the same verdicts. The difference is that if in the merged architecture a verdict is emitted, then, in the non-merged architecture the same verdict will be produced with an additional delay. 
\begin{lemma}
\label{lem:twocomponentverdicts}
 In a two-component architecture, if in the centralised case a verdict is produced for some trace $u$, then, in the decentralised case, one of the monitor will produce the same verdict. Formally:
\[
 \forall \varphi\in\LTL.\forall u\in\Sigma^+.\ P(\varphi,u) = \top/\bot \Rightarrow \forall\sigma\in\Sigma^\ast.\exists i\in [1,2].\ \lo(i,|u\cdot\sigma|,\varphi) = \top/\bot.
\]
\end{lemma}
\begin{proof}
Let us consider a formula $\varphi\in\LTL$ and a trace $u\in\Sigma^+$ s.t. $|u|=t$. We shall only consider the case where $P(\varphi,u)=\top$. The other case is symmetrical.  Let us suppose that $\lo(1,t,\varphi)\neq\top$ and $\lo(2,t,\varphi)\neq\top$ (otherwise the results holds immediately). Because of the correctness of the algorithm (Theorem~\ref{theo:soundness}), we know that $\lo(1,t,\varphi)\neq\bot$ and $\lo(2,t,\varphi)\neq\bot$. Moreover, according to Lemma~\ref{lem:removingpastinlocaloblig}, we have necessarily that $\lo(1,t,\varphi)$ and $\lo(2,t,\varphi)$ are urgent formulas: $\Upsilon(\lo(1,t,\varphi))>0$ and $\Upsilon(\lo(2,t,\varphi))>0$. Since, there are only two components in the considered architecture, we have $\bigcup_{\varphi'\in\sus(\lo(1,t,\varphi))} \Prop(\varphi')\subseteq\AP{2}$ and $\bigcup_{\varphi'\in\sus(\lo(2,t,\varphi))}\Prop(\varphi')\subseteq\AP{1}$. According to Algorithm~L, we have then $\send(1,t-1,2) = \true$ and $\send(2,t-1,\varphi)=\true$. Then $\inlo(1,t,\varphi) = \lo(2,t-1,\varphi) \wedge \# = \lo(2,t-1,\varphi)$. Hence: $\lo(1,t,\varphi)= P(\lo(2,t-1,\varphi), u_1(t),\AP{1})$. According to Lemma~\ref{lem:proprp} item 4, we have $\lo(1,t,\varphi)= P(\rp(\lo(2,t-1,\varphi),u(0)\cdots u(t),1),u_1(t),\AP{1})$. Since \[\bigcup_{\varphi'\in\sus(\lo(2,t,\varphi))}\Prop(\varphi')\subseteq\AP{1},\] we have $\rp\big(\lo(2,t-1,\varphi),u(0)\cdots u(t),1\big) = \rp\big(\lo(2,t-1,\varphi),u(0)\cdots u(t)\big)$. It follows that:
\newline
\[
\begin{array}{lll}
\lo(1,t,\varphi)& = P(\rp(\lo(2,t-1,\varphi),u(0)\cdots u(t)), u_1(t),\AP{1}) & \\
&= P( P(\varphi,u(0)\cdots u(t)), u_1(t), \AP{1}) & \text{(Lemma~\ref{lem:removingpastinlocaloblig})}\\
&= P(\top,u_1(t),\AP{1}) = \top
\end{array}
\]
Symmetrically, we can find that $\lo(2,t,\varphi) = \top$.
\qed
\end{proof}
Given two components $C_1$ and $C_2$ with two monitors attached $M_1$ and $M_2$ observing respectively two partial traces $u_1$ and $u_2$ of some global trace $u$. The alphabets of $C_1$ and $C_2$ are $\Sigma_1$ and $\Sigma_2$ respectively. Consider the architecture ${\cal C}= \{C_1,C_2\}$ with the set of monitors ${\cal M} = \{M_1,M_2\}$. Let us define the new component $\merge(C_1,C_2)$ that produces events in $\Sigma_1\cup\Sigma_2$. To the component $\merge(C_1,C_2)$ is attached a monitor $M$ observing events in the same alphabet. Now let us consider the architecture ${\cal C}'= \{\merge(C_1,C_2)\}$ which is a one-component architecture with the set of monitors ${\cal M}' =\{\merge(M_1,M_2)\}$.

We can parameterise the satisfaction relation of $\LTL$ formula according to the considered architecture. The relation $\models_D$ becomes $\models_D^{\cal M}$ where ${\cal M}$ is the considered architecture. The definition of $\models_D^{\cal M}$ is the same as the definition of $\models_D$ (Definition~\ref{def:dltl}).

\begin{lemma}
\label{lem:mergingarchi1}
 For a monitoring architecture ${\cal M} = \{M_1,M_2\}$ and the monitoring architecture ${\cal M}' =\{\merge(M_1,M_2)\}$ where monitors of ${\cal M}$ have been merged, we have:
\[
 \forall u\in\Sigma^+.\forall \varphi\in\LTL.\ u\models_D^{\cal M}\varphi = \top/\bot \Rightarrow \forall\sigma\in\Sigma^+.\ u\cdot\sigma\models_D^{{\cal M}'} \varphi = \top/\bot.
\]
\end{lemma}
\begin{proof}
 This is a direct consequence of Lemma~\ref{lem:twocomponentverdicts} and Corollary~\ref{cor:reducedone}. Indeed, ${\cal M}'$ is a one-component architecture, thus $u\models_D^{{\cal M}'}\varphi = \top/\bot$ implies $u\models_3\varphi = \top/\bot$, i.e., $P(\varphi,u) = \top/\bot$. Now, since ${\cal M}$ is a two-component architecture, using Lemma~\ref{lem:twocomponentverdicts}, for all $\sigma\in\Sigma$,  there exists $i\in [1,|{\cal M}|]$ s.t. $\lo(i,|u\cdot\sigma|,\varphi) = \top/\bot$. That is $u\cdot\sigma\models_D^{\cal M} \varphi = \bot/\top$.
\qed
\end{proof}
The following lemma relates verdict production in a $n$-component architecture and in the same architecture where the two components with the lowest priority have been merged.
\begin{lemma}
\label{lem:mergingarchi2}
 Let ${\cal M}$ be a $n$-component architecture, with $n\geq2$ such that the priority between components is $M_1 < M_2<\ldots<M_n$, i.e., $M_1$ and $M_2$ are the two components with the lowest priority\footnote{Here, without loss of generality, we assume that monitors have been sorted according to their index. If this hypothesis does not hold initially, the indexes of components can be re-arranged so that this hypothesis holds.}. Let us consider the architecture ${\cal M}'=\{\merge(M_1,M_2),M_3,\ldots,M_n\}$, then we have:
\[
 \forall u\in\Sigma^+.\forall\varphi\in\LTL.\ u\models_D^{{\cal M}'}\varphi = \top/\bot \Rightarrow \forall\sigma\in\Sigma.\ u\cdot\sigma\models_D^{\cal M} = \top/\bot.
\]
\end{lemma}
\begin{proof}
We give a proof for the case where the verdict is $\top$ (the other case is symmetrical). Let us consider $u\in\Sigma^+, \varphi\in\LTL$ s.t. $u\models_D^{{\cal M}'}\varphi = \top$. Let $u'$ be the smallest prefix of $u$ s.t. $P(\varphi,u' )= \top$. From the theorem about the maximal delay (Theorem~\ref{theo:maxdelay}, we have that $|u|-|u'|\leq (n-1)$. Now each of the local obligations in the architecture ${\cal M}'$ will transit through at most $n$ monitors following the ordering between components. That is, in the worst case (i.e., if a verdict is not produced before time $|u|$), any obligation will be progressed according to all components. More precisely, each time a local obligation is progressed on some component $C_i$, past obligations w.r.t. component $C_i$ are removed (Lemma~\ref{lem:proprp} - item 3). Using the compositionality of $\rp$ and the progression function on conjunction, in the worst case the local obligation at time $|u'| +n$ will be a conjunction of formulas of the form 
\begin{small}
\[
\begin{array}{l}
P(\\
\quad\cdots \\
\quad\cdots P( \\
\quad\quad \quad\quad P( \rp (\cdots \rp(\rp(\varphi,u',i),u',i_1)\cdots,u',i_n),  u_i(|u'|), \AP{i})\\
\quad\quad, u_{i_1}(|u'|+1, \AP{i_1}),\\
\quad\cdots,\\
u_{i_n}(|u'|+n), \AP{i_n})\\
\end{array}
\]
\end{small} where $\varphi$ is a local obligation at time $|u'|$ and $\{i_1,\ldots,i_n\} \supseteq [1,|{\cal M'}|]$ (because of the ordering between components). Now according to Lemma~\ref{lem:proprp} - item 5:
\[
\rp (\cdots \rp(\rp(\varphi,u',i),u',i_1)\cdots,i_n) = \rp(\varphi,u') = \top.
\] 
Following the definition of the progression function for $\top$, we have that necessarily, the resulting local obligation at time $|u'|+n$ is $\top$.
\qed
\end{proof}
\begin{lemma}
\label{lem:mergingarchi3}
 Let ${\cal M}$ be a $n$-component architecture, with $n\geq2$ such that the priority between components is $M_1 < M_2<\ldots<M_n$. Let us consider the architecture ${\cal M}'=\{\merge(M_n, \merge(\ldots,\merge(M_2,M_1)\}$, then we have:
\[
 \forall u\in\Sigma^+.\forall\varphi\in\LTL.\ u\models_D^{{\cal M}'}\varphi = \top/\bot \Rightarrow \forall u'\in\Sigma^+.\ |u'|\geq n\Rightarrow u\cdot u'\models_D^{\cal M} = \top/\bot.
\]
\end{lemma}
\begin{proof}
 By an easy induction on the number of components merged using Lemma~\ref{lem:mergingarchi2}.
\qed
\end{proof}

\paragraph{Back to the proof of Theorem~\ref{theo:completeness}.} Based on the previous results, we can easily show Theorem~\ref{theo:completeness}.
\begin{proof}
 Let us consider an $n$-component architecture ${\cal M} = \{M_1,\ldots,M_n\}$, a trace $u\in\Sigma^+$ and a formula $\varphi\in\LTL$. Let us suppose that $u\models_3\varphi = \top/\bot$. As the alphabets of monitors are respectively $\Sigma_1,\ldots\Sigma_n$ and each monitor $M_i$ is observing a sub-trace $u_i$ of $u$ where the hypothesis about alphabets partitionning mentioned in Section~\ref{sec:prelim} holds, we can consider the architecture ${\cal M}'=\{\merge(M_n, \merge(\ldots,\merge(M_2,M_1)\}$ where there is a unique monitor $M$ observing the same trace $u$. Now, since ${\cal M}'$ is a one-component architecture, from $u\models_3\varphi = \top/\bot$, by Corollary~\ref{cor:reducedone}  we get $u\models_D\varphi = \top/\bot$. Using Lemma~\ref{lem:mergingarchi2}, we obtain that $\forall u'\in\Sigma^+.\ u\cdot u'\models_D^{\cal M} = \top/\bot$.
\qed
\end{proof}

\end{document}